\documentclass{article}

\usepackage{color}
\usepackage{graphicx}
\usepackage{amsmath}
\usepackage{amsfonts}
\usepackage{amssymb}
\usepackage{bm}
\usepackage{pdflscape}
\usepackage{url}
\usepackage{a4wide}

\usepackage[version=3]{mhchem}
\usepackage{chemarr}

\usepackage[version=3]{mhchem}

\title{Approximation and inference methods for stochastic biochemical kinetics - a tutorial review}
\usepackage{authblk}
\author[1,2,3]{David Schnoerr}
\author[2,3]{Guido Sanguinetti}
\author[1,3,*]{Ramon Grima}

\affil[1]{School of Biological Sciences, University of Edinburgh, Edinburgh EH9 3JH, UK}
\affil[2]{School of Informatics, University of Edinburgh, Edinburgh EH8 9AB, UK}
\affil[3]{SynthSys, University of Edinburgh, Edinburgh EH9 3JD, UK}
\affil[*]{ramon.grima@ed.ac.uk}

\let\OLDthebibliography\thebibliography
\renewcommand\thebibliography[1]{
  \OLDthebibliography{#1}
  \setlength{\parskip}{0pt}
  \setlength{\itemsep}{0pt plus 0.3ex}
}

\begin{document}

\maketitle

\begin{abstract}

Stochastic fluctuations of molecule numbers are ubiquitous in biological systems. Important examples include gene expression and enzymatic processes in living cells. Such systems are typically modelled as chemical reaction networks whose dynamics are governed by the Chemical Master Equation. Despite its simple structure, no analytic solutions to the Chemical Master Equation are known for most systems. Moreover, stochastic simulations are computationally expensive, making systematic analysis  and statistical inference a challenging task. Consequently, significant effort has been spent in recent decades on the development of efficient approximation and inference methods. This article gives an introduction to basic modelling concepts as well as an overview of state of the art methods. First, we motivate and introduce deterministic and stochastic methods for modelling chemical networks, and give an overview of simulation and exact solution methods. Next, we discuss several approximation methods, including the chemical Langevin equation, the system size expansion, moment closure approximations, time-scale separation approximations and hybrid methods. We discuss their various properties and review recent advances and remaining challenges for these methods. We present a comparison of several of these methods by means of a numerical case study and highlight some of their respective advantages and disadvantages. Finally, we discuss the problem of inference from experimental data in the Bayesian framework and review recent methods developed the literature. In summary, this review gives a self-contained introduction to modelling, approximations and inference methods for stochastic chemical kinetics.\\

%

\end{abstract}

\tableofcontents

\section{Introduction}

Understanding the functioning of living cells and biological organisms at the system level has gained increasing attention in recent years and defines a key research programme for the next decades. Experimental techniques are developing at breathtaking speed producing a wealth of data at finer and finer resolutions. However, such experimental data does not by itself reveal the function of such biological systems. The underlying processes typically involve large numbers of interacting components giving rise to highly complex behaviour \cite{eldar2010functional}. Moreover, 
experimental data are generally corrupted by measurement noise and incomplete, thus posing the mathematical and statistical challenge to infer the relevant biological information from such measurements.

We focus here on mathematical and statistical modelling of chemical reaction networks in biological systems in which \emph{random fluctuations} of molecule numbers play an important role. Recent experiments have shown this to be the case in many biological processes, gene expression being a prominent example \cite{Elowitz2002}. Such random fluctuations or \emph{stochastic effects} in gene expression have been found to lead to dramatically differing behaviours of genetically identical cells. From a modelling point of view, stochastic systems are considerably harder to analyse than their deterministic counterparts. 

The \emph{rate equations} give a macroscopic, deterministic description of the dynamics of chemical reaction networks. They consist of a set of ordinary differential equations governing the dynamics of the mean concentrations of the species in a system, thereby ignoring stochastic fluctuations. The rate equations have been successfully applied to various problems and have the advantage of being relatively straightforward to analyse. They typically give an accurate description of systems with large numbers of molecules, which is generally the case in \emph{in vitro} experiments. However, the rate equations do no longer provide a valid description in cases where the effects of stochastic fluctuations become significant. This is typically the case when some species in a system occur at low molecule numbers, a common feature of chemical networks in cells. In this case, the
\emph{Chemical Master Equation} (CME) constitutes the accepted probabilistic description of the resulting \emph{stochastic process} \cite{Gillespie1992}. In this framework, the state of a system is given by the molecule numbers of different species, and the CME governs the single-time probability of being in a certain state at a given time. 

Despite its simple structure, no analytic solutions to the CME are known for all but the simplest systems. It is however possible to simulate exact sample paths of the underlying stochastic process, by means of the \emph{stochastic simulation algorithm} \cite{Gillespie1976,Gillespie1977}. The latter allows us to draw exact samples from the process. However, since the stochastic simulation algorithm simulates each and every chemical reaction event in a system, it is computationally expensive and quickly becomes infeasible for larger systems. Accordingly, significant effort has been spent in recent decades on the development of approximation methods of the CME, and a large variety of different methods has emerged. Similarly, due to the growing amount of experimental data becoming available, a considerable amount of work has been devoted to the development of statistical inference methods for such data, i.e., methods that allow to calibrate a model to observational data. 

There exists a significant amount of literature on modelling of stochastic chemical kinetics. However, most reviews are mainly concerned with simulation based methods \cite{turner2004stochastic,el2005stochastic, li2008algorithms,pahle2009biochemical,Mauch2011, Gillespie2013,szekely2014stochastic}. Others are rather technical and require a high level of pre-existing and/or mathematical knowledge \cite{nicolis1977self, VanKampen2007,Gardiner2009,goutsias2013markovian,weber2016master}. None of these references gives a thorough introduction into modelling of stochastic chemical kinetics in the biological context that is accessible for non-experts, neither do they give an overview or comparison of state of the art approximation and inference methods.

With this article, we aim at filling these gaps in several ways: 
\begin{enumerate}
\item We give a self-contained introduction to deterministic and stochastic modelling techniques for chemical reaction networks.  First, we introduce and  give a historic motivation to the deterministic rate equations. Next, we derive and give a detailed discussion of the CME. We review exact solution methods for the CME as well as simulation methods.  
\item We give a detailed derivation and discussion of the following approximation methods of the CME: the chemical Langevin equation and its associated chemical Fokker-Planck equation; the system size expansion; moment closure approximations. Moreover, we give an introduction and overview of other types of approximations, including time-scale separation based methods and  hybrid approximations.
\item We perform a numerical case study comparing the various approximation methods mentioned before. 
\item We give an introduction to inference methods in the Bayesian framework and review existing methods from the literature. 
\end{enumerate}
The presentation is written to be accessible for non-experts that are new to the field of stochastic modelling.

Even though this review is motivated by stochastic effects in systems biology, it is important to stress that many systems in other scientific fields are frequently modelled by means of Master Equations. The methods discussed in this article can therefore readily be applied to such systems. Examples include ecology \cite{blythe2007stochastic, datta2010jump, black2012stochastic},  epidemiology \cite{bartlett1949some, bailey1950simple,keeling2008methods, black2010stochasticity, jenkinson2012numerical, pastor2015epidemic}, social sciences \cite{weidlich1991physics, weidlich2002sociodynamics, bonabeau2002agent} and neuroscience \cite{ohira1993master, el2009master, benayoun2010avalanches, buice2010systematic, wallace2011emergent, leen2012stochastic, goychuk2015stochastic}. 

This article is structured as follows. We start by discussing the  importance of stochasticity in biological systems in Section \ref{sec_pre_bio}. We describe the underlying mechanisms giving rise to stochastic fluctuations in cells (Section \ref{sec_emergence_stoch}) and discuss key experimental studies that measured such fluctuations (Section \ref{sec_exp_evidence}). Next in Section \ref{sec_stoch_chem} we discuss deterministic and stochastic modelling methods for chemical kinetics. We start by introducing the concept of chemical reaction networks and deterministic descriptions in terms of macroscopic rate equations in Section \ref{sec_chemical_networks}. Next, we  introduce stochastic modelling techniques in terms of the CME and stochastic simulation algorithms in Sections  \ref{sec_stochkin_cme} and \ref{sec_ssa}, respectively. We discuss analytic solution methods for certain classes of reaction systems in Section \ref{sec_exact_solutions}. Section \ref{sec_approximations} is devoted to approximation methods of the CME. We give detailed introductions and discussions of the chemical Langevin equation (Section \ref{sec_stochkin_cle}), the system size expansion (Section \ref{sec_sse}) and moment closure approximations (Section \ref{sec_stochkin_ma}). Next, we discuss how approximate moment values obtained from these methods can be used to construct probability distributions by means of the maximum entropy principle in Section \ref{sec_max_entropy}. In Section \ref{sec_software} we review existing software packages implementing the approximation methods discussed in Sections \ref{sec_stochkin_cle}-\ref{sec_stochkin_ma}. We give an introduction to other approximation methods in Section \ref{sec_other_approximations}, including the finite state projection algorithm, time-scale separation based approximations and hybrid methods. In Section \ref{sec_comparison} we perform a numerical case study and compare the chemical Langevin equation, the system size expansion and moment closure approximations.  Section \ref{sec_inference} deals with the problem of inference for CME type systems from observational data. We introduce the Bayesian approach to this problem and review existing methods from the literature. Finally, we summarise and conclude in Section \ref{sec_conclusions}.

\section{Stochasticity in biological systems}\label{sec_pre_bio}

Stochastic effects play an important role in many chemical reaction networks in living cells. Examples are enzymatic/catalytic processes, transduction of external signals to the interior of cells or the process of gene expression, to name just a few.  Here we discuss different sources of stochasticity and illustrate their  emergence in biochemical networks using gene expression as an example in Section \ref{sec_emergence_stoch}. Subsequently in Section \ref{sec_exp_evidence} we explain how the different sources of stochasticity can be measured and distinguished experimentally and highlight the importance of stochasticity for living cells.

\subsection{Emergence of stochasticity}\label{sec_emergence_stoch}

\begin{figure} [t]
\centering
  \includegraphics[scale=0.7]{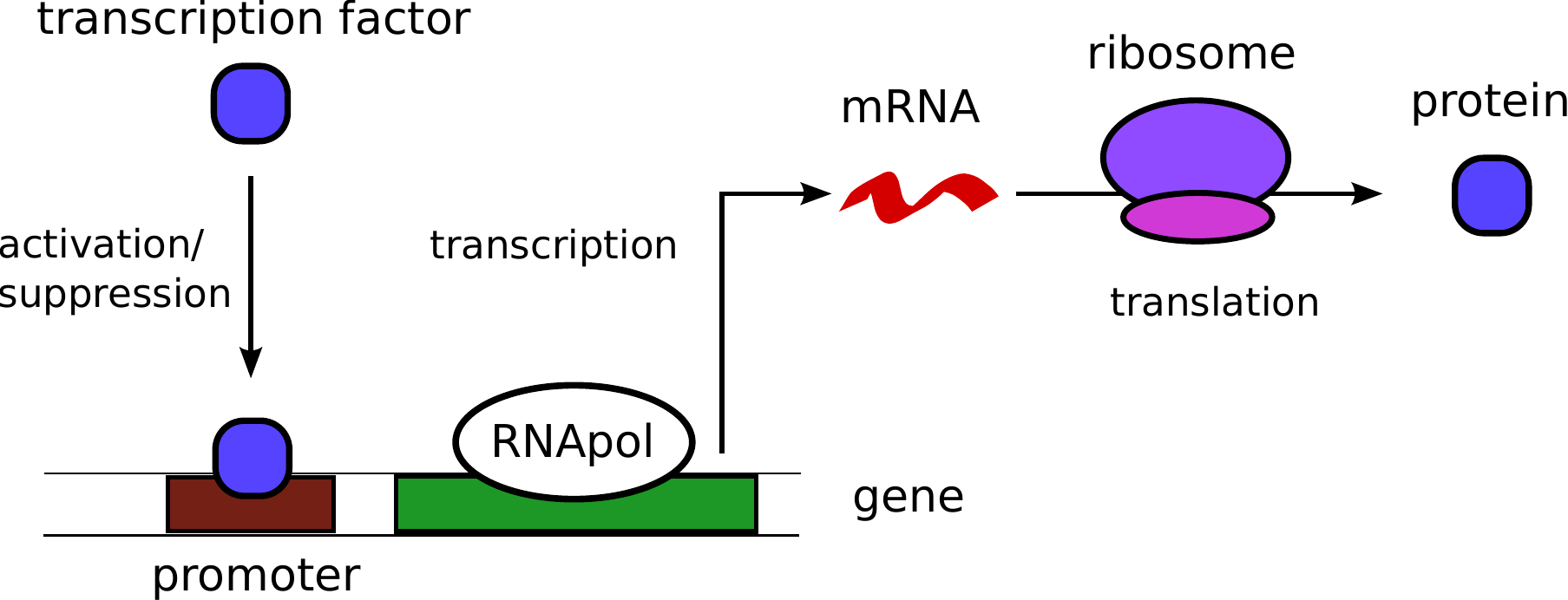} 
  \caption{Illustration of gene expression in prokaryotic cells. Transcription is conducted by a RNA polymerase enzyme (RNApol) that binds to the gene and produces an mRNA molecule from the gene's DNA. The mRNA is then translated into proteins by ribosomes. Transcription factors may bind to the promoter of the gene and thereby influence the recruitment of RNA polymerase and hence the transcriptional rate of the gene. Gene expression in eukaryotes happens similarly but is compartmentalised in the nucleus (transcription) and the cytosol (translation).}
  \label{fig_gene_expression}
\end{figure}

The term ``gene expression'' denotes the process of synthesis of functional gene products such as proteins. The mechanism is illustrated in Figure \ref{fig_gene_expression} for prokaryotic cells, namely cells lacking in-membrane-bound organelles (e.g. nucleus, mitochondria), such as bacteria. The process includes two main steps: \emph{transcription} during which mRNA molecules are produced, and \emph{translation} during which protein molecules are synthesised \cite{alberts1995molecular}. The transcription process involves the enzyme RNA polymerase. For the mechanism to initiate, an RNA polymerase enzyme must bind to the initial sequence of a gene. It then slides along the gene and produces an mRNA molecule that reproduces the DNA code of the gene. The RNA polymerase molecules move around randomly in the cell, a process which can be approximately described as Brownian motion \cite{Gardiner2009}. This means that the RNA polymerase binding to the gene is a stochastic event that happens randomly in time. As it turns out, not only the binding of the RNA polymerase to the gene, but also  the sliding along the gene happens stochastically. Therefore, the production of mRNA molecules is a stochastic process.

The production of protein molecules from mRNA during translation is conducted by ribosomes, which are RNA and protein complexes. The ribosomes and mRNA diffuse in the cell and hence meet randomly before translation can occur. Translation is thus also a stochastic process. 
Similarly, the degradation of mRNA molecules and proteins is conducted by certain enzymes and hence happens stochastically. 

The transcriptional regulation of gene expression is frequently modulated by regulatory proteins named  \emph{transcription factors}. Transcription factors are gene specific and bind to the \emph{promoter region}, which is located upstream of the gene's encoding region. Upon binding, transcription factors tune the affinity of the RNA polymerase molecules  for the promoter, thereby modulating the rate at which transcription initiation events occur and hence  the overall transcription rate. A transcription factor can either increase or decrease the binding rate of RNA polymerase and thus either enhance or suppress gene expression.

In eukaryotes, i.e., cells possessing in-membrane-bound organelles such as a nucleus, gene expression happens in a similar but somewhat more complicated way involving more steps. 
For instance, genes are generally located in the nucleus and transcription hence happens in the nucleus. Processes such as mRNA maturation and diffusion of mRNA to the cytoplasm need to happen before the mature mRNA can be translated into protein. 

\begin{figure} [t]
\centering
\includegraphics[scale=0.54]{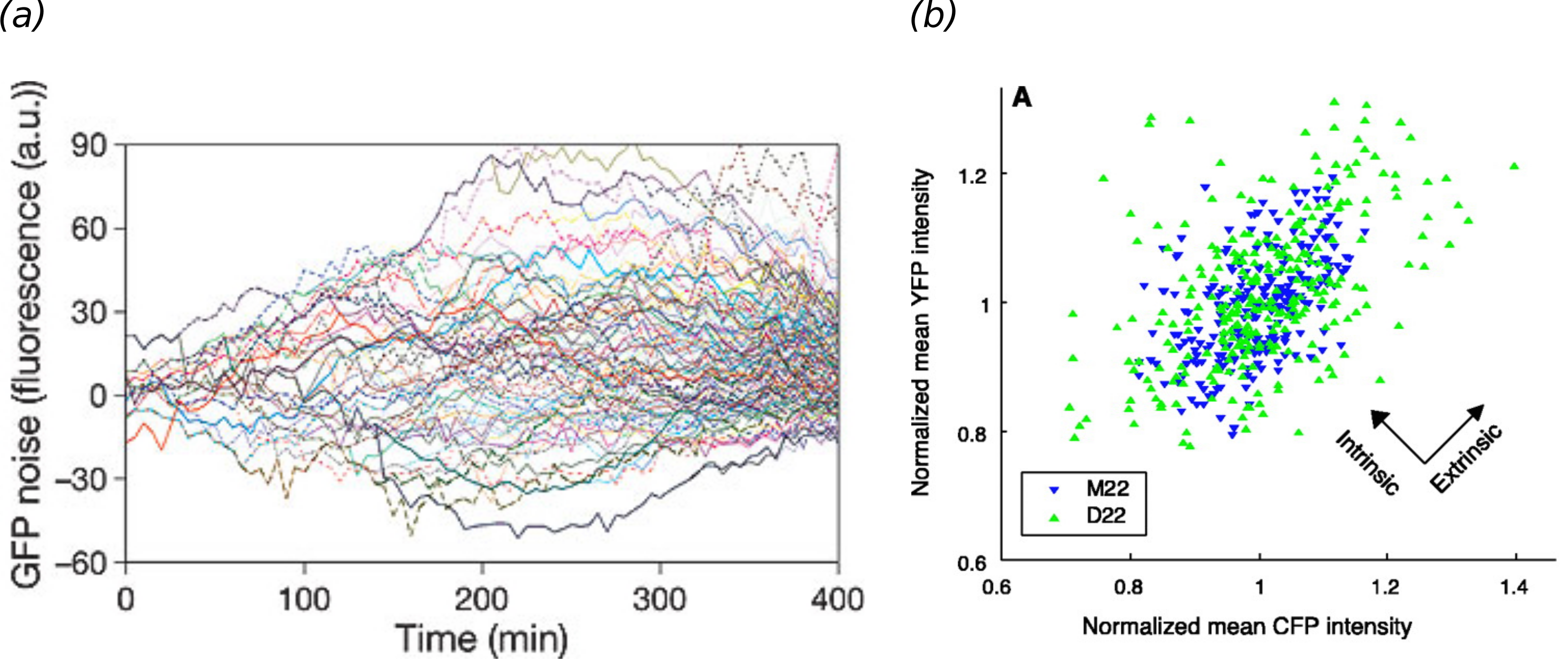}
  \caption{Measurements of intrinsic and extrinsic noise. $(a)$  Fluorescent time series acquired on \emph{Escherichia coli} cells expressing a green fluorescent protein (GFP) encoding gene \cite{austin2006gene}. The time trajectories correspond to the difference of the measured intensities in individual cells and the population average. We observe that the protein numbers  fluctuate strongly over time \cite{austin2006gene}. Permission kindly granted by the publisher. $(b)$ Fluorescence intensities for the dual reporter technique measured in \cite{Elowitz2002} for the two \emph{Escherichia coli}  strains M22 and D22. Permission kindly granted by the publisher. Each triangle corresponds to the measured fluorescence intensity in one cell. Variations along and perpendicular to the diagonal correspond to extrinsic and intrinsic noise, respectively. We observe that in this system both intrinsic and extrinsic noise contribute significantly to the overall fluctuations. }
  \label{fig_intrinsic_noise}
\end{figure}

The inherent stochasticity of chemical processes leads to fluctuations of molecule numbers in time. As an example Figure \ref{fig_intrinsic_noise} (a) shows  fluorescence intensity measurements of a fluorescent protein in individual \emph{Escherichia coli} cells, indicating strong fluctuations of protein numbers in time. Temporal fluctuations that are due to the stochasticity of chemical processes is what we call \emph{intrinsic noise} \cite{Elowitz2002}. Differences in molecule numbers of a certain species, say proteins, between different cells can originate from this type of fluctuations. However, such differences can also stem from other effects, such as physiological differences between cells or differing environmental conditions. For example, the numbers of RNA polymerase or ribosomes may differ between different cells, or different cells may be exposed to varying nutrient concentrations due to environmental fluctuations. 
Such differences that are not due to the stochasticity of chemical reactions are referred to as \emph{extrinsic noise} \cite{swain2002intrinsic}.

\subsection{Experimental evidence}\label{sec_exp_evidence}

As explained above, stochastic fluctuations are inherent to biochemical processes such as gene expression. The question hence arises: what effects do these fluctuations have on the functioning of cells and which processes dominate the emergence of these fluctuations? To answer such questions it is crucial to be able to empirically distinguish and quantify intrinsic and extrinsic noise.

One of the first experiments that aimed at explicitly separating the effects of intrinsic and extrinsic noise on gene expression was conducted by Elowitz et al. in 2002 \cite{Elowitz2002} by means of the \emph{dual reporter technique}. In this study, the authors integrated two distinguishable but identically regulated fluorescent reporter genes into the chromosome of \emph{Escherichia coli} cells, one expressing  cyan fluorescent protein (CFP) and one yellow fluorescent protein (YFP). The two genes were controlled by identical promoters, which means that they experienced the same external effects, i.e., the same extrinsic noise. Assuming that the two promoters are independent, fluctuations of the two proteins originating from the stochasticity of chemical reactions, i.e., intrinsic noise, should be completely uncorrelated. On the other hand, extrinsic effects should influence both expression systems equally, and the corresponding fluctuations should therefore be strongly correlated. This can be visualised as in Figure \ref{fig_intrinsic_noise} (b) which is taken from \cite{Elowitz2002}. The figure shows the measured intensities of individual cells for two different strains, with the YFP intensity on the $y$-axis and the CFP on the $x$-axis. The width of the cloud along the diagonal corresponds to correlated fluctuations and hence extrinsic noise, while the width orthogonal to the diagonal corresponds to uncorrelated fluctuations, i.e., intrinsic noise. The figure indicates that for this particular system both intrinsic and extrinsic noise significantly contribute to the overall noise. 

Many other experimental studies have been conducted investigating the origins and role of intrinsic and extrinsic noise in cells (for example \cite{Ozbudak2002, colman2005regulated, rosenfeld2005gene, bar2006noise, austin2006gene, Taniguchi2010}).  For an overview of experimental techniques see \cite{brehm2004single, maheshri2007living}. In \cite{bar2006noise}, for example, the authors measured the expression levels of $43$ genes in \emph{Escherichia coli} under various different environmental conditions. They observed that the variance of protein numbers scales roughly like the mean protein numbers and that for intermediate abundance proteins the intrinsic fluctuations are comparable or larger than extrinsic fluctuations. 
In another large scale study on \emph{Escherichia coli} in \cite{Taniguchi2010}, the authors found that fluctuations are dominated by extrinsic noise at high expression levels.

 The reported fluctuations can have significant influence on the functional behaviour of cells \cite{eldar2010functional}. A particularly important example are \emph{stochastic cell fate decisions} \cite{Balazsi2011}, where genetically identical cells under the same environmental conditions stochastically differentiate into functionally different states.  Such differing cell fate decisions of genetically identical cells are believed to be beneficial for a population of cells experiencing  fluctuating environments \cite{Balazsi2011}.

Most of the mentioned experimental studies measure certain statistics of gene expression systems, such as the equilibrium mean and variance of protein numbers. While experimental methods such as the \emph{dual reporter technique} allow to estimate the magnitude of intrinsic and extrinsic noise, they do not explain where the stochasticity originates from and what the underlying processes are. In order to answer such questions, mathematical methods are needed that allow researchers to model the dynamics and stochasticity of such biochemical reaction systems.

\section{Stochastic chemical kinetics}\label{sec_stoch_chem}

After having given a biological motivation for the importance of stochasticity in chemical reaction networks in the previous section, we consider next the question of how such systems can be described mathematically. First, we introduce the concept of chemical reaction networks in Section \ref{sec_chemical_networks} and describe the classical mathematical description of such systems in terms of deterministic rate equations. Next, we discuss the stochastic description of such systems in Section \ref{sec_stochkin_cme}. We derive the Chemical Master Equation and discuss its validity and solutions. Next, we discuss the stochastic simulation algorithm in Section \ref{sec_ssa}. Finally, we review methods for exact analytic solutions of the CME for certain classes of systems in Section \ref{sec_exact_solutions} and give an overview of available analytic solutions of specific systems that do not fall under any of these categories. 

\subsection{Chemical reaction networks and deterministic rate equations}\label{sec_chemical_networks}

Biological processes such as gene expression generally consist of complicated mechanisms involving several different types of molecules and physical operations. For a mathematical description of certain processes, one typically does not model all these mechanisms explicitly, but rather replaces them by an effective single chemical reaction event. In the context of gene expression, for example, transcription or translation may be modelled as single chemical reactions. A finite set of chemical species that interact via a set of such chemical reactions constitutes what we call a \emph{chemical reaction network}. Given a set of chemical species $X_i, ~i = 1,...,N,$ we define $R$ chemical reactions by the notation
\begin{align} \label{chemical_reactions}
  \sum_{i=1}^N  s_{ir} X_i \xrightarrow{\quad k_r \quad } \sum_{i=1}^N s'_{ir} X_i, 
   \quad r = 1, \ldots,  R,
\end{align}
where the \emph{stoichiometric coefficients} $s_{ir}$ and $s'_{ir}$ are non-negative integer numbers denoting numbers of reactant and product molecules, respectively. We say that the $r^{\text{th}}$ reaction is ``of order $m$'' if $\sum_{i=1}^N  s_{ir} = m$, i.e., if it involves $m$ reactant molecules. We further call a reaction ``unimolecular'' if $m=1$, ``bimolecular'' if $m=2$ and a system ``linear" if $m\leq 1$ for all reactions occurring in the system. 
We further call a system ``open'' if it contains a chemical process that generates molecules, and ``closed'' otherwise. 
The quantity $k_r$ in Equation \eqref{chemical_reactions}  is called the \emph{reaction rate constant} of the $r^{\text{th}}$ reaction. 

Classically, the dynamics of a chemical reaction system as in Equation \eqref{chemical_reactions} has been modelled by \emph{the Law of Mass Action} which was developed by Guldberg and Waage in its first version in the 1860s in the context of macroscopic \emph{in vitro} experiments, i.e., macroscopic amounts of chemical molecules in solutions \cite{guldberg1864studies, waage1864experiments, guldberg1864concerning, guldberg1879concerning}. The Law of Mass Action states that the rate of a reaction is proportional to the product of the concentrations of reactant molecules. Specifically, if we define $\boldsymbol{\phi}=(\phi_1, \ldots, \phi_N)$, where $\phi_i$ denotes the concentration of species $X_i$, the rate $g_r$ of a reaction as in Equation \eqref{chemical_reactions} is given by 
\begin{equation}\label{re_propensity}
  g_r(\boldsymbol{\phi}) = k_r \prod_{i=1}^N\phi_i^{s_{ir}}.
\end{equation}
We call $g_r$ the \emph{macroscopic rate function} of the $r^{\text{th}}$ reaction. Let us further define the \emph{stoichiometric matrix} $\mathbf{S}$ as 
\begin{equation}\label{stoi_mat}
  S_{ir}  =  s'_{ir} - s_{ir}, \quad i=1, \ldots, N, \quad r=1,\ldots,R. 
\end{equation}
The entry $S_{ir}$ corresponds to the net change of $X_i$ molecules when the $r^{\text{th}}$ reaction occurs. Consequently, $S_{ir} g_r(\boldsymbol{\phi})$ is the rate at which the concentration $\phi_i$ of species $X_i$ changes due to the $r^{\text{th}}$ reaction. Summing up the contributions from all reactions in a given system one obtains \cite{klipp2008systems}
\begin{equation}\label{res}
  \frac{d}{d t} \phi_i = \sum_{r=1}^R S_{ir} g_r(\boldsymbol{\phi}), \quad i=1,\ldots, N.
\end{equation}
This set of ordinary differential equations is called the \emph{rate equations}. The Law of Mass Action and the associated rate equations assume continuous concentrations, which means that they ignore the discreteness of molecule numbers. Moreover, they constitute a deterministic method, i.e., for a fixed initial condition the state of the system is exactly determined for all times. However, in the previous section we have seen that chemical reactions occur stochastically leading to fluctuations of molecule numbers in time. The reason why the rate equations give an accurate description in experiments such as the ones of Guldberg and Waage is that they studied chemical systems in solutions which typically contain a large number of substrate molecules. 
For large molecule numbers, it has been found experimentally that the relative fluctuations, i.e., the standard deviation divided by the mean value of molecule concentrations, scales like the inverse square root of the mean concentration and hence becomes small for systems with large molecule numbers \cite{bar2006noise}. One can therefore expect the rate equations to be an accurate description whenever a system has large molecule numbers for all species.

\begin{figure} [t]
\centering
  \includegraphics[scale=0.5]{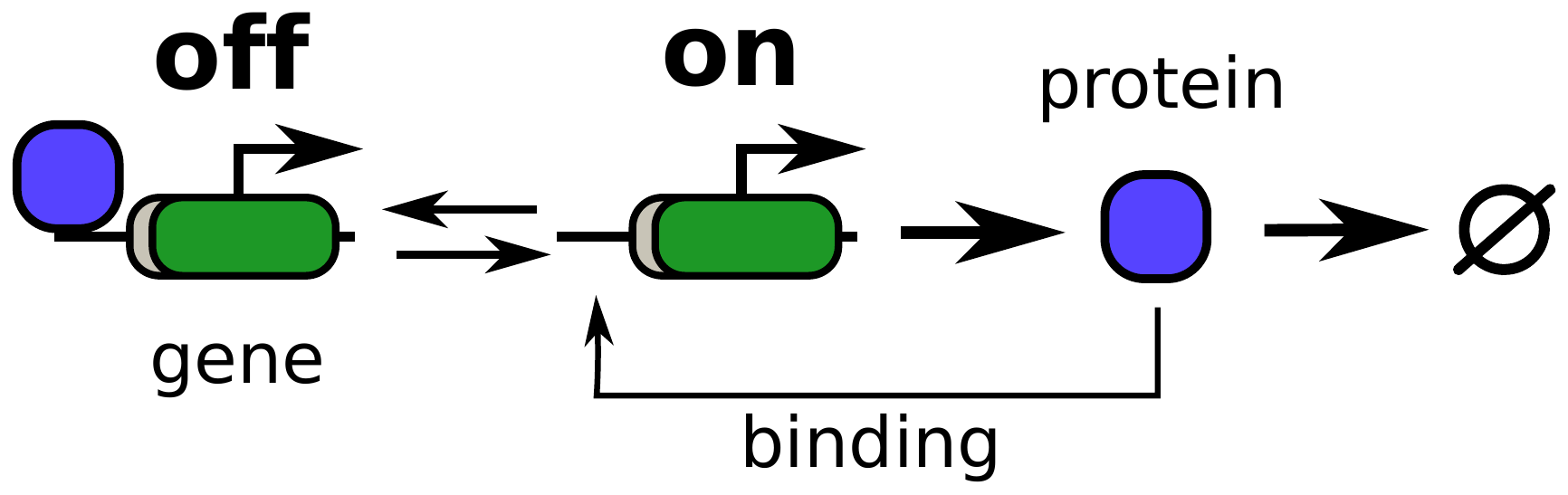} 
  \caption{Illustration of a gene expression system with negative feedback loop. When the gene is in the ``on'' state it produces proteins of a certain type. The protein can decay or bind to the gene's promoter. In the bound state the gene is in the ``off'' state, i.e., no protein is produced. The corresponding reactions are given in Equation \eqref{gene_feedback_reactions}. The protein hence suppresses its own production.}
  \label{fig_gene_system}
\end{figure}

\textbf{Example. }
As an example, consider the gene system in Figure \ref{fig_gene_system}. This system can be viewed as a simplified version of the gene expression process described in Figure \ref{fig_gene_expression}: we replace the process of transcription (gene produces an mRNA molecule) and translation (mRNA molecule produces a protein) by a single effective reaction in which the gene directly produces the protein.

The example in Figure \ref{fig_gene_system} depicts negative autoregulation. The latter is one of the most common regulatory mechanisms: the gene's product downregulates its own expression. In the bound state $G^{\text{off}}$, the gene does not produce any protein. The protein thus suppresses its own production, which means the system is an example of a \emph{negative feedback loop}. The corresponding reactions in the notation of Equation \eqref{chemical_reactions} read
\begin{equation}\label{gene_feedback_reactions}
  \begin{split}
  G^{\text{on}} & \xrightarrow{\quad k_1 \quad } G^{\text{on}}  + P, 
  \quad G^{\text{on}}  + P \xrightleftharpoons[\quad k_3 \quad]{\quad k_2 \quad } G^{\text{off}} ,
    \quad   P  \xrightarrow{\quad k_4 \quad } \varnothing,
\end{split}
\end{equation}
where we call the gene in the bound and unbound state $G^{\text{on}}$ and $G^{\text{off}}$, respectively and  the protein $P$. In our nomenclature, $G^{\text{on}}  + P \to G^{\text{off}}$ is a second-order or bimolecular reaction, while the other three reactions are first order or linear. By ``$P  \to \varnothing$'' we indicate that $P$ leaves the system under consideration. This could for example mean that $P$  becomes converted into different types of chemical species  or degraded in its constitutive elements that are not included in our model. 

Let us order reactions according to the rate constants in Equation \eqref{gene_feedback_reactions}, and species as $G^{\text{on}}, P$ and $G^{\text{off}}$, respectively. Consider the stoichiometric coefficients $s_{ir}$ and $s_{ir}'$ defined in Equation  \eqref{chemical_reactions}, which correspond to the number of reactant and product molecules, respectively, of species $X_i$ in the $r^\text{th}$ reaction. For the reaction $G^{\text{on}}  \to G^{\text{on}}  + P$, for example, we have $s_{11}=1, s_{21}=0$ and  $s_{31}=0$, since there is one $G^{\text{on}}$ on the l.h.s. of the reaction but no $P$ or $G^{\text{off}}$, and $s'_{11}=1, s'_{21}=1$ and  $s'_{31}=0$ since there is one $G^{\text{on}}$, one $P$ and no $G^{\text{off}}$ molecules on the r.h.s. 
Proceeding similarly for the other reactions in Equation \eqref{gene_feedback_reactions},  we find
\begin{align}\label{gene_s_and_s_matrix_non_red}
  \mathbf{s}
  & =
    \begin{pmatrix}
      1 & 1 & 0 & 0 \\
      0 & 1 & 0 & 1 \\
      0 & 0  & 1 & 0
    \end{pmatrix},
  \quad  \quad  \mathbf{s}'
   =
    \begin{pmatrix}
      1 & 0 & 1 & 0 \\
      1 & 0 & 1 & 0 \\
      0 & 1  & 0 & 0
    \end{pmatrix}. 
\end{align}
Accordingly, the stoichiometric matrix $\mathbf{S} = \mathbf{s}' - \mathbf{s}$ reads
\begin{align}\label{res_stoi_unreduced}
  \mathbf{S}
  & =
    \begin{pmatrix}
      0 & -1 & 1 & 0 \\
      1 & -1 & 1 & -1 \\
      0 & 1  & -1 & 0
    \end{pmatrix}.
\end{align}
Let $\boldsymbol{\phi}=(\phi_1, \phi_2, \phi_3)$, where $\phi_1, \phi_2$ and $\phi_3$ denote the concentrations of $G^{\text{on}}, P$ and $G^{\text{off}}$, respectively. The macroscopic rate vector $\mathbf{g}^{(0)}(\boldsymbol{\phi})=(g_1^{(0)}(\boldsymbol{\phi}), \ldots, g_r(\boldsymbol{\phi}))^T$, with the $g_r$ defined in Equation \eqref{re_propensity}, is obtained by using $\mathbf{s}$ in Equation \eqref{gene_s_and_s_matrix_non_red} and reads
\begin{align}\label{res_prop_unreduced}
  \quad \mathbf{g}(\boldsymbol{\phi})
  =
    (k_1 \phi_1, k_2 \phi_1 \phi_2, k_3 \phi_3, k_4 \phi_2)^T.
\end{align}
Using Equations \eqref{res}, \eqref{res_stoi_unreduced} and \eqref{res_prop_unreduced} it is easy to write down the corresponding rate equations. However, note that the system has a conservation law in particle numbers which we can use to find a simplified description by reducing the number of variables: the total number of genes in the ``on'' state and genes in the ``off'' state is constant, i.e., $\phi_1+ \phi_3= const. \equiv g_0$. We can thus reduce the system to a two species system by using $\phi_3= g_0 - \phi_1$. The matrices $\mathbf{s}$ and $\mathbf{s}'$  for the reduced system are obtained from Equation \eqref{gene_s_and_s_matrix_non_red} by dropping the last row,
\begin{align}\label{gene_s_and_s_matrix_red}
  \mathbf{s}
  & =
    \begin{pmatrix}
      1 & 1 & 0 & 0 \\
      0 & 1 & 0 & 1
    \end{pmatrix},
  \quad  \quad  \mathbf{s}'
   =
    \begin{pmatrix}
      1 & 0 & 1 & 0 \\
      1 & 0 & 1 & 0 
    \end{pmatrix},
\end{align}
and the stoichiometric matrix and propensity vector of the reduced system read accordingly
\begin{align}\label{gene_stoichiometric}
  \mathbf{S}
  & =
    \begin{pmatrix}
      0 & -1 & 1 & 0 \\
      1 & -1 & 1 & -1
    \end{pmatrix},
  \quad \mathbf{g}(\boldsymbol{\phi})
  =
    (k_1 \phi_1, k_2 \phi_1 \phi_2, k_3 (g_0 - \phi_1), k_4 \phi_2)^T.
\end{align}
Using Equation \eqref{res} we hence obtain the rate equations
\begin{equation}
\begin{split}\label{gene_res}
  \frac{\partial}{\partial t} \phi_1
  & =
    - k_2 \phi_1 \phi_2 + k_3 (g_0 - \phi_1), \\
  \frac{\partial}{\partial t} \phi_2
  & =
    k_1 \phi_1 -  k_2 \phi_1 \phi_2 + k_3 (g_0 - \phi_1) - k_4 \phi_2.
\end{split}
\end{equation}
The gene system in Figure \ref{fig_gene_system} will be used to showcase several different methods in this article, and we will throughout use the reduced system.

\subsection{Stochastic methods}\label{sec_stochkin_cme}

The rate equations discussed in the previous section constitute a deterministic description of chemical reaction systems in terms of continuous molecule concentrations. As we have seen in Section \ref{sec_pre_bio}, however, fluctuations become important in many biological processes such as gene expression. The reason for this is that often some species occur at low molecule counts. A more general, stochastic description would keep the discrete nature of the $X_i$ molecules  in Equation \eqref{chemical_reactions} intact. One general approach is to explicitly model the spatial positions of molecules and to model their movement as Brownian motion, with chemical reactions happening stochastically under certain rules. Exact analytic results for such systems are generally not known. Simulations, on the other hand, have to keep track of every single particle and quickly become computationally unfeasible. However, under certain conditions, which we discuss in the following section, simplified descriptions can be employed making spatial models and the simulation of single particles unnecessary.

\subsubsection{The Chemical Master Equation}\label{sec_sub_cme}

We now consider a chemical reaction system as in Equation \eqref{chemical_reactions} in a closed compartment of volume $\Omega$. We seek a stochastic description of the system under well-mixed and dilute conditions. By ``well-mixed'' we mean that the diffusion of particles in the compartment constitutes the fastest time scale of the system, in the sense that the expected distance travelled by each particle between successive reactive collisions is much larger then the length scale of the compartment. This implies that the spatial positions of molecules can be ignored and the dynamics of the system only depends on the total molecule numbers. By ``dilute'' we mean that the combined volume of all the considered molecules is much smaller than the total volume, which means that the molecules can be considered as point particles. 

If these two conditions are met, it can be shown \cite{Gillespie1992} that the state of the system at any time is fully determined by the state vector $\mathbf{n}=(n_1, \ldots, n_N)$, where $n_i$ is the molecule number of species $X_i$ in the compartment. In particular, the spatial locations and diffusion of molecules does not have to be modelled, and the system corresponds to a \emph{continuous-time Markov jump process}. It can further be shown that the probability for the $r^{\text{th}}$ reaction to happen in an infinitesimal time step $dt$ is given by $f_r (\mathbf{n}) dt$  where  $f_r (\mathbf{n})$ is the  \emph{propensity function} of the $r^{\text{th}}$ reaction and  proportional to the number of combinations of reactant molecules in $\mathbf{n}=(n_1, \ldots, n_N)$. Consider for example a bimolecular reaction of the form $A +B \to \varnothing$. The number of pairs with one $A$ and one $B$ molecule is $n_A n_B$, where $n_A$ and $n_B$ are the molecule numbers of $A$ and $B$, respectively. The corresponding propensity function is hence given by $k_r n_A n_B / \Omega$. The scaling with the volume $\Omega$ stems from the fact that the probability for two molecules to collide is proportional to $1/ \Omega$.  Generalising these arguments to reactions as in Equation \eqref{chemical_reactions} leads to
\begin{equation}\label{general_propensity}
  f_r(\mathbf{n}) = k_r \Omega \prod_{i=1}^N \frac{n_i!}{(n_i - s_{ir})! \Omega^{s_{ir}}}.
\end{equation}
Propensity functions of this form are called \emph{mass-action kinetics type} \cite{VanKampen2007}. {Although Equation \eqref{re_propensity} is often, due to historical reasons, stated to be the Law of Mass Action, from a microscopic perspective it is more accurate to state Equation \eqref{general_propensity} as the Law of Mass Action. Equation \eqref{re_propensity}  can be viewed as the macroscopic version of \eqref{general_propensity} obtained in the  limit of large molecule numbers and small fluctuations. }
Specifically, for reactions of up to order two Equation \eqref{general_propensity} becomes
\begin{equation}\label{general_propensity_second_order}
 \begin{matrix}
      \text{zeroth order} \quad  &  \varnothing \to  \quad &  f_r(\mathbf{n}) = k_r \Omega,  \\
      \text{first order}  \quad &  A \to   \quad &   f_r(\mathbf{n}) = k_r n_A, \\
      \text{second  order} \quad & A+B \to   \quad &   f_r(\mathbf{n}) = \frac{k_r}{\Omega} n_A n_B, \\
      \text{second  order} \quad & A+A \to   \quad &   f_r(\mathbf{n}) = \frac{k_r}{\Omega} n_A (n_A -1). 
    \end{matrix}
\end{equation}
Let us now consider how we can mathematically describe the dynamics of such a system. To this end, consider the probability distribution $P(\mathbf{n},t | \mathbf{n}_0,t_0)$ for the system to be in state $\mathbf{n}$ at time $t$ given that it was in state $\mathbf{n}_0$ at time $t_0$. We will use the shorthand $P(\mathbf{n},t) = P(\mathbf{n},t | \mathbf{n}_0,t_0)$ in the following and implicitly assume conditioning on an initial state. The probability $P(\mathbf{n},t + dt)$ after an infinitesimal time step $dt$ is  given by $P(\mathbf{n},t)$ plus the probability to transition into state $\mathbf{n}$ from a different state $\mathbf{n}^*$ minus the probability to leave state $\mathbf{n}$, which leads us to
\begin{align}\label{cme_diff}
  P(\mathbf{n},t+dt) 
  & = 
    P(\mathbf{n},t)  + dt \left(\sum_{r=1}^R f_r (\mathbf{n} - \mathbf{S}_r) P(\mathbf{n} - \mathbf{S}_r, t) 
    - \sum_{r=1}^R f_r (\mathbf{n}) P(\mathbf{n}, t) \right),
\end{align}
where we used the fact that probability of the $r^{\text{th}}$ reaction to happen in an infinitesimal time interval $dt$ is given by $f_r (\mathbf{n}) dt$ and
where $\mathbf{S}_r$ is the $r^{\text{th}}$ column of the stoichiometric matrix $\mathbf{S}$.
Subtracting $P(\mathbf{n},t)$, dividing by $dt$ and taking the limit $dt \to 0$ gives the \emph{Chemical Master Equation} (CME) \cite{Gillespie1992},
\begin{align}\label{cme}
  \partial_t P(\mathbf{n},t) 
  & = 
    \sum_{r=1}^R f_r (\mathbf{n} - \mathbf{S}_r) P(\mathbf{n} - \mathbf{S}_r, t) 
    - \sum_{r=1}^R f_r (\mathbf{n}) P(\mathbf{n}, t).
\end{align}
Since $\mathbf{n}$ is a discrete-valued vector, Equation \eqref{cme} is a coupled system of linear ordinary differential equations. Note that this system is infinite whenever $\mathbf{n}$ is unbounded. Despite its simple structure, there are generally no analytic solutions known to the CME. We call a distribution $P(\mathbf{n},t) $ a \emph{steady-state solution} of the CME if it solves the CME and fulfils $\partial_t P(\mathbf{n},t)  = 0$.

In the context of chemical reactions, one of the first applications of the CME was done by Delbr\"uck in 1940 for describing the dynamics of an autocatalytic reaction \cite{Delbruck1940}. It has later been applied to linear reactions by Bartholomay \cite{bartholomay1958stochastic} and McQuarrie \cite{mcquarrie1963kinetics}, and bimolecular reactions by Ishida \cite{ishida1964stochastic} and McQuarrie et al \cite{mcquarrie1964kinetics}.  Gillespie derived the CME from molecular physics of a dilute and well-mixed gas in 1992 \cite{Gillespie1992}, which he extended to liquids in 2009 \cite{gillespie2009diffusional}.

\textbf{Example. }
Consider again the gene system in Figure \ref{fig_gene_system} with reactions given in Equation \eqref{gene_feedback_reactions}. The corresponding stoichiometric matrix $\mathbf{S}$ is given in Equation  \eqref{gene_stoichiometric}. Let $n_1$ and $n_2$ be the molecule numbers of the gene in the unbound state $G^{\text{on}}$ and the protein $P$, respectively. Let us assume that there is only one gene in the system, which implies that the number of bound genes $G^{\text{off}}$ is $1-n_1$.
 Using Equation \eqref{general_propensity} we find the propensity vector
\begin{align}\label{gene_prop}
  \quad \mathbf{f}(\mathbf{n})
  =
    (k_1 n_1, \frac{k_2}{\Omega} n_1 n_2, k_3 (1-n_1), k_4 n_2)^T.
\end{align}
The corresponding CME becomes (c.f.~Equation \eqref{cme})
\begin{equation}\label{gene_cme}
\begin{split}
  \partial_t P(n_1,n_2,t)
  & =
    k_1 n_1 P(n_1, n_2-1, t) +  \frac{k_2}{\Omega} (n_1+1) (n_2+1) P(n_1+1, n_2+1, t) \\
  & \quad
    + k_3 (2 - n_1) P(n_1-1,n_2-1,t) + k_4 (n_2+1) P(n_1, n_2+1,t) \\
  & \quad 
    - (k_1 n_1 +  \frac{k_2}{\Omega} n_1 n_2 + k_3 (1-n_1) + k_4 n_2) P(n_1,n_2,t).
\end{split}
\end{equation}
Despite having a relatively simple system here with effectively only two species, no time-dependent solution for its CME in Equation \eqref{gene_cme} is known to our knowledge.  
A solution in steady state has been derived in \cite{GrimaNewman2012}, but for most other systems not even a steady state solution is available. Therefore, one generally needs to rely on stochastic simulations or approximations to study the behaviour of such systems.

\subsubsection{Moment equations}\label{sec_moment_equations}

Suppose we are not interested in the whole distribution solution of the CME but only in its first few moments, say the mean and variance. Starting from the CME in Equation \eqref{cme} one can derive time evolution equations for the moments of its solution as follows.
To obtain the time evolution equation for the moment $\langle n_i \ldots n_l \rangle$ we multiply Equation \eqref{cme} by $n_i \ldots n_l$ and sum over all molecule numbers, leading to
\begin{align}\label{cme_moms}
  \partial_t \langle n_i \ldots n_l \rangle
  & = 
    \sum_{r=1}^R \langle (n_i+S_{ir})\ldots(n_l+S_{lr}) f_r (\mathbf{n}) \rangle
    - \sum_{r=1}^R \langle n_i \ldots n_l f_r (\mathbf{n}) \rangle.
\end{align}
Here, $\langle \cdot \rangle$ denotes the expectation with respect to the solution $P(\mathbf{n},t)$ of the CME in Equation \eqref{cme}.
For moments of up to order two Equation \eqref{cme_moms} becomes 
\begin{align}\label{cme_closure_order_one}
  \partial_t \langle n_i \rangle
  & = 
     \sum_{r=1}^R S_{ir} \langle f_r(\mathbf{n}) \rangle, \\
\label{cme_closure_order_two}
  \partial_t \langle n_i n_j \rangle
  & =
    \sum_{r=1}^R  \big[ S_{jr} \langle n_i f_r(\mathbf{n})  \rangle + S_{ir} \langle  f_r(\mathbf{n}) n_j \rangle
    + S_{ir} S_{jr} \langle  f_r(\mathbf{n})   \rangle \big].
\end{align}
We see that if all $f_r(\mathbf{n})$ are zeroth or first-order polynomials in $\mathbf{n}$, i.e.,  the system is linear without any bimolecular or higher order reactions, the equation of a moment of order $m$ depends only on moments of order $m$ or lower, i.e., the equations are not coupled to higher order equations. The equations up to a certain order hence constitute a finite set of linear ordinary differential equations which can be readily solved numerically or by matrix exponentiation as will be described in the context of the CME in Section \ref{sec_exact_solutions}. Note that no approximation has been made here which means that the equations describe the exact moments of the CME's solution. This means that \emph{for linear systems, the exact moments up to a finite order of the process can be obtained by solving a finite set of linear ordinary differential equations}. 

If the systems is non-linear, i.e., contains bimolecular or higher order reactions, the equation of a certain moment depends on higher order moments. This means that the moment equations of different orders are coupled to each other, leading to an infinite hierarchy of coupled equations. This can obviously not be solved directly but gives the basis of approximation methods such as \emph{moment closure approximations} which we will discuss in Section \ref{sec_stochkin_ma}.

\textbf{Example. }
Let us again consider the gene system in Figure \ref{fig_gene_system} with reactions given in Equation \eqref{gene_feedback_reactions}. The corresponding stoichiometric matrix and propensity vector are given in Equation \eqref{gene_stoichiometric}.
Using these in Equations \eqref{cme_closure_order_one} and \eqref{cme_closure_order_two} one obtains
\begin{align}
\label{moms_example1}
  \partial_t y_1
  & = 
     - \frac{k_2}{\Omega} y_{1,2} + k_3 (1-y_1), \\
\label{moms_example2}
  \partial_t y_2
  & = 
     k_1 y_1 - \frac{k_2}{\Omega} y_{1,2} + k_3 (1-y_1)
     - k_4  y_2, \\
\label{moms_example3}
  \partial_t y_{1,1}
  & =
    \frac{k_2}{\Omega} (- 2 y_{1,1,2} + y_{1,2}) + k_3( -2 y_{1,1} + y_1 + 1), \\
\label{moms_example4}
  \partial_t y_{1,2}
  & =
    k_1 y_{1,1}  + \frac{k_2}{\Omega} (-y_{1,1,2} - y_{1,2,2} + y_{1,2})
    + k_3 (-y_{1,1} -y_{1,2} +y_2 + 1) - k_4 y_{1,2}, \\
\nonumber
  \partial_t y_{2,2}
  & =
    k_1 (2y_{1,2}+y_{1})  + \frac{k_2}{\Omega} (-2 y_{1,2,2} + y_{1,2}) \\
\label{moms_example5}
  & \quad \quad  + k_3 (-2y_{1,1} -y_{1} + 2 y_2 + 1) + k_4 (-2y_{2,2}+y_2),
\end{align}
where we introduced the shorthand
\begin{align}
  y_{i_1,\ldots,i_k}
  &  = 
    \langle n_{i_1} \ldots n_{i_k} \rangle.
\end{align}
Note that the equations for the first order moments in Equations \eqref{moms_example1} and \eqref{moms_example2} depend on the second moment $y_{1,2}$, and that the equations for the second moments \eqref{moms_example3}-\eqref{moms_example5} depend on the third moments $y_{1,1,2}$ and $y_{1,2,2}$. Similarly, it is easy to see that moment equations of any order depend on higher order moments, which means that we have an infinite system of coupled equations. Note that all terms in Equations \eqref{moms_example1}-\eqref{moms_example5} depending on higher order moments are proportional to the rate constant $k_2$ of the bimolecular reaction in Equation \eqref{gene_feedback_reactions}, illustrating that the moment equations decouple in the absence of bimolecular (and higher order) reactions. This could be achieved here by setting $k_2=0$ for which the moment equations would decouple and could thus be solved numerically.

\subsubsection{Non-mass-action propensity functions}

So far we only considered propensity functions of mass-action type defined in Equation \eqref{general_propensity}. However, in the literature propensity functions that are not of mass-action kinetics type are frequently used, such as Michaelis-Menten or Hill functions to model the dependence of the production rate of a product on the substrate concentration in an enzymatic catalysis or the dependence of the expression level of a gene on its transcription factor. Such propensity functions typically arise in reduced models where an effective reaction replaces several microscopic reactions. For a gene that becomes regulated by a transcription factor, for example, the binding of the transcription factor to the promoter of the gene is not modelled explicitly, but the effect of its concentration included in the modified propensity function of the expression reaction. Such non-mass-action type reactions should thus normally be seen as an effective reaction replacing a set of mass-action type reactions.

A possible reduced model of the gene expression system in Figure \ref{fig_gene_system}, for example, eliminates the gene from the system and combines the binding, unbinding and protein production reactions into one effective reaction $\varnothing \to P$ with a Michaelis-Menten type propensity function $k_1 k_3 n_P /(k_2 n_P + k_3)$, which means that the reaction producing protein now depends on the protein number $n_P$. Such reductions often emerge from separations of time-scales in a system and become exact in certain limits, see Section \ref{sec_other_approximations} for more details.

Independently of the origin of a non-mass-action propensity function, as long as it is interpreted as the firing rate of the corresponding reaction, the CME description remains valid.
Unless explicitly stated otherwise, we will assume mass-action kinetics in this article.

\subsection{Stochastic simulations}\label{sec_ssa}

As mentioned above, there are generally no analytic solutions known to the CME. However, it is possible to directly simulate the underlying process. The stochastic simulation algorithm (SSA) is a popular Monte Carlo method that allows one to simulate exact sample paths of the stochastic process described by the CME.

The stochastic simulation algorithm was first proposed in the context of chemical kinetics by Gillespie \cite{Gillespie1976,Gillespie1977}, and several variants have been proposed in the literature, see \cite{pahle2009biochemical,Mauch2011,Gillespie2013} for reviews. The basic idea is to simulate reaction events explicitly in time and to update the time and state vector accordingly. The stochastic process described by the CME is a continuous-time Markov jump process, which has the important property that waiting times, i.e., time intervals between successive reaction events, are exponentially distributed \cite{Gardiner2009}.  Since it is easy to sample from exponential distributions, it is straightforward to simulate the occurrences of chemical reactions. 

One example is the so-called \emph{direct method} \cite{Gillespie1976}, which samples the time step $\tau$ for the next reaction to happen and subsequently which of the different reactions occurs. Specifically, let $p(\tau | \mathbf{n}, t)$ be the probability for the next reaction to happen in an infinitesimal time interval $dt$ around $\tau+t$, given that the state of the system is $\mathbf{n}$ at time $t$, and $p(r | \mathbf{n}, t)$ the probability that the next reaction is a reaction of type $r$. Using that $f_r(\mathbf{n}) dt$ is the probability for the $r^{\text{th}}$ reaction to happen in $dt$, it can be shown that \cite{Gillespie1976}
\begin{align}\label{ssa_direct1}
  p(\tau | \mathbf{n}, t) & = \lambda \exp \left(- \tau \lambda \right), \quad \lambda = \sum_{r=1}^R f_r(\mathbf{n}), \\
\label{ssa_direct2}
  p(r | \mathbf{n}, t)
  & =
    \frac{f_r(\mathbf{n})}{\lambda}.
\end{align}
{ Samples from Equations \eqref{ssa_direct1} and \eqref{ssa_direct2} can be respectively obtained as
\begin{align}\label{ssa_direct1_sample}
  \tau 
  & = - \ln(u_1)/\lambda, \\
\label{ssa_direct2_sample}
  r
  & =
    \text{smallest integer satisfying}~ \sum_{i=1}^r f_r(\textbf{n}) > u_2 \lambda,
\end{align}
where $u_1$ and $u_2$ are uniform random numbers between $0$ and $1$.
The direct SSA method iteratively updates the state vector and time of the process by first sampling the time point of the next reaction event according to Equation \eqref{ssa_direct1_sample} and subsequently sampling which reaction happens according to Equation \eqref{ssa_direct2_sample}.}

Unfortunately, the applicability of the SSA is severely limited due to its computational cost. Since each and every reaction is simulated explicitly, the SSA becomes computationally expensive even for systems with few species. This is particularly the case if the molecule numbers have large fluctuations or if many reactions happen per unit time. In the first case a large number of samples have to be simulated to obtain statistically accurate results, whereas in the second case single simulations become expensive since the time between reaction events becomes small.

\subsubsection{Extrinsic noise}\label{sec_extrinsic_noise}

So far we have only discussed stochastic simulations of systems with propensity functions that depend solely on the system's molecule numbers and the system's volume. This implies that the propensity functions are constant between reactions events, leading to inter-reaction intervals being exponentially distributed. However, as already mentioned in Section \ref{sec_sub_cme}, it is frequently of interest to consider stochastically varying propensity functions. This could for example represent the effect of a randomly fluctuating environment on a cell. 

To model the effect of extrinsic noise, one typically includes a stochastic variable in one or several propensity functions, whose dynamics may for example be governed by a stochastic differential equation. This means that the corresponding propensities become explicit (stochastic) functions of time. While the CME in Equation \eqref{cme} is still valid in this case, stochastic simulation algorithms discussed above are not since the inter-reaction times are now not exponentially distributed anymore. 

The simplest approach is to assume the propensity functions to be constant between reaction events, which allows one to use the standard simulation algorithm \cite{Gillespie1976, Guerriero2011}. However, this is of course an approximation and can lead to inaccuracies if the extrinsic process (and hence the corresponding propensity functions) fluctuates strongly between consecutive reaction times. The most straightforward way to simulate such processes exactly is to integrate the propensity functions step-wise over time until a certain target value is reached \cite{Anderson2007, Shahrezaei2008b, caravagna2013interplay}. Apart from numerical integration errors this class of algorithms is exact. However, due to the numerical integration it becomes computationally highly expensive. More recently an alternative exact method has been proposed \cite{voliotis2016stochastic}. This method introduces an additional reaction channel in such a way that the total propensity (i.e., $\lambda$ in Equation \eqref{ssa_direct1}) is constant between successive reactions. This in turn allows one to use a standard SSA algorithm on the augmented system. It is hence generally more efficient than the aforementioned integral methods.

\subsection{Exact results}\label{sec_exact_solutions}

As pointed out before, for most systems no analytic solutions of the CME are known. However, for certain classes of systems analytic solutions do exist. For some classes the general time-dependent case can be solved while others can be solved only in steady state. Moreover, analytic solutions have been derived for various simple example systems that do not fall under any of these classes. We give here first an overview of general classes of systems that can be solved analytically, and subsequently list some additional systems for which analytic solutions are known.

\subsubsection{Finite state space}

Suppose the state space of $\mathbf{n}$ is finite with $M$ elements. Let us associate with each state a probability $p_i(t), i=1, \ldots, M$. In Section \ref{sec_stochkin_cme} we have noted that in this case the CME in Equation \eqref{cme} is a finite system of coupled ordinary differential equations. If we write $\mathbf{p}(t)= (p_1(t), \ldots, p_M(t))^T$, the CME can be written in the matrix form as
\begin{align}\label{cme_mat}
  \partial_t \mathbf{p}(t)
  & = 
    \mathbb{E} \mathbf{p}(t),
\end{align}
where $\mathbb{E}$ is a $M \times M$ matrix whose elements can be easily derived for a given system using Equation \eqref{cme}. The solution of Equation \eqref{cme_mat} is simply given by 
\begin{align}\label{cme_mat_sol}
  \mathbf{p}(t)
  & = 
    \exp (\mathbb{E} t)  \mathbf{p}(0).
\end{align}
Therefore, for systems with finite state space a solution of the CME for all times can be in principle computed using Equation \eqref{cme_mat_sol}. However, even for systems with finite state space, the dimension of Equation \eqref{cme_mat_sol} is often quite large in practice, making matrix exponentiation computationally expensive. Efficient numerical methods for matrix exponentiation have been developed in recent years \cite{al2011computing, moler2003nineteen}, but for many chemical reaction systems of practical interest it remains an intractable task.  For systems with large or infinite state space Equation \eqref{cme_mat_sol} is hence not of direct practical use. It forms the basis for certain approximation methods, however, see Section \ref{sec_other_approximations}.

\subsubsection{Linear systems}\label{sec_cme_sol_linear}

In Section \ref{sec_moment_equations} we found that the moment equations decouple from higher order equations for linear reaction systems, i.e., systems without bimolecular or higher order reactions. The moment equations up to a finite order can hence be directly solved numerically for such systems. For non-linear systems, in contrast, this is not possible since the moment equations couple to higher orders. 
Similarly, exact solutions for the whole distribution solution of the CME are easier to obtain for linear systems. Often this can be done by means of the \emph{generating function method} which transforms the CME into a partial differential equation \cite{Gardiner2009}. The form of the latter becomes particularly simple for linear systems and can often be solved analytically. 

In 1966 Darvey and Staff used the generating function approach to derive a multinomial solution for $P(\mathbf{n},t)$ for multinomial initial conditions and for systems with $\sum_{i=1}^N s_{ir} = \sum_{i=1}^Ns'_{ir} = 1$ for all reactions $r$, i.e., exactly one reactant and exactly one product molecule for all reactions \cite{Darvey1966}. Note that these are closed systems. Later Gardiner used the so-called ``Poisson representation'' to show that for  systems with $\sum_{i=1}^N s_{ir}, \sum_{i=1}^N s'_{ir} \leq 1$ for all reactions $r$ and Poisson product initial conditions, the solution of the CME is a Poisson product for all times \cite{gardiner1977poisson}.   The auto- and cross-correlation functions for such systems have more recently been computed in \cite{heuett2006grand}. These results have been generalised to systems with $\sum_{i=1}^N s_{ir}, \sum_{i=1}^N s'_{ir} \leq 1$  for all reactions $r$, with arbitrary initial conditions in \cite{Jahnke2007}. In this case the solution can be written as a convolution of multinomial and product Poisson distributions. In all these cases, the solutions can be constructed from the solution of the deterministic rate equations discussed in Section \ref{sec_chemical_networks} and can hence be obtained efficiently by (numerically) solving a finite set of ordinary differential equations. 

{We would like to stress that the methods described here do not apply to all linear reactions, but only to those with less than two product molecules ($\sum_{i=1}^N s'_{ir} \leq 1$). This excludes  reactions of the type $A \to A + B$  which are common in biology, such as translation of a protein $B$ from its mRNA, $A$. }

\subsubsection{Non-linear systems}\label{sec_exact_sol_detailed_balance}

 As discussed in the previous section, for linear systems where each reaction has at most one product molecule, an analytic solution of the CME can be derived. In contrast, for non-linear systems or linear systems including reactions with two product molecules no analytic solutions are known in general. However, for certain subclasses of systems it is possible to derive steady-state solutions, i.e., solutions satisfying $\partial_t P(\mathbf{n}, t) = 0$. One example are \emph{reversible systems} that obey \emph{detailed balance}, which is also called \emph{thermodynamic equilibrium}. By ``reversible'' we mean that each reaction possesses a corresponding reversing reaction. Note that the steady-state condition $\partial_t P(\mathbf{n}, t) = 0$ in the CME in Equation \eqref{cme} merely means that the probability flow out of a state (second term in the CME) equals the probability flow into a state (first term in the CME). These probabilities consist of sums over all reactions. Detailed balance is a stronger condition, as it requires the flow of each single reaction in each state to be balanced by its corresponding reversible reaction. Specifically, let $r$ and $r'$ be the indices of two reactions that revert each other and let $P(\mathbf{n})$ be a steady-state solution. The detailed balance condition reads \cite{Gardiner2009}
\begin{align}\label{detailed_balance}
  f_r(\mathbf{n}) P(\mathbf{n}) = f_{r'}(\mathbf{n} + \mathbf{S}_{r}) P(\mathbf{n}+\mathbf{S}_{r}),
\end{align}
which needs to be fulfilled for all reactions and all $\mathbf{n}$. Note that $\mathbf{S}_{r}=-\mathbf{S}_{r'}$. The left side of Equation \eqref{detailed_balance} is the flow out of state $\mathbf{n}$ due to the $r^{\text{th}}$ reaction, while the right side is the flow into state $\mathbf{n}$ due to the backward reaction $r'$. An analogue detailed balance condition can be formulated for the deterministic rate equations. It can be shown that for a reversible reaction system with mass-action kinetics, the stochastic system is in detailed balance if and only if the deterministic system is in detailed balance \cite{Whittle1986}. If the detailed balance condition is fulfilled, the solution of the CME is a product of Poisson distributions times a function accounting for conservation laws in molecule numbers, where the mean values of the Poisson distributions are simply given by the detailed balance solutions of the rate equations \cite{van1976equilibrium, Gardiner2009}.  

As an example, consider a closed, bimolecular system with reactions ${A+A \xrightleftharpoons[k_2] {\quad k_1 \quad } B}$. Let $n_A$ and $n_B$ be the molecule numbers of $A$ and $B$ respectively. The quantity $n_A+ 2n_B = C$ is conserved. The corresponding steady-state solution of the CME in detailed balance reads
\begin{align}\label{detailed_balance_sol}
  P(n_A,n_B) = \frac{e^{- \alpha_1} \alpha_1^{n_A}}{n_A !}  \frac{e^{- \alpha_2}  \alpha_2^{n_B}}{n_B !} \delta (n_A+ 2 n_B -C),
\end{align}
where $\alpha_1 = \Omega \phi^0_1$ and $\alpha_2 = \Omega\phi^0_2$, $\phi^0_1$ and $\phi^0_2$ are the steady-state solutions of the rate equations of $A$ and $B$, respectively, $\Omega$ is the system volume, and $\delta$ is the delta function accounting for the conservation law. 

These results have more recently been generalised to \emph{weakly reversible} reaction systems \cite{Anderson2010}. A system is called weakly reversible when for each reaction the change in state space can be reversed by a chain of other reactions. For example, a system with the reactions $X_1 \to X_2 \to X_3 \to X_1$ is not reversible but it is weakly reversible. 
Weak reversibility is a generalisation of reversibility since each reversible system is also weakly reversible. Correspondingly, the concept of \emph{complex balance} has been employed which is a generalisation of detailed balance \cite{Horn1972}. 
It can be shown that if the rate equations of a system with mass-action kinetics are complex balanced, the CME possesses a steady-state solution given by a product of Poisson distributions times a function accounting for conservation laws \cite{Anderson2010}. The mean values of the Poisson distributions are given by the complex-balance solutions of the rate equations. Moreover, it can be shown that for any mass-action kinetics reaction system
with deficiency zero (see \cite{Feinberg1995} for a definition), the rate equations have a complex-balance solution if and only if the system is weakly reversible \cite{Feinberg1995}.

\subsubsection{Exact results for special cases}

Most reaction systems of practical interest are neither linear nor do they satisfy the detailed/complex balance conditions. However, the CME has been solved for several specific systems that do not fall under any of these categories. In particular for systems with only a single reaction it is often straightforward to obtain a time-dependent solution of the CME. Some examples are given below.
{Most of these solutions are obtained
either by means of the aforementioned generating function method or the Poisson repre-
sentation. While the former transforms the CME into a partial differential equation, the
latter maps the discrete process underlying the CME to a continuous diffusion process
governed by a Langevin equation defined in a complex-valued space \cite{gardiner1977poisson}. We would like to stress that both approaches
are exact. However the Poisson representation has not been used much in the literature, probably because of its relative complexity compared to other methods and hence we do
not cover it in detail here. We refer the reader to the following references for specific
applications \cite{voituriez2005corrections, gerstung2009noisy, thomas2010stochastic, petrosyan2014nonequilibrium, iyer2014mixed,sugar2014self, Schnoerr2016}. For more details on the Poisson representation
and the generating function approach we refer the reader to \cite{Gardiner2009}.} \\


\noindent
\textbf{Single reaction systems.} 

\noindent
Time-dependent solutions have been derived for
\begin{itemize}
\item $A \to A+A$ \cite{Delbruck1940}
\item $A+B \to C$ \cite{ishida1964stochastic}
\item $A+A \to C$ \cite{ishida1964stochastic}
\item $A+B \leftrightarrow C$ \cite{laurenzi2000analytical}
\end{itemize}

\noindent
\textbf{Gene regulatory networks.}
\begin{itemize}
\item $G^{\text{on}}  \to G^{\text{on}} + P, 
   G^{\text{on}}  + P \leftrightarrow G^{\text{off}} ,
       P  \to \varnothing, G^{\text{off}}  \to G^{\text{off}} + P, G^{\text{off}} \to G^{\text{on}}$: a steady-state solution has been derived if there is one gene in the system \cite{GrimaNewman2012}. Note that if we ignore the last two reactions this system corresponds to the gene system in Figure \ref{fig_gene_system}.
\item A system similar to the previous one but with geometrically distributed bursts of protein production has been solved in steady state in  \cite{kumar2014exact}.
\end{itemize}

\noindent
\textbf{Enzyme systems.}
\begin{itemize}
\item $E+S \leftrightarrow C \to E + P$: A time-dependent solution has been derived for the case of a single enzyme in the system \cite{aranyi1976full}.
\item Adding the reaction $E+P \to C$ to the previous system makes it reversible and a corresponding steady-state solution has been derived in \cite{staff1970stochastic}.
\item The system with added substrate input reaction $\varnothing \to S$ has been solved in steady state in \cite{Schnoerr2014}.
\item A similar system but with multiple substrate species competing for the enzyme has been solved in steady state in \cite{mather2010correlation}.
\end{itemize}
Most of these examples only consider steady-state solutions and all of them consider systems with few species. The lack of solutions for more complicated systems makes the development of approximation methods necessary.

\section{Approximation methods}\label{sec_approximations}

As discusses in the previous Section, analytic solutions of the CME are known only for very restrictive classes of systems and few simple special cases. For most systems of practical interest, no analytic solutions are known to date. Stochastic simulation algorithms introduced in Section \ref{sec_ssa} allow to simulate exact samples of the underlying stochastic process, but they quickly become computationally infeasible for larger systems. For these reasons significant effort has been spent in the literature on the development of approximation methods. We give here an introduction to a wide range of such methods. First, we give a detailed introduction to a few approximation methods that can be applied to (almost) arbitrary systems without any pre-knowledge of the system necessary. We derive the approximations, give examples and discuss their properties.  Subsequently, we give an overview of other types of approximation methods developed in the literature. 

The first method aims at approximating the solution of the CME, namely the \emph{chemical Fokker-Planck equation} (CFPE) and the associated \emph{chemical Langevin equation} (CLE), which we introduce in Section \ref{sec_stochkin_cle}. The CFPE/CLE define an approximating stochastic process.  
An alternative method for approximating the solution of the CME is given by the \emph{system size expansion}, which constitutes a systematic expansion of the CME in the inverse volume size. The system size expansion includes the popular \emph{linear noise approximation} (LNA) and also provide an efficient way of approximating the moments of a process. We discuss the system size expansion in Section \ref{sec_sse}. Next, in Section \ref{sec_stochkin_ma} we introduce a certain class of \emph{moment closure approximations}, which approximate the moments of a process. In Section \ref{sec_max_entropy} we show how such approximate moments can be used to construct distributions using the maximum entropy principle. Next, in Section \ref{sec_software} we review software packages implementing the discussed approximation methods.  Finally, in Section \ref{sec_other_approximations} we give a brief overview of other approximation methods found in the literature. As we shall see, many of these methods use the CLE, the system size expansion or moment closure approximations as building blocks.

\subsection{The chemical Langevin equation}\label{sec_stochkin_cle}

The \emph{chemical Langevin equation} (CLE) and the corresponding \emph{chemical Fokker-Planck equation} (CFPE) constitute a popular diffusion approximation of the CME. Kramers and Moyal derived the CFPE by applying a Taylor expansion to the CME which upon truncation leads to a partial differential equation approximation of the CME \cite{Kramers1940,Moyal1949}: 
Suppose we let the variables in the CME in Equation \eqref{cme} become continuous and let $\mathbf{x}=(x_1,\ldots,x_N)$, where $x_i$ is the continuous variable denoting the molecule number of species $X_i$. 
Performing a Taylor expansion to second order around $\mathbf{x}$ in the first term of the r.h.s.~of Equation \eqref{cme} gives (with $\mathbf{n}$ replaced by $\mathbf{x}$)
\begin{align}\nonumber
   f_r (\mathbf{x} - \mathbf{S}_r) P(\mathbf{x} - \mathbf{S}_r, t) 
     & \approx 
       f_r (\mathbf{x}) P(\mathbf{x}, t) - 
       \sum_{i=1}^N S_{ir}  \frac{\partial}{\partial x_i} [f_r (\mathbf{x}) P(\mathbf{x}, t)]
       \\
       \label{cfpe_taylor}
    & \quad \quad   + \sum_{i,j=1}^N S_{ir} S_{jr}  \frac{\partial}{\partial x_i}\frac{\partial}{\partial x_j} [f_r (\mathbf{x}) P(\mathbf{x}, t)].
\end{align}
Inserting this into the CME in Equation \eqref{cme}, we see that the first term on the r.h.s. of Equation \eqref{cfpe_taylor} cancels the last term on the r.h.s. of Equation \eqref{cme} leading to the CFPE:
\begin{align}\label{cfpe}
   \partial_t P(\mathbf{x},t)
   = &
     - \sum_{i=1}^N \frac{\partial}{\partial x_i} \left[ A_i(\mathbf{x}) P(\mathbf{x},t) \right]
     + \frac{1}{2} \sum_{i,j=1}^N \frac{\partial}{\partial x_i}\frac{\partial}{\partial x_j}
         \left[ B_{ij}(\mathbf{x})  P(\mathbf{x},t) \right], 
\end{align}
where the drift vector $\mathbf{A}$ and diffusion matrix $\mathbf{B}$ are respectively given by
\begin{align}\label{cle_drift}
  A_i(\mathbf{x})
  & = 
    \sum_{r=1}^R S_{ir} f_r(\mathbf{x}),  \\
\label{cle_diff}
  B_{ij}(\mathbf{x})
  & = 
    \sum_{r=1}^R S_{ir} S_{jr} f_r(\mathbf{x}).
\end{align}
Note that the drift vector $\mathbf{A}(\mathbf{x})$ and diffusion matrix $\mathbf{B}(\mathbf{x})$ do not depend on time. 
Note also that whereas the state variables denote discrete molecule numbers in the CME, they denote continuous real numbers in the CFPE. 
The CFPE in Equation \eqref{cfpe} is equivalent to the CLE
\begin{align}\label{cle}
  d \mathbf{x} 
  & = 
    \mathbf{A}(\mathbf{x}) dt + \mathbf{C}(\mathbf{x}) d\mathbf{W}, \quad \quad \mathbf{C}(\mathbf{x}) \mathbf{C}(\mathbf{x})^T = \mathbf{B}(\mathbf{x}),
\end{align}
which is an \emph{Ito stochastic differential equation} \cite{Gardiner2009}. $\mathbf{W}$ in Equation \eqref{cle} is a multi-dimensional Wiener process. It can be shown \cite{Gardiner2009} that the distribution of a process described by Equation \eqref{cle} agrees exactly with the solution of the corresponding Fokker-Planck equation in \eqref{cfpe}. One can thus interpret the CLE in Equation \eqref{cle} as a generator of realisations of the stochastic process described by the corresponding CFPE. In this sense, the CLE and CFPE are considered to be equivalent to each other.
An alternative derivation, which leads directly to the CLE in Equation \eqref{cle}, was given by Gillespie in \cite{gillespie2000chemical}.

Generally there exist different choices for $\mathbf{C}(\mathbf{x})$ in Equation \eqref{cle}  corresponding to different factorisations of the matrix $\mathbf{B}(\mathbf{x})$; these lead to as many different representations of the CLE. One possibility is given by
\begin{align}\label{cle_standard_form}
  d x_i
  & = 
    \sum_{r=1}^R S_{ir} f_r (\mathbf{x}) dt
    + \sum_{r=1}^R S_{ir} \sqrt{f_r (\mathbf{x})} dW_r, \quad i=1,\ldots, N.
\end{align}
This representation of the CLE is the one most commonly used in the literature \cite{gillespie2000chemical,Melykuti2010}.

\textbf{Example.}
Let us come back to the gene expression system in Figure \ref{fig_gene_system} with reactions in Equation \eqref{gene_feedback_reactions} and consider the corresponding CFPE and CLE. Using the stoichiometric matrix in Equation \eqref{gene_stoichiometric} and propensity vector in Equation  \eqref{gene_prop} we obtain for the drift vector and diffusion matrix defined in Equation \eqref{cle_drift} and Equation \eqref{cle_diff}, respectively,
\begin{align}
  \mathbf{A}(\mathbf{x})
  & = 
        \begin{pmatrix}
      - \frac{k_2}{\Omega} x_1 x_2 + k_3 (1-x_1) \\
      k_1 x_1 - \frac{k_2}{\Omega} x_1 x_2 + k_3 (1-x_1) - k_4 x_2
    \end{pmatrix},  \\
  \mathbf{B}(\mathbf{x})
  & = 
    \begin{pmatrix}
      \frac{k_2}{\Omega} x_1 x_2 + k_3 (1-x_1)  &  \frac{k_2}{\Omega} x_1 x_2 + k_3 (1-x_1) \\
      \frac{k_2}{\Omega} x_1 x_2 + k_3 (1-x_1) &  k_1 x_1 + \frac{k_2}{\Omega} x_1 x_2 + k_3 (1-x_1) + k_4 x_2
    \end{pmatrix},
\end{align}
where $x_1$ and $x_2$ are the (continuous) particle numbers of $G^{\text{on}}$ and $P$, respectively.
To obtain the CLE in Equation \eqref{cle}, we have to compute $\mathbf{C}$, which is the square root of $\mathbf{B}$ and thus generally not uniquely defined. One possibility as given in Equation \eqref{cle_standard_form} reads
\begin{align}\label{gene_cle_c_matrix}
  C_{ij}(\mathbf{x})
  & = 
    \begin{pmatrix}
      0  &  - \sqrt{\frac{k_2}{\Omega} x_1 x_2}  &  \sqrt{k_3 (1-x_1)}  &  0 \\
     \sqrt{ k_1 x_1} &  - \sqrt{\frac{k_2}{\Omega} x_1 x_2}  &  \sqrt{k_3 (1-x_1)}  &  - \sqrt{k_4 x_2}
    \end{pmatrix}.
\end{align}
This gives rise to the CLE
\begin{align}\label{gene_cle}
  d x_1
  & = 
    (- \frac{k_2}{\Omega} x_1 x_2 + k_3 (1-x_1)) dt - \sqrt{\frac{k_2}{\Omega} x_1 x_2}  dW_2
    +  \sqrt{k_3 (1-x_1)} dW_3, \\
\label{gene_cle2}
  d x_2 
  & =
    (k_1 x_1 - \frac{k_2}{\Omega} x_1 x_2 + k_3 (1-x_1) - k_4 x_2) dt \\
\nonumber
  &  \quad 
    + \sqrt{ k_1 x_1} dW_1  - \sqrt{\frac{k_2}{\Omega} x_1 x_2} dW_2  + \sqrt{k_3 (1-x_1)} dW_3  - \sqrt{k_4 x_2} dW_4,
\end{align}
where the $W_i$ are independent Wiener processes. Note that it does not make a difference if one changes the signs in front of the square roots in Equation \eqref{gene_cle_c_matrix}, or equivalently in front of the noise terms in Equation \eqref{gene_cle}, as long as one does so simultaneously for each occurrence of a specific term, i.e., changes the sign of whole columns in Equation \eqref{gene_cle_c_matrix}. To see that such changes are equivalent note that the diffusion matrix $\mathbf{B}= \mathbf{C} \mathbf{C}^T$ of the CFPE is invariant under such changes. This can also be seen directly from the CLE in Equations \eqref{gene_cle} and \eqref{gene_cle2} since the Wiener processes are symmetric.

\subsubsection{Stochastic simulations}\label{sec_euler_maruyama}

As for the CME, there are no analytic solutions known for the CLE for most systems. However, the computational cost of CLE simulations scales with the number of species, rather than with the rate of reaction events as CME simulations. This means that CLE simulations are often more efficient than simulations of the CME, typically  if the molecule counts in a system are not too small. 

A popular method to simulate the CLE is the \emph{Euler-Maruyama algorithm} which discretises time into intervals $d t$ and simulates the process iteratively as \cite{Kloeden1992}
\begin{align}\label{euler_maruyama}
  x(t+d t)
  & =
    x(t) + A (x(t)) d t + \sqrt{B(x(t)) d t} ~ dw, \quad dw \sim \mathcal{N}(0,1),
\end{align}
where $\mathcal{N}(0,1)$ is a normal distribution with mean $0$ and variance $1$. The smaller $d t$, the better the true process is approximated by Equation \eqref{euler_maruyama} but also the slower the algorithm becomes. The right choice of $d t$ is therefore a tradeoff between accuracy and efficiency.  
It is important to point out that more efficient simulation methods exist, see for example \cite{Kloeden1992,burrage2000numerical, higham2001algorithmic,burrage2004numerical}.

\subsubsection{Properties and recent developments}

The CLE is typically a good approximation whenever the molecule numbers of the system are not too small and is particularly useful if one is interested in time trajectories or distributions of a process. It can be shown that the differences between the CLE and the CME tend to zero in the limit of large molecule numbers \cite{Kurtz}. In other words, the CLE becomes exact in the thermodynamic limit. 

By multiplying the CFPE in Equation \eqref{cfpe}  by $x_i \ldots x_l$ and integrating over the whole state space, one obtains ordinary differential equation  for the moment $\langle x_i \ldots x_l \rangle$ of the process described by the CLE. Importantly, it turns out that the equations for moments of up to order two are \emph{exactly the same as the corresponding equations derived from the CME.} These are given in Equations \eqref{cme_closure_order_one} and \eqref{cme_closure_order_two}. Note however that since they are generally coupled to higher order moments for which the evolution equations derived from the CLE and CME do not agree, the first two moments (and higher order moments) of the CLE do not generally agree with the ones of the CME. However, since the moment equations decouple for linear systems as shown in Section \ref{sec_moment_equations}, we obtain the important result that \emph{the moments up to order two of the processes described by the CLE and CME agree exactly for linear reaction systems}.

\begin{figure} [t]
\centering
  \includegraphics[scale=0.45]{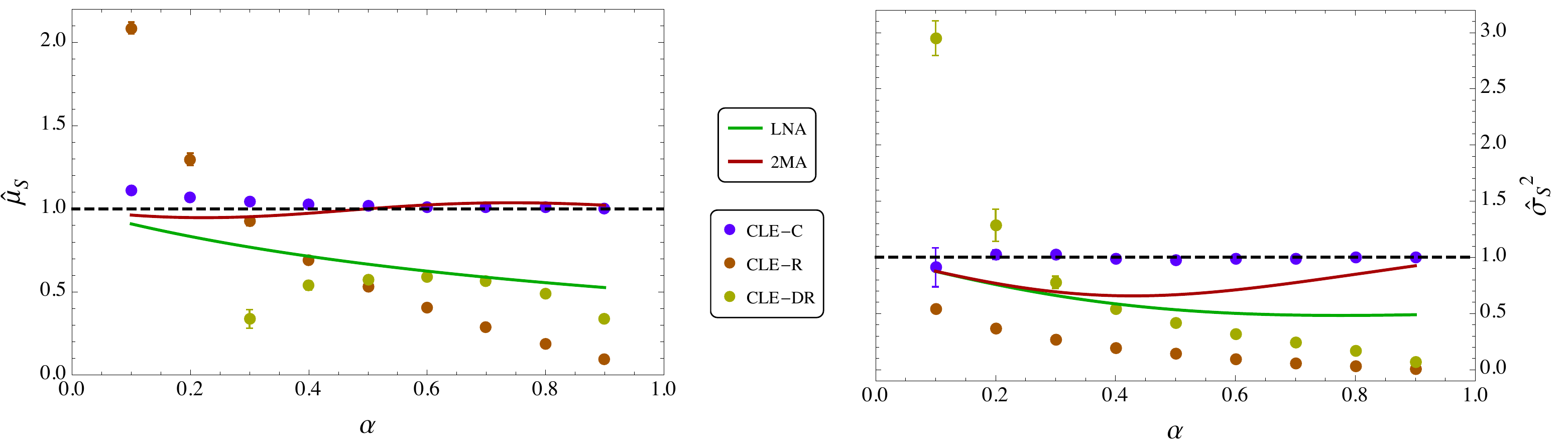} 
  \caption{\textbf{Comparison of different CLE implementations.} The figures show the steady-state mean (left) and variance (right) of a substrate in a Michaelis-Menten reaction system as a function of the strength of the substrate-to-enzyme binding strength. The results are normalised by the exact CME result, which means that the horizontal dashed line corresponds to the exact value. CLE-C corresponds to the complex-valued CLE, CLE-R to a real-valued implementation with rejecting boundaries at zero molecules, and CLE-DR to the real-valued version proposed in \cite{Dana2011}. LNA and 2MA correspond to the linear noise approximation and the second-order normal moment closure, which we will introduce in Sections \ref{sec_sse} and \ref{sec_stochkin_ma}, respectively. We observe that both real-valued implementations give large deviations from the CME result, significantly more than the LNA and 2MA. The complex CLE on the other hand is significantly more accurate than all the other methods.
  The figure is taken from \cite{Schnoerr2014} with permission kindly granted by the publisher. }
  \label{fig_cle_comparisons}
\end{figure}

Note that the CME has a natural boundary at zero molecule numbers, i.e., for a sensible initial condition with zero probability to have a negative number of molecules, this probability remains zero for all times. Until recently it has not been clear how this boundary condition behaves when approximating the discrete process underlying the CME by a continuous process using the CLE. As it turns out, this boundary issue leads to an ill-definedness of the CLE due to occurrences of square roots of negative expressions in finite time with finite probability \cite{Wilkie2008, Szpruch2009}. Since traditionally the domain of the CLE is (implicitly) assumed to be that of real numbers, the CLE is not well-defined in this case.  
More recently, it has been shown that this problem is independent of the chosen factorisation of the CFPE's noise matrix $\mathbf{B}$ (c.f.~Equation \eqref{cle}), which means that the CLE is not well-defined for real variables \cite{Schnoerr2014}. In this study it has been shown that the same can be expected for the majority of reaction systems.

Several modified versions of the CLE have been proposed that try to keep the state space real, for example \cite{Wilkie2008, Dana2011}. However, these are \emph{ad hoc} modifications and have been found to  introduce high inaccuracies for some non-linear reaction systems \cite{Schnoerr2014}. Importantly, they also have been found to violate the CLE's exactness for the moments up to order two for linear systems. 
Alternatively, the ill-defined problem can be solved by extending the state space of the CLE to complex variables. It has been proven that this leads to real-valued moments, real-valued autocorrelation functions and real-valued power spectra, and to restore the CLE's exactness  for the moments up to order two for linear systems \cite{Schnoerr2014}. This complex CLE
was found to be highly accurate for some non-linear systems in comparison to other modified versions. For one example the results from \cite{Schnoerr2014} are shown in Figure \ref{fig_cle_comparisons}.
The complex-valued CLE has other drawbacks, however. For example, it does not directly give approximations of the process or of distributions. Rather, the results have to be projected to real space. 

In recent years the CLE has been frequently used for approximating the dynamics of intermediate or high abundance species in so-called \emph{hybrid methods}, see Section \ref{sec_other_approx_hybrid}. In this case, the probability of negative concentrations becomes very small. If the partitioning into low and high abundance species is done adaptively, the problem of square roots of negative expressions is avoided completely. In this case the complex-valued CLE automatically reduces to the real-valued CLE. The CLE is particularly useful in simulation based hybrid methods since it gives an approximation of the whole process (rather than for example only the process's moments). 

Recently, a tensor-based method has been proposed in \cite{liao2015tensor} for the direct numerical solution of the CFPE. This is a promising approach since it avoids computationally expensive ensemble averaging and has been used for sensitivity and bifurcation analysis \cite{liao2015tensor}.
Recall that the CLE becomes exact in the limit of large system sizes. As one may expect, the CLE therefore generally captures the multimodality of the CME's solution if the multimodality persists for large volumes, since this is the case if the deterministic rate equations are multistable. Surprisingly, however, it has recently been found that the CLE is also able to reproduce \emph{noise-induced multimodality} of the CME's solution, i.e., multimodality that occurs only if the system volume is decreased below a certain critical value \cite{biancalani2014noise}. However, it was found that this is not the case for all reaction systems \cite{Duncan2015b}. In \cite{beccuti2014analysis}, the CLE as a diffusion approximation was applied to Petri nets.

\subsection{The system size expansion}\label{sec_sse}

Suppose we are not interested in approximating the whole process but only in its distribution or first few moments. Running stochastic simulations of the CLE seem like  unnecessary computational effort in this case and the question arises if there exist more efficient approximation methods to achieve this. We next discuss the \emph{system size expansion} which aims at approximating the distribution or the first few moments of a process.  The system size expansion is a perturbative expansion of the CME in the inverse system size originally developed by van Kampen \cite{VanKampen1976,VanKampen2007}. The idea is to separate concentrations into a deterministic part, given by the solution of the deterministic rate equations, and a part describing the fluctuations about the deterministic mean.

\subsubsection{Derivation}

The system size expansion splits the instantaneous particle numbers $n_i$ of the CME into a deterministic part and a fluctuating part as 
\begin{align}\label{sse_ansatz}
  \frac{n_i}{\Omega}
  & = 
    \phi_i + \Omega^{-1/2} \epsilon_i, \quad i=1, \ldots, N,
\end{align}
where $\Omega$ is the volume of the system, $\phi_i$ is the solution of the deterministic rate equations in \eqref{res} and we introduced the new variables $\epsilon_i$ representing fluctuations about the deterministic mean. Following \cite{Thomas2012b}, the system size expansion proceeds by transforming the master equation to these new variables $\epsilon_i$. To this end, we rewrite the CME in Equation \eqref{cme} as
\begin{align}\label{cme_step_ops}
  \partial_t P(\mathbf{n},t) 
  & = 
    \sum_{r=1}^R \left( \prod_{i=1}^N E_i^{-S_{ir}} - 1 \right) f_r (\mathbf{n}) P(\mathbf{n}, t),
\end{align}
where we introduced the step operators $E_i^{-S_{ir}}$, which are defined in terms of their action on a general function of the state space, $h(\mathbf{n})=h(n_1, \ldots, n_N)$ as
\begin{align}\label{sse_def_step_op}
  E_i^{-S_{ir}} h(n_1, \ldots, n_i, \ldots,  n_N)
  & =
    h(n_1, \ldots, n_i - S_{ir}, \ldots,  n_N).
\end{align}
The system size expansion now proceeds in three steps: 
\begin{enumerate}
\item \textbf{Time derivative.} Due to the change of variables in Equation \eqref{sse_ansatz} we need to transform the distribution $P(\mathbf{n},t)$ into a distribution $\Pi(\bm{\epsilon},t)$ of the new variables. Note that the time derivate in the CME in Equation \eqref{cme_step_ops} is taken at constant $\mathbf{n}$, which implies $\partial \epsilon_i / \partial_t = - \Omega^{1/2} \partial \phi_i / \partial_t$. This in turn leads to 
\begin{align}\label{sse_time_derivative}
  \partial_t P(\mathbf{n},t)
  & =
    \frac{\partial \Pi(\bm{\epsilon},t)}{\partial t} - \Omega^{1/2} \sum_{i=1}^N \frac{\partial \phi_i} {\partial t} \frac{\partial \Pi(\bm{\epsilon},t)}{\partial \epsilon_i}.
\end{align}
\item \textbf{Step operator.} Using Equation \eqref{sse_ansatz} in the definition of the step operator in Equation \eqref{sse_def_step_op}, we obtain 
\begin{align}
  \prod_{i=1}^N E_i^{-S_{ir}} h(\epsilon_1, , \ldots,  \epsilon_N)
  & =
    h(\epsilon_1 - \Omega^{-1/2} S_{1r}, \ldots,  \epsilon_N - \Omega^{-1/2} S_{Nr}).
\end{align}
Expanding the r.h.s. in powers of $\Omega^{-1/2}$ we can expand the term of the CME involving the step operators as 
\begin{align}\label{sse_step_op_expansion}
 \nonumber
  \prod_{i=1}^N E_i^{-S_{ir}} -1
  & =
    - \Omega^{-1/2} \sum_{i=1}^N S_{ir}  \frac{\partial}{\partial \epsilon_i} 
    + \frac{\Omega^{-1}}{2} \sum_{i,j=1}^N S_{ir}  S_{jr} \frac{\partial^2}{\partial \epsilon_i \partial \epsilon_j} \\
  & \quad 
    - \frac{\Omega^{-3/2}}{6} \sum_{i,j,l=1}^N S_{ir}  S_{jr} S_{lr} \frac{\partial^3}{\partial \epsilon_i \partial \epsilon_j \partial \epsilon_l} 
    + O(\Omega^{-2}).
\end{align}
\item \textbf{Propensity functions.} We assume that the propensity functions in the CME can be expanded as 
\begin{align}\label{sse_prop_expansion}
  f_r(\mathbf{n})
  & = 
    \Omega \sum_{i=0}^{\infty} \Omega^{-i} f_r^{(i)}(\tfrac{\mathbf{n}}{\Omega}).
\end{align}
If we assume mass-action kinetics for which the propensity functions take the form in Equation \eqref{general_propensity}, the propensity functions are polynomials and it is easy to see that such an expansion always exists. The same is also true for many propensity functions that are not of mass-action kinetics type, such as Michaelis-Menten and Hill-type propensity functions. Using Equation \eqref{sse_ansatz} and expanding the r.h.s. of Equation \eqref{sse_prop_expansion} in $\Omega^{-1/2}$ we obtain
\begin{align}\label{sse_prop_expansion2}
    \nonumber
  f_r(\mathbf{n})
  & = 
    g_r(\bm{\phi}) + \Omega^{-1/2} \sum_{i=1}^N \epsilon_i \frac{\partial g_r(\bm{\phi})}{\partial \phi_i}
    \\
    & \quad 
      + \Omega^{-1} \left( \frac{1}{2} \sum_{i,j=1}^N \epsilon_i \epsilon_j \frac{\partial^2 g_r(\bm{\phi})}{\partial \phi_i \partial \phi_j} 
      - f_r^{(1)} \right)
      + O(\Omega^{-3/2}),
\end{align}
where we have identified $f_r^{(0)} (\bm{\phi}) = g_r (\bm{\phi})$ and $g_r (\bm{\phi})$ is the macroscopic rate function of the $r^{\text{th}}$ reaction introduced in Equation \eqref{re_propensity} in context of the deterministic rate equations.
\end{enumerate}
Applying Equations \eqref{sse_time_derivative}, \eqref{sse_step_op_expansion} and  \eqref{sse_prop_expansion2} to Equation \eqref{cme_step_ops} we obtain 
\begin{equation}\label{sse_emre_cme}
\begin{split}
  \partial_t \Pi(\bm{\epsilon}, t)
  & =
      [ \mathcal{L}^{(0)} + \Omega^{-1/2} \mathcal{L}^{(1)} + \Omega^{-1} \mathcal{L}^{(2)} + O(\Omega^{-3/2}) ] \Pi(\bm{\epsilon}, t),
\end{split}
\end{equation}
where we defined 
\begin{align}
\label{sse_op_0}
  \mathcal{L}^{(0)}
  & =
    - \partial_i {J^{(0)}}_i^j  \epsilon_j + \frac{1}{2} D^{(0)}_{ij} \partial_i \partial_j, \\
\label{sse_op_1}
  \mathcal{L}^{(1)}
      & \quad
    + \Omega^{-1/2} \Big(
    \tfrac{1}{2}{J^{(0)}}_{i,j}^{q} \partial_i \partial_j \epsilon _q 
    -\tfrac{1}{2}  {J^{(0)}}_{i}^{q,r} \partial_i\epsilon _q \epsilon _r
    -\tfrac{1}{6}  D^{(0)}_{i,j,k} \partial_i \partial_j \partial_k
    - D^{(1)}_i  \partial_i \Big) \Big] \Pi, \\
\label{sse_op_2}
   \nonumber
  \mathcal{L}^{(2)}
      & \quad
      + \Omega^{-1} \Big(
      -\tfrac{1}{6} {J^{(0)}}_{i,j,k}^{q} \partial_i \partial_j \partial_k \epsilon _q
   +\tfrac {1}{4} {J^{(0)}}_{i,j}^{q,r} \partial_i \partial_j \epsilon _q \epsilon _r
   -\tfrac {1}{6} {J^{(0)}}_{i}^{q,r,s} \partial_i \epsilon _q \epsilon _r \epsilon _s \\
   & \quad \quad
   -{J^{(1)}}_{i}^{q} \partial_i \epsilon _q
   + \tfrac{1}{2} D^{(1)}_{i,j} \partial_i \partial_j
   +\tfrac{1}{24} D^{(0)}_{i,j,k,l}  \partial_i \partial_j \partial_k \partial_l 
    \Big).
\end{align}
Here we use the shorthand $\partial_i = \partial/\partial \epsilon_i$ and assume summations over doubly occurring indices for notational simplicity. We note that in Equation \eqref{sse_emre_cme} also two terms of order $\Omega^{1/2}$ occur. However, these terms cancel each other because they just correspond to the deterministic rate equations. In Equations \eqref{sse_op_0}-\eqref{sse_op_2} we defined 
\begin{align}\label{lna_d_matrix}
  D^{(n)}_{i, \ldots k}
  & = 
    \sum_{r=1}^R S_{ir} \ldots S_{kr} f_r^{(n)}(\phi), \\
\label{lna_j_matrix}
  {J^{(n)}}_{i,\ldots, k}^{l,\ldots, m}
  & = 
     \partial_{\phi_l} \ldots \partial_{\phi_m}D^{(n)}_{i, \ldots k}.
\end{align}
Note that both the $\mathbf{J}^{(i)}$ and $\mathbf{D}^{(i)}$ matrices depend on the solution $\bm{\phi}$ of the rate equations in \eqref{res} and are thus generally time dependent. Comparing with Equation \eqref{res} we find that $\mathbf{J}^{(0)}$ in Equation \eqref{lna_j_matrix} is just the Jacobian of the rate equations.

\subsubsection{The linear noise approximation}\label{subsec_lna}

The popular \emph{linear noise approximation} (LNA) \cite{VanKampen1976,VanKampen2007} is obtained if we truncate the expansion in Equation \eqref{sse_emre_cme} to zeroth order, leading to
\begin{align}\label{lna}
  \partial_t \Pi(\bm{\epsilon}, t)
  & =
      \Big[ -  \sum_{i=1}^N \frac{\partial}{\partial \epsilon_i} \sum_{j=1}^N {J^{(0)}}_i^j  \epsilon_j + \frac{1}{2} \sum_{i,j=1}^ND^{(0)}_{ij} \frac{\partial}{\partial \epsilon_i} \frac{\partial}{\partial \epsilon_j}  \Big] \Pi(\bm{\epsilon}, t) + O(\Omega^{-1/2}).
\end{align}
Equation \eqref{lna} is a Fokker-Planck equation with drift and diffusion linear and constant in $\bm{\epsilon}$, respectively,  and hence has a multivariate normal solution under appropriate initial conditions. 
By multiplying Equation \eqref{lna} with $\epsilon_i$ and $\epsilon_i \epsilon_j$ and integrating over all $\bm{\epsilon}$, one obtains ordinary differential equations for the first and second-order moments, $\langle \epsilon_i \rangle$ and $\langle \epsilon_i \epsilon_j \rangle$, respectively. By doing so one finds that if the mean is initially zero, $\langle \epsilon_i \rangle |_{t=0} = 0$, it remains zero for all times. This is normally the case since otherwise the initial condition of the rate equations would not agree with the initial mean value of the stochastic system. If we additionally assume deterministic initial conditions, we also have  $\langle \epsilon_i \epsilon_j \rangle |_{t=0} = 0$. 
The solution of Equation \eqref{lna} is thus a multivariate normal distribution with zero mean. Since $\bm{n}$ and $\bm{\epsilon}$ are related by the linear transformation given in Equation \eqref{sse_ansatz}, the distribution of $\bm{n}$ is also given by a multivariate normal distribution. The mean of the latter satisfies the rate equations in Equation \eqref{res} and the covariance $\bm{\Sigma}$ defined as $\Sigma_{ij} = \langle n_i n_j \rangle - \langle n_i  \rangle \langle n_j \rangle$  fulfils
\begin{align}\label{lna_eqs_covariance}
  \partial_t \bm{\Sigma}
  & = 
    \bm{J}^{(0)} \bm{\Sigma} + \bm{\Sigma} {\bm{J}^{(0)} }^T + \Omega^{-1} \bm{D}^{(0)} .
\end{align}
The LNA describes the lowest order fluctuations of the system size expansion about the deterministic mean and is valid in the limit of large volumes, i.e., the thermodynamic limit. 

An alternative derivation of the LNA is given in \cite{Gardiner2009}, where it is derived from the chemical Langevin equation given in Equation \eqref{cle}.

\textbf{Example. }
Let us consider the rate equations and LNA for the gene system in Figure \ref{fig_gene_system}. The corresponding stoichiometric matrix, the macroscopic rate vector and the rate equations were already derived in Section \ref{sec_chemical_networks} and are given in Equations \eqref{gene_stoichiometric} and \eqref{gene_res}, respectively. 
The LNA defined in Equation \eqref{lna} requires matrices $\mathbf{J}^{(0)}$ and $\mathbf{D}^{(0)}$ defined in Equations \eqref{lna_d_matrix} and \eqref{lna_j_matrix}, respectively, for which we obtain
\begin{align}\label{gene_lna_d}
  \mathbf{D}^{(0)}
  & =
        \begin{pmatrix}
      k_2 \phi_1 \phi_2 + k_3 (\frac{1}{\Omega} - \phi_1)  &  k_2  \phi_1 \phi_2 + k_3 (\frac{1}{\Omega}-\phi_1) \\
      k_2  \phi_1 \phi_2 + k_3 (\frac{1}{\Omega}-\phi_1) &  k_1 \phi_1 + k_2  \phi_1 \phi_2 + k_3 (\frac{1}{\Omega}-\phi_1) + k_4 \phi_2
    \end{pmatrix}, \\
\label{gene_lna_j}
  \mathbf{J} ^{(0)}
  & =
    \begin{pmatrix}
      - k_2 \phi_2 - k_3  &  - k_2  \phi_1 \\
      k_1 - k_2 \phi_1 - k_3  &  - k_2 \phi_1 - k_4
    \end{pmatrix}.
\end{align}
Note that $\mathbf{D}^{(0)}$ and $\mathbf{J}^{(0)}$ are functions of the time-dependent solutions $\phi_1$ and $\phi_2$ of the rate equations in \eqref{gene_res}. The solution of the LNA is a normal distribution in $\mathbf{n}=(n_1,n_2)$. Its mean is obtained by (numerically) solving the rate equations in \eqref{gene_res}, and  the covariance by subsequently solving Equation \eqref{lna_eqs_covariance} using Equations \eqref{gene_lna_d} and \eqref{gene_lna_j}.

\subsubsection{Higher order corrections}\label{subsec_emre}

Suppose we want to include higher orders in $\Omega^{-1/2}$ in Equation \eqref{sse_emre_cme} beyond the LNA. 
In contrast to the LNA, now the PDE in Equation \eqref{sse_emre_cme} can generally not be solved analytically. However, we can derive ordinary differential equations for the moments of the system accurate to the corresponding order in $\Omega^{-1/2}$, as follows. Assuming an expansion of the distribution $\Pi(\bm{\epsilon}, t)$ as  
\begin{align}\label{sse_pdf_expansion}
  \Pi(\bm{\epsilon}, t)
  & =
    \sum_{i=0}^\infty \Omega^{-i/2} \Pi^{(i)}(\bm{\epsilon}, t),
\end{align}
one obtains an expansion of the moments
\begin{align}\label{sse_moment_expansion}
  \langle \epsilon_j \ldots \epsilon_k \rangle
  & = 
    \sum_{i=0}^{\infty} [\epsilon_j \ldots \epsilon_k]^{(i)} \Omega^{-i/2}, 
    \quad \quad [\epsilon_j \ldots \epsilon_k]^{(i)} = \int d \bm{\epsilon}~\epsilon_j \ldots \epsilon_k \Pi^{(i)}(\bm{\epsilon}, t).
\end{align}
Multiplying Equation \eqref{sse_emre_cme} with $\epsilon_j \ldots \epsilon_k$, using Equations \eqref{sse_pdf_expansion} and \eqref{sse_moment_expansion} and integrating over all $\bm{\epsilon}$, we obtain
\begin{align}
  \partial_t [\epsilon_i]_0
  & = 
    {J^{(0)}}_i^q [ \epsilon_q]_0, \\
  \partial_t [\epsilon_i]_1
  & = 
    {J^{(0)}}_i^q [ \epsilon_q]_1 + D_i^{(1)} + \tfrac{1}{2} {J^{(0)}}_i^{q,r} [\epsilon_q \epsilon_r]_0 , \\
  \partial_t [\epsilon_i]_2
  & = 
    {J^{(0)}}_i^q [ \epsilon_q]_2 + \tfrac{1}{2} {J^{(0)}}_i^{q,r} [\epsilon_q \epsilon_r]_1
    + \tfrac{1}{6} {J^{(0)}}_i^{q,r,s} [\epsilon_q \epsilon_r \epsilon_s]_0 
    + {J^{(1)}}_i^{q} [\epsilon_q]_0, \\
  \partial_t [ \epsilon_i \epsilon_j ]_0
  & = 
    {J^{(0)}}_i^q [ \epsilon_j \epsilon_q]_0 + {J^{(0)}}_j^q [ \epsilon_i \epsilon_q]_0 + D^{(0)}_{ij}, \\
\nonumber
  \partial_t [ \epsilon_i \epsilon_j ]_1
  & = 
    {J^{(0)}}_i^q [ \epsilon_j \epsilon_q]_1 + {J^{(0)}}_j^q [ \epsilon_i \epsilon_q]_1 +    D^{(1)}_i [\epsilon_j]_0 + D^{(1)}_j [\epsilon_i]_0  \\
  & \quad
    + \tfrac{1}{2} {J^{(0)}}_i^{q,r} [ \epsilon_j \epsilon_q  \epsilon_r ]_0 + \tfrac{1}{2} {J^{(0)}}_j^{q,r} [ \epsilon_i \epsilon_q  \epsilon_r ]_0
    + {J^{(0)}}_{i,j}^{q} [\epsilon_q]_0, \\
\nonumber
  \partial_t [ \epsilon_i \epsilon_j ]_2
  & = 
    {J^{(0)}}_i^q [ \epsilon_j \epsilon_q]_2 + {J^{(0)}}_j^q [ \epsilon_i \epsilon_q]_2 +    D^{(1)}_i [\epsilon_j]_1 + D^{(1)}_j [\epsilon_i]_1 \\
\nonumber
  & \quad
    + \tfrac{1}{2} {J^{(0)}}_i^{q,r} [ \epsilon_j \epsilon_q  \epsilon_r ]_1 + \tfrac{1}{2} {J^{(0)}}_j^{q,r} [ \epsilon_i \epsilon_q  \epsilon_r ]_1
    + {J^{(0)}}_{i,j}^{q} [\epsilon_q]_1      \\
\nonumber
  & \quad
    + D^{(1)}_{ij} 
    + \tfrac{1}{6} {J^{(0)}}_i^{q,r,s} [ \epsilon_j \epsilon_q  \epsilon_r \epsilon_s]_0 + \tfrac{1}{6} {J^{(0)}}_j^{q,r,s} [ \epsilon_i \epsilon_q  \epsilon_r \epsilon_s]_0 \\
  & \quad
    + \tfrac{1}{2} {J^{(0)}}_{i,j}^{q,r} [\epsilon_q \epsilon_r]_0
    + {J^{(1)}}_i^q [\epsilon_j \epsilon_q]_0 + {J^{(1)}}_j^q [\epsilon_i \epsilon_q]_0, \\
  \partial_t [ \epsilon_i \epsilon_j \epsilon_k ]_0
  & = 
    D_{ij} [\epsilon_k]_0 + (i \leftrightarrow j \leftrightarrow k)
    + {J^{(0)}}_i^q [ \epsilon_j \epsilon_k \epsilon_q ]_0 + (i \leftrightarrow j \leftrightarrow k) , \\
\nonumber
  \partial_t [ \epsilon_i \epsilon_j \epsilon_k ]_1
  & = 
    D_{ij} [\epsilon_k]_1+ (i \leftrightarrow j \leftrightarrow k)
    + {J^{(0)}}_i^q [ \epsilon_j \epsilon_k \epsilon_q ]_1 + (i \leftrightarrow j \leftrightarrow k)  \\
\nonumber
  & \quad
    + D^{(1)}_i [ \epsilon_j \epsilon_k]_0 + (i \leftrightarrow j \leftrightarrow k)
    + D_{ijk}
    + \tfrac{1}{2} {J^{(0)}}_i^{q,r} [ \epsilon_j \epsilon_k \epsilon_q \epsilon_r]_0 \\
  & \quad
    + {J^{(0)}}_{i,j}^q [ \epsilon_k \epsilon_q]_0  + (i \leftrightarrow j \leftrightarrow k), \\
  \partial_t [ \epsilon_i \epsilon_j \epsilon_k \epsilon_l]_0
  & = 
    D_{ij} [\epsilon_k \epsilon_l]_0 + (i \leftrightarrow j \leftrightarrow k \leftrightarrow l)
    + {J^{(0)}}_i^q [ \epsilon_j \epsilon_k \epsilon_l  \epsilon_q ]_0 + (i \leftrightarrow j \leftrightarrow k \leftrightarrow l).
\end{align}
Here, $(i \leftrightarrow j \leftrightarrow k)$ denotes a sum over the previous term over all cyclic permutations of the indices   $i,j$ and $k$, and similarly for four indices. For example, we have $D_{ij} [\epsilon_k]_1+ (i \leftrightarrow j \leftrightarrow k) = D_{ij} [\epsilon_k]_1 + D_{ki} [\epsilon_j]_1 + D_{jk} [\epsilon_i]_1$.
We find that for deterministic initial conditions $[\epsilon_i]_0=[\epsilon_i]_2=[\epsilon_i \epsilon_j]_1 = [\epsilon_i \epsilon_j \epsilon_k]_0 = 0$ holds for all times.
The moments in molecule numbers $n_i$ can now be obtained from the moments of the $\epsilon_i$ variables by using the ansatz in Equation \eqref{sse_ansatz}. For the mean and covariance this leads to 
\begin{align}\label{sse_emre_moments}
  &\left< \frac{n_i}{\Omega} \right>
   =
    \phi_i + \Omega^{-1} [\epsilon_i]_1 + O(\Omega^{-2}), \\
\label{sse_ios}
   & \left< \frac{n_i}{\Omega} \frac{n_j}{\Omega} \right> - \left< \frac{n_i}{\Omega} \right> \left< \frac{n_j}{\Omega} \right>
   =
    \Omega^{-1} [\epsilon_i \epsilon_j]_0 + \Omega^{-2} ([\epsilon_i \epsilon_j]_2 - [\epsilon_i]_1 [\epsilon_j]_1) + O(\Omega^{-3}).
\end{align}
To order $\Omega^0$, Equation \eqref{sse_emre_moments} corresponds to the macroscopic rate equations, while Equation \eqref{sse_ios} to order $\Omega^{-1}$ corresponds to the LNA estimate for the covariance. Including terms of order $\Omega^{-1}$ in Equation \eqref{sse_emre_moments} gives the leading order corrections to the mean given by the rate equations. To this order, the system size expansion equations have been called ``effective mesoscopic rate equations'' in the literature \cite{Grima2010}.
The next leading order corrections to the covariance, which have been called ``Inverse Omega Square'' in the literature \cite{Thomas2012b},  are obtained by keeping terms of order $\Omega^{-2}$ in Equations  \eqref{sse_ios}.

\subsubsection{Properties and recent developments}

As long as one is only interested in the LNA or higher order corrections to the moments, rather than the distributions of higher order truncations, the system size expansion amounts to the solution of finite sets of ordinary differential equations. It is thus generally significantly more efficient than stochastic simulations of the CME or of the CLE.

Since the system size expansion is an expansion around the deterministic mean, it cannot be used for deterministically multistable systems, i.e., systems whose rate equations have two or more positive stable steady states, unless one is only interested in the short-time behaviour of a process (for example for the purpose of inference from time-series data, c.f.~Section \ref{sec_inference}).

\begin{figure} [t]
\centering
  \includegraphics[scale=0.65]{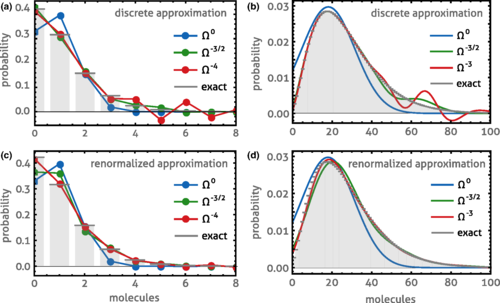} 
  \caption{\textbf{Distribution of non-linear birth-death process.}  The figures show the steady-state distribution for a non-linear birth-death process for two parameter sets, one corresponding to a small mean number of molecules (left panel) and one corresponding to a large mean number of molecules (right panel). Results are shown for different orders of the system size expansion with discrete support (top panel) as well as the renormalised approximation (see main text for a definition). The studied system corresponds to a birth-death process where the propensity of the death process is of Michaelis-Menten type. Such a system is obtained when reducing a Michaelis-Menten system under time-scale separation. This will be demonstrated in Section \ref{sec_other_approx_time}. 
  The figure is taken from \cite{thomas2015approximate} with permission kindly granted by the publisher.  }
  \label{fig_sse_renormalized}
\end{figure}

In many cases, the lowest order of the system size expansion, i.e., the LNA, already gives remarkably accurate results and has been used succesfully in various applications \cite{mckane2007amplified,pahle2012biochemical,challenger2013synchronization, bianca2015evaluation, hufton2016intrinsic}. However, for some systems it gives rise to significant deviations from the exact result \cite{thomas2013reliable}. Higher order approximations beyond the LNA have frequently been found to be highly accurate for such systems. Examples include oscillations in networks of coupled autocatalytic reactions \cite{dauxois2009enhanced, cianci2012analytical}, predator-prey systems
\cite{scott2012non}  and circadian oscillators \cite{thomas2013signatures}. 

We note that for truncations of higher orders than zero, the resulting PDEs involve higher order derivatives than two and hence have no probabilistic interpretation due to non positive-definite solutions \cite{Pawula1967}. Moreover,  the solutions to the PDEs are generally not longer known in this case. However, for systems with only one species a general solution to all orders has recently been derived \cite{thomas2015approximate}. A similar approach has been developed in \cite{leen2011perturbation} for discrete-time models in neuroscience. Note that the variables described by the system size expansion are typically assumed to be continuous. In \cite{thomas2015approximate} it has been found that if the support is assumed to be continuous, the first few leading orders of the system size expansion lead to oscillations in the tail of the Poisson distribution for a linear birth-death process. In contrast, if one assumes a discrete support, the distribution is captured accurately. In the same work, a modified system size expansion has been employed. Instead of expanding the variables of the CME  around the deterministic mean value given by the rate equations, one expands the variables around the mean value given by the system size expansion itself to the considered order (which is not known a priori). The authors call this method ``renormalised approximation''.
This approach has been found to give an improved approximation of the distribution for a non-linear process as shown in Figure \ref{fig_sse_renormalized}.

\subsection{Moment closure approximations}\label{sec_stochkin_ma}

Another popular class of methods that approximate the first few moments of a process are so-called \emph{moment closure approximations}. We present a popular class of such methods here that are particularly easy to derive and implement. 

\subsubsection{General formulation}

In Section \ref{sec_moment_equations} we showed that the CME gives rise to ordinary differential equations for the moments of a process. For linear reaction systems, these equations can be solved numerically up to a certain desired order. For non-linear systems, however, we found that the moment equations of a certain order couple to higher order equations, leading to an infinite hierarchy of equations which can hence not be solved directly. Moment closure approximations truncate this infinite set of equations at a certain order $M$ in some approximate way.

One popular class of moment closure approximations close the moment equations by expressing all moments above a certain order $M$ as functions of lower-order moments. One way to achieve this is for example by assuming the distribution of the system to have a particular functional form, for example a normal distribution. This decouples the equations of the moments up to order $M$ from higher-order moments, and hence leads to a finite set of coupled ordinary differential equations which can then be solved (numerically). We refer to such a moment closure as ``$M$th order moment closure''.

Let
\begin{align}\label{def_notation_moments}
  y_{i_1,\ldots,i_k}
  &  = 
    \langle n_{i_1} \ldots n_{i_k} \rangle, \\
\label{def_notation_central_moments}
  z_{i_1,\ldots,i_k} 
  & = 
    \left\{\def\arraystretch{1.2}%
	  \begin{array}{@{}c@{\quad}l@{}}
  	  	\langle (n_{i_1}-y_{i_1}) \ldots (n_{i_k}-y_{i_k}) \rangle & \quad \text{if} \quad k \geq 2, \\
  	 	y_{i_1}  & \quad \text{if} \quad k=1, \\
	  \end{array}\right. \\
\label{def_cum}
  c_{i_1,\ldots,i_k}
  & =
    \partial_{s_{i_1}} \ldots \partial_{s_{i_k}} g(s_1,\ldots, s_N)|_{s_1,\ldots, s_N=0} ,
\end{align}
denote raw  moments, central moments and cumulants of order $k$, respectively. 
We call $y_{i_1,\ldots,i_k}$ a ``diagonal moment'' if $i_l = i_m$ for all  $l,m \in \{1,\ldots, k \}$, and a ``mixed moment'' otherwise, and similarly for central moments and cumulants.
The function $g(s)$ in Equation \eqref{def_cum} is the \emph{cumulant generating function} and is defined as 
\begin{align}
  g(s_1,\ldots, s_N) = \log \langle \exp (s_1 n_1 + \ldots + s_N n_N) \rangle.
\end{align}
We note that all three types of moments are respectively invariant under permutations of their indices. Therefore, only one representative combination of each permutation class has to be considered. Taking this symmetry into account significantly reduces the number of variables and moment equations. 

Some  popular moment closure methods can be defined as
\begin{itemize}
\item ``Normal moment closure" \cite{Goodman1953, Whittle1957, mcquarrie1964kinetics,Gomez2007}(also called ``cumulant neglect moment closure" in the literature): all cumulants above order $M$ are set to zero, i.e.,
\begin{align}\label{ma_def_normal}
  c_{i_1, \ldots, i_k} = 0, \quad \text{for} \quad k>M.
\end{align}
Note that the cumulants of order higher than two are zero for a normal distribution, hence the name ``normal moment closure''. We refer to the normal moment closure approximations to second and third order as ``2MA'' and ``3MA'', respectively.

\item ``Poisson moment closure" \cite{Nasell2003}: the cumulants of a one-dimensional Poisson distribution are all equal to the mean value. We assume here a multivariate distribution to be a product of uni-variate Poisson distributions. Accordingly, for the Poisson moment closure approximation of order $M$ we set all diagonal cumulants to the corresponding mean and all mixed cumulants to zero, i.e.,
\begin{align}\label{ma_def_poisson_1}
  c_{i_1, \ldots, i_k} 
  & = 
    y_i, \quad \text{for} \quad k>M \quad \text{and} \quad  i_1, \ldots, i_k=i, \quad \text{for some} \quad i \in \{1,\ldots, N \}, \\
\label{ma_def_poisson_2}
  c_{i_1, \ldots, i_k} 
  & = 
    0, \quad  \text{for} \quad k>M \quad \text{and} \quad i_m \neq i_n \quad \text{for some} \quad m,n \in \{1,\ldots, K \}.
\end{align}
\item ``Log-normal moment closure" \cite{Keeling2000}: let $\mathbf{m}$ and $\mathbf{S}$ be the mean vector and covariance matrix of a multi-dimensional normal random variable. Then the logarithm of this random variable has a multivariate log-normal distribution whose moments can be expressed in terms of $\mathbf{m}$ and $\mathbf{S}$  as \cite{Edwin1988}
\begin{align}
  y_{i_1, \ldots, i_k} = \exp \left( \mathbf{v}^T \mathbf{m} + \frac{1}{2} \mathbf{v}^T \mathbf{S} \mathbf{v}\right), \quad \text{for} \quad k>M,
\end{align}
where $\mathbf{v} = (g_1, \ldots, g_N)$, and $g_m$ is the number of $i_j$'s having the value $m$. This allows one to express $\mathbf{m}$ and $\mathbf{S}$  in terms of the first two moments $y_i$ and $y_{i,j}$ which then in turn allows one to express higher-order moments in terms of $y_i$ and $y_{i,j}$, too.

\item ``Central-moment-neglect moment closure" (also called ``low dispersion moment closure" in the literature) \cite{Hespanha2008}: all central moments above order $M$ are set to zero:
\begin{align}\label{ma_def_cmn}
  z_{i_1, \ldots, i_k} = 0, \quad \text{for} \quad k>M.
\end{align}

\item ``Derivative matching'': the idea of this method is to express moments above order $M$  by lower order moments in such a way that the time derivatives of the moments of the closed system approximate the time derivatives of the moments up to order $M$ of the exact system at some initial time point. In \cite{singh2006lognormal} a method is derived that allows one to produce the corresponding expressions.

\end{itemize}
Each of the closure methods allows one to express all raw moments above a specified order $M$ in terms of lower order moments and hence to close the moment equations. Note that third order central moments and third order cumulants are identical. Therefore, whenever the moment equation of up to order two do not depend on moments of order higher than three, the normal and central-moment-neglect moment closure are identical. Similarly, the normal and Poisson moment closure approximations can be equivalent for certain systems.

\textbf{Example.}
As an example, consider again the gene system in Figure \ref{fig_gene_system}. The moment equations up to order two for this system are given in Equations \eqref{moms_example1}-\eqref{moms_example5}. To close these we need to express the third order moments $y_{1,1,2}$ and $y_{1,2,2}$ in terms of lower order moments. We do so by means of the normal moment closure defined in Equation \eqref{ma_def_normal} to second order, i.e., we set the third order cumulants $c_{1,1,2}$ and $c_{1,2,2}$ to zero:{
\begin{align}
c_{1,1,2}
 & =
  y_{1,1,2}
  - 
   2 y_1 y_{1,2} - y_2 y_{1,1} + 2 y_2 y_1^2, \\
c_{1,2,2}
 & =
  y_{1,2,2}
  - 
   2 y_2 y_{1,2} - y_1 y_{2,2} + 2 y_1 y_2^2.
\end{align}
Setting these} to zero and rearranging gives
\begin{align}
  y_{1,1,2}
  & = 
   2 y_1 y_{1,2} + y_2 y_{1,1} - 2 y_2 y_1^2, \\
  y_{1,2,2}
  & = 
   2 y_2 y_{1,2} + y_1 y_{2,2} - 2 y_1 y_2^2.
\end{align}
Using these expressions in Equations \eqref{moms_example1}-\eqref{moms_example5} the equations decouple from higher order moments. We give here the resulting equations in terms of central rather than raw moments (c.f.~Equation \eqref{def_notation_central_moments})
\begin{align}
  \partial_t z_1
  & = 
     - \frac{k_2}{\Omega} (z_{1,2}+z_1 z_2) + k_3 (1-z_1), \\
  \partial_t z_2
  & = 
     k_1 z_1 - \frac{k_2}{\Omega} (z_{1,2}-z_1 z_2) + k_3 (1-z_1)
     - k_4  z_2, \\
  \partial_t z_{1,1}
  & =
    \frac{k_2}{\Omega} (- 2 z_{2} z_{1,1} - 2 z_1 z_{1,2} + z_{1,2} + z_1 z_2) + y_{1,2}) + k_3(-2 z_{1,1} - z_1 + 1), \\
\nonumber
  \partial_t z_{1,2}
  & =
    k_1 z_{1,1}  + \frac{k_2}{\Omega} (-z_2 z_{1,1} -z_1 z_{1,2} - z_2 z_{1,2} - z_1 z_{2,2}  +
     + z_{1,2} + z_1 z_2 ) \\
  & \quad
    + k_3 (z_{1,2} - z_{1,1}  - z_1 +1) - k_4 z_{1,2}, \\
\nonumber
  \partial_t z_{2,2}
  & =
    k_1 (2 z_{1,2}+z_{1})  + \frac{k_2}{\Omega} (- 2 z_1 z_{2,2} - 2 z_2 z_{1,2} + z_{1,2} + z_1 z_2) \\
  & \quad
    + k_3 (-2 z_{1,2} -z_{1} + 1) + k_4 (-2z_{2,2}+z_2).
\end{align}
Note that the equations do not depend on third or higher order moments and  hence are closed and can be integrated. 

\subsubsection{Properties and recent developments}

Moment closures are a popular class of approximations of the moments of the CME with many useful properties. First of all, they are easy to derive and implement. Moreover, since they amount to solving a finite set of ordinary differential equations no ensemble averaging is needed, which means they are computationally significantly more efficient than stochastic simulations of the CME or the CLE, and comparable to the system size expansion. However, this computational efficiency comes at a cost: moment closure methods only give approximate moments and not an approximation of the process or distributions. One advantage of moment closures over the system size expansion is that they can be applied to deterministically multistable systems. However, care must be taken for such system since moment closures can lead to unphysical results, see below. 

In contrast to the system size expansion, moment closures are not an expansion in any small parameter, and one can therefore generally not expect that increasing the closure order leads to a higher accuracy. Numerical case studies suggest that this is often  indeed the case, however \cite{Grima2012, Ale2013, Schnoerr2014b}. One explanation for this was given in \cite{Grima2012} where it has been shown that increasing orders of the normal moment closure agree with increasing orders of the system size expansion in the limit of large volumes. These results can to some extent be generalised to other moment closure methods. For monostable systems, one may therefore expect higher order closures to become more accurate for large enough system sizes. For small system sizes on the other hand, the closures can generally not be expected to converge.

\begin{figure} [t]
\centering
  \includegraphics[scale=0.55]{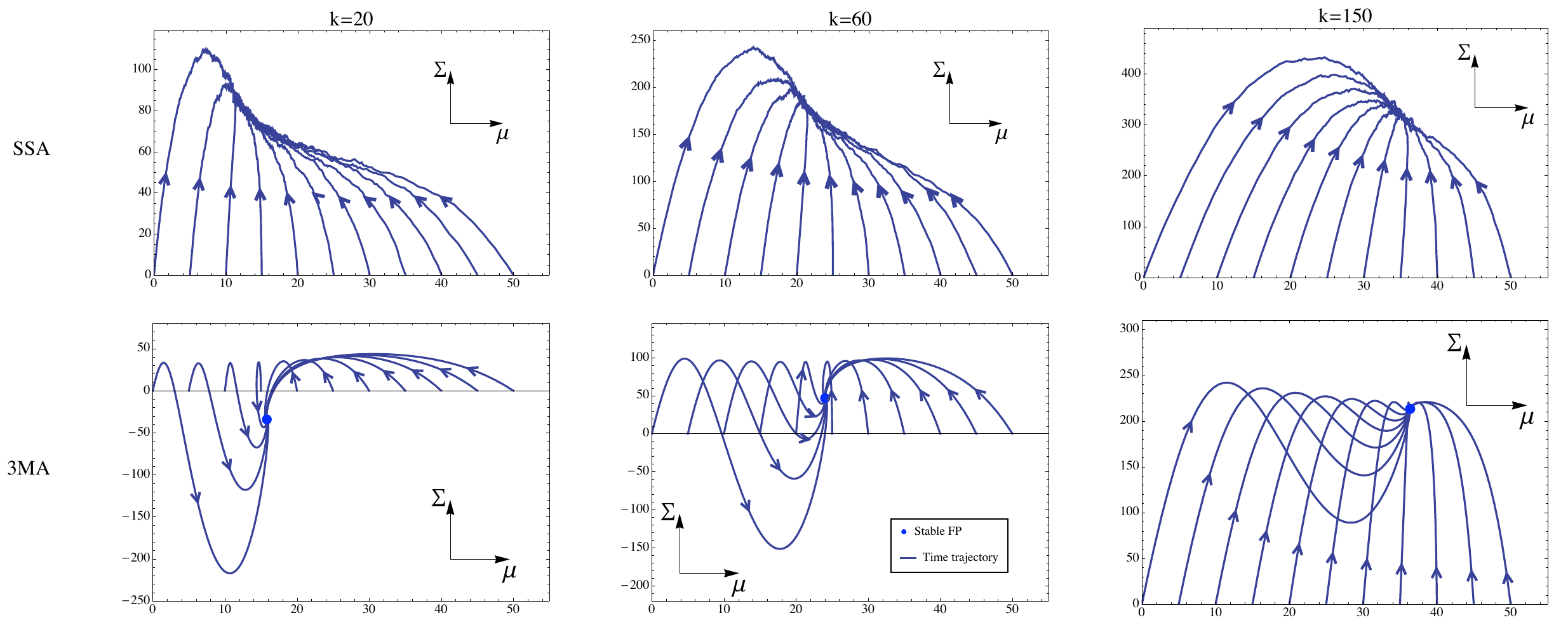} 
  \caption{\textbf{Time trajectories of moment closure with negative variance.} The figures show time trajectories in the mean-variance ($\mu-\Sigma$) plane of a protein $P$ in a system with bursty protein production. The reactions are 
  $\varnothing \xrightarrow{} m P, \quad m \in \mathbb{N}\, \backslash \{0 \}, \quad
  P + P \xrightarrow{} \varnothing$, where the burst size $m$ is a geometrically distributed random number. The top panel shows results obtained from exact stochastic simulations (SSA) and the lower panel results obtained using the normal moment closure to third order (3MA). The parameter $k$ corresponds to the system size. We find that the results from stochastic simulations converge to a positive fixed point for all system sizes and all initial conditions, and that the mean and variance remain non-negative for all times. The 3MA, in contrast, converges to a fixed point with negative variance for the smallest system size $k=20$. For $k=60$, the fixed point is now positive, but some initial conditions still lead to trajectories that have a negative variance for some time and are hence physically not meaningful. Only for the largest system size $k=150$ do all trajectories lead to non-negative trajectories. 
  The figure is taken from \cite{Schnoerr2014b} with permission kindly granted by the publisher. }
  \label{fig_ma_trajectories}
\end{figure}

Moreover, since moment closure approximations are an \emph{ad hoc} approximation, it is not even clear if they always give physically meaningful results. As it turns out, this is not always the case, but they sometimes give rise to unphysical behaviour, such as negative mean values, negative variances or negative higher central moments, as well as diverging trajectories \cite{Nasell2003, Schnoerr2014b}. In numerical case studies, it has been found for several non-linear reaction systems that the normal, log-normal, Poisson and central-moment neglect moment closure show such unphysical behaviour for system volumes \emph{below a certain critical volume} \cite{Schnoerr2014b, Schnoerr2015}. An example of such unphysical  behaviour is shown in Figure \ref{fig_ma_trajectories}. This may be expected, since smaller volumes correspond to stronger non-linearity and hence stronger fluctuations. However, it has been found for deterministically oscillatory and deterministically multistable systems, that the moment closure methods give rise to non-physical oscillations and non-physical multistability, respectively, for system volumes \emph{above a certain critical volume} \cite{Schnoerr2014b, Schnoerr2015}. In \cite{Schnoerr2015}, the normal moment closure was found to give physically meaningful results for larger ranges of system volumes than the other three methods. In terms of accuracy, the four methods performed similar to each other. We emphasise that these were numerical case studies, and it remains open to what extent these results can be generalised to arbitrary reaction systems. In \cite{Hespanha2011} and \cite{Lakatos2015}, for example, the log-normal moment closure was found to be significantly more accurate than the normal and central-moment-neglect moment closure for several reaction systems.

\subsection{Construction of distributions from moments}\label{sec_max_entropy}

The system size expansion and moment closure approximations generally only provide approximations of the moments of a process. Suppose we are however interested in approximating distributions of a process. This can be achieved by running Monte Carlo simulations using the SSA or simulations of the chemical Langevin equation. These methods are computationally quite expensive, however. An alternative approach is to first compute approximations of the moments up to a certain order by means of the system size expansion or moment closure methods, and subsequently constructing a distribution from these moments. If the second step can be done efficiently, this can provide a significantly more efficient method than stochastic simulations since one avoids ensemble averaging.  One popular method to construct distributions from moments relies on the \emph{maximum entropy principle}. 

\begin{figure} [t]
\centering
  \includegraphics[scale=0.7]{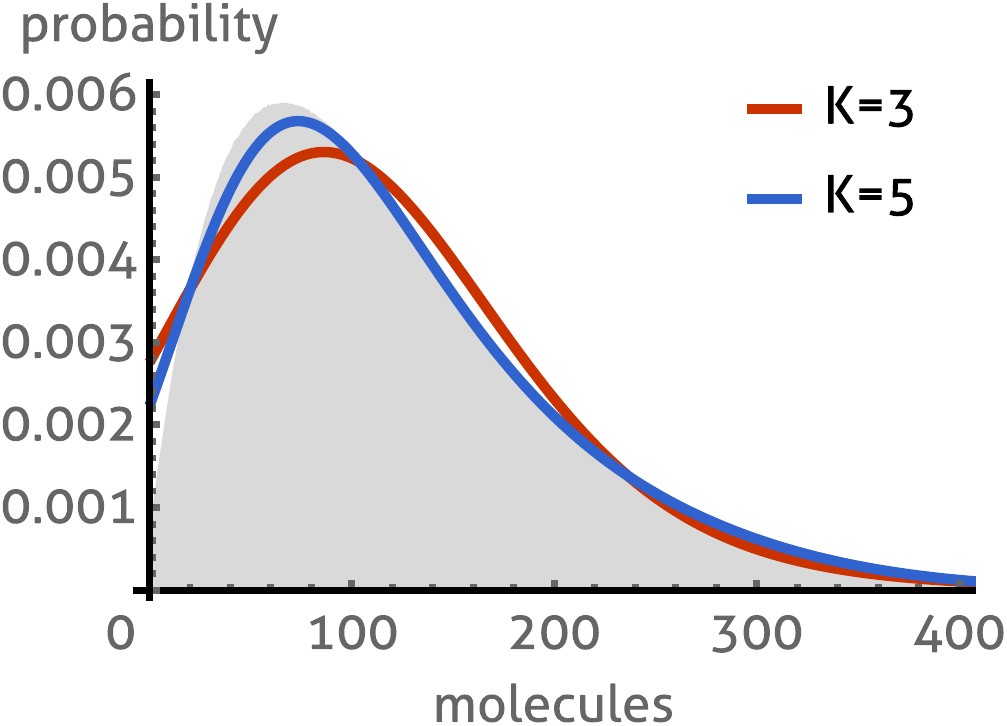} 
  \caption{\textbf{Construction of distribution via maximum entropy.} The figure shows the marginal steady-state distribution of a protein in a chemical reaction system with bursty protein production. The grey area shows the exact result computed using the SSA, while the coloured lines show the results obtained using the central-moment-neglect moment closure to order $K=3$ (red) and $K=5$ (blue) in combination with maximum entropy. We observe that the approximations accurately capture the skewed distribution, with increasing accuracy at higher order.
  The figure is taken from \cite{andreychenko2015distribution}.}
  \label{fig_max_ent}
\end{figure}

For simplicity, assume we have a one-dimensional problem with discrete-valued variable $n$ and that we have approximate values for the first $K$ moments $\mu_1,\ldots, \mu_K$, for example obtained by means of the system size expansion. The goal is to construct a distribution $P(n)$ matching these moments. The entropy $H$ of a distribution $P(n)$ is defined as \cite{shannon2001mathematical}
\begin{align}
  H(P) = -  \sum_n P(n) \log (P(n)).
\end{align}
The maximum entropy method aims at finding a distribution $P(n)$ in a certain family of distributions that maximises the entropy. In our case, the family of distributions is given by the constraint that the first $K$ moments of the distribution should match $\mu_1,\ldots, \mu_K$. This is a non-linear constrained optimisation problem, which can be solved by means of Lagrange multipliers as follows. We define the Lagrangian
{ 
\begin{align}\label{max_ent_langrangian}
  \mathcal{L}
  & =
    H(P) - \sum_{i=0}^K \lambda_i \left( \sum_n n^i P(n) - \mu_i \right).
\end{align}
}Optimising $\mathcal{L}$ with respect to $P$ and the $\lambda_i$ corresponds to solving the constrained optimisation problem.  The variation of $\mathcal{L}$ with respect to $P(n)$ reads
\begin{align}
  \frac{\delta \mathcal{L}}{\delta P(n)}
  & =
    - \log (P(n)) - 1 - \sum_{i=0}^K \lambda_i  n^i.
\end{align}
Setting this to zero we obtain
\begin{equation}\begin{split}\label{max_ent_distribution}
  P(n) 
  & =
    \frac{1}{Z} \exp (- \sum_{i=1}^K \lambda_i  n^i), \\
  \quad Z 
  & = \exp(1+\lambda_0)  
  = 
    \sum_n \exp( - \sum_{i=1}^K \lambda_i  n^i),
\end{split}\end{equation}
with normalisation constant $Z$. Inserting Equation \eqref{max_ent_distribution} into Equation \eqref{max_ent_langrangian} one obtains an unconstrained optimisation problem which can be efficiently solved using standard numerical methods. The maximum entropy method for the construction of the marginal distributions of single species has recently been used  in combination with moment closure methods and the system size expansion in \cite{andreychenko2015model, andreychenko2015distribution}. Figure \ref{fig_max_ent} shows the result for one example system.

A related approach has been developed in \cite{smadbeck2013closure}, where the maximum entropy principle has been used directly to close the moment equations. For time-dependent approximations this method is computationally expensive because the moment equations have to be solved iteratively in small time steps, and in each time step a multi-variate optimisation problem has to be solved. For steady-state approximations, however, the method was found to be efficient and accurate.

\subsection{Software}\label{sec_software}

Several software packages for using the discussed approximation methods, as well as for deterministic rate equations and stochastic simulation algorithms, are freely available. For exact stochastic simulations, for example, available packages include the Java-based program Dizzy \cite{ramsey2005dizzy}, stand-alone COPASI \cite{hoops2006copasi}, stand-alone StochKit \cite{sanft2011stochkit2} and the python package StochPy \cite{maarleveld2013stochpy}. 
The system size expansion for orders beyond the LNA is implemented in the stand-alone package iNA  \cite{thomas2012intrinsic}, which also allows one to perform exact stochastic simulations. 
Various moment closure approximations are available in the Matlab toolbox StochDynTools \cite{Hespanha2007}, and the normal moment closure in Python package MomentClosure \cite{Gillespie2009}. These packages require programming knowledge and are only applicable to mass-action propensity functions. The mathematica package MOCA \cite{Schnoerr2015} extends the applicability to non-polynomial and time-dependent propensity functions and does not require any programming skills. Similarly, the python package MEANS \cite{fan2016means} extends moment closure methods to non-mass-action propensity functions. 
The Matlab package CERENA \cite{kazeroonian2016cerena} implements several of the mentioned methods, including exact stochastic simulations, the system size expansion and moment closure methods. For a detailed overview of available software packages and their capabilities see Figure 1 in \cite{kazeroonian2016cerena}.

\subsection{Other approximations}\label{sec_other_approximations}

One reason why the CLE, the system size expansion and moment closure approximations discussed above are so popular is that they are easy to implement, do not require any pre-knowledge of the system (except monostability for the system size expansion), are generally efficient computationally, and often give accurate approximations. The three methods have been frequently applied successfully in the literature. However, there are many scenarios where the three methods give quite inaccurate results. In particular if one or several of the species occur in very low copy numbers, the three methods often perform poorly. A large number of other approximation methods have been developed in the literature, and we give an overview of these methods here. Many of these methods are more sophisticated but only apply to certain classes of systems and require pre-knowledge and/or fine-tuning. As we shall see, the CLE, the system size expansion or moment closure approximations form building blocks of many of these methods.

\subsubsection{State space truncation}\label{sec_other_approx_state_space}

In Section \ref{sec_stochkin_cme} we saw that the CME can be solved exactly by matrix exponentiation whenever the state space is finite (c.f.~Equation \eqref{cme_mat_sol}). However, for many chemical systems of interest the state space is infinite and the solution in Equation \eqref{cme_mat_sol} can not be computed. The idea of the \emph{finite state projection algorithm} is to truncate the state space to a finite subspace and use matrix exponentiation to obtain an approximation of the distribution on this subspace \cite{munsky2006finite}. It also provides an estimate of the error, which can be systematically reduced by increasing the truncated space.

While this is an efficient method in some cases, often a large truncated space has to be chosen to achieve a reasonable accuracy, making matrix exponentiation  intractable, despite recent progress on numerical algorithms, see \cite{moler2003nineteen, goutsias2013markovian} for overviews.  Similarly, sometimes the state space of a system is finite but too large to compute the matrix exponential. Accordingly, several modified versions of the finite state projection algorithm have been developed, see \cite{dinh2016understanding} for a review.


\subsubsection{Tau-leaping}\label{sec_other_approx_tau}

{ 
\emph{Tau-leaping methods} are approximate ways of simulating a chemical network with the goal of being more efficient than exact stochastic simulations \cite{gillespie2001approximate}. The basic idea is to ``leap'' along time in certain steps $\tau$ during which several reaction events occur, thereby avoiding the simulation of each single reaction event as exact stochastic simulations do. The time step has to be chosen small enough such that the propensity functions do not change significantly during one time step. In that case different reactions become independent of each other, and the number of reaction events $N_r(\tau; \mathbf{n}, t)$ of the $r^{\text{th}}$ reaction during the time interval $\tau$ given that the state at time $t$ is $\mathbf{n}$   becomes a Poisson random variable $\mathcal{P}(f_r(\mathbf{n}) \tau)$ with mean $f_r(\mathbf{n}) \tau$. Accordingly, the tau-leaping algorithm updates the state $\mathbf{n}(t)$ of the system iteratively as
\begin{align}\label{tau-leap}
  \mathbf{n}(t+\tau) = \mathbf{n}(t) + \sum_{r=1}^R \mathbf{S}_r \mathcal{P}(f_r(\mathbf{n}) \tau), 
\end{align}
where $\mathbf{S}_r$ is the $r^{\text{th}}$ row of the stiochiometric matrix. If many reactions are happening per time step the algorithm is more efficient than exact simulations. Increasing $\tau$ increases the algorithm's efficiency but obviously decreases its accuracy. If a system contains species with very low particle numbers $\tau$ may have to be chosen smaller than the average inter-reaction time to ensure approximately constant propensity functions, which would lead to the algorithm becoming less efficient than exact simulations. Roughly speaking, tau-leaping therefore works best for systems with not too small average molecule numbers. 

Despite its simplicitly, implementing the tau-leaping method bears several difficulties. Most importantly, the trade-off between accuracy and efficiency makes the choice of the step size $\tau$ non-trivial. A bad choice can lead to highly inefficient or inaccurate results and also to other problems, such as negative molecule numbers. Recent years have therefore seen a wide range of studies developing modified versions of the tau-leaping method that aim at solving these problems, including implicit methods \cite{rathinam2003stiffness, cao2007adaptive}, binomial methods \cite{tian2004binomial, chatterjee2005binomial, peng2007efficient, leier2008generalized}, multinomial methods \cite{pettigrew2007multinomial}, K/R-leaping \cite{cai2007k,leier2008generalized} and a post-hoc correcting method \cite{anderson2008incorporating}.  See \cite{li2008algorithms, pahle2009biochemical,gillespie2008simulation} for overviews of these methods.

}
\subsubsection{Time-scale separation}\label{sec_other_approx_time}

Many biochemical systems involve processes with highly varying time scales. Given a certain separation of time scales in a system, it is often possible to derive a reduced model of the system which either allows more efficient simulations or more accurate approximations. Such methods have been well-known for deterministic ordinary differential equation models of biochemical kinetics  for several decades \cite{bowen1963singular} but have only more recently been developed for stochastic methods \cite{janssen1989elimination}. We start here by describing the deterministic setting.

\textbf{The deterministic case.} In the following we describe two popular methods for reducing deterministic rate equations under time-scale separation. 
The first method assumes that the chemical reactions in a given system can be divided into ``slow'' and ``fast'' \emph{reactions}, i.e., reactions that happen very infrequently or very frequently on a certain time scale of interest. In this case it is sometimes possible to derive a reduced model by eliminating the fast reactions, typically by assuming that certain reactions balance each other. Such methods are often called \emph{quasi-equilibrium approximations} (QEA) \cite{heinrich2012regulation}. 

The second method separates a system into ``slow'' and ``fast'' \emph{species}, rather than reactions. The fast species are assumed to be asymptotically in steady state on the time scale of the slow species. This is known as the \emph{quasi-steady-state approximation} (QSSA). The idea is to eliminate the fast species from the system and to include their steady-state effect on the slow species.

\textbf{Example.} Let us illustrate the QEA and QSSA in the deterministic setting by means of a simple example, the well-studied Michaelis-Menten system: 
\begin{align}\label{michaelis_menten}
  S+E  \xrightleftharpoons[k_2] {\quad k_1 \quad } C \xrightarrow{\quad k_3 \quad } E+P.
\end{align}
A substrate $S$ reversibly binds to an enzyme $E$ to form the substrate-enzyme complex $C$, from which a product molecule $P$ becomes catalysed. The corresponding rate equations are 
\begin{equation}\label{michaelis_menten_res}
\begin{split}
  \partial_t \phi_S
  & = 
    - k_1 \phi_S \phi_E + k_2 \phi_C, \\
  \partial_t \phi_E
  & = 
    - k_1 \phi_S \phi_E + (k_2+k_3) \phi_C, \\
  \partial_t \phi_C
  & = 
    k_1 \phi_S \phi_E - (k_2+k_3) \phi_C, \\
  \partial_t \phi_P
  & = 
    k_3 \phi_C, 
\end{split}
\end{equation}
where $\phi_S, \phi_E, \phi_C$ and $\phi_P$ denote the concentrations of $S, E, C$ and $P$, respectively. Let us first make the  assumption that the  binding and unbinding of the substrate and the enzyme in Equation \eqref{michaelis_menten} are fast reactions and that the catalysis reaction is slow. This is the case whenever $k_3 \ll k_2$. On the time scale of the slow reaction, this means that the two fast reactions balance each other, i.e.,
\begin{align}\label{michaelis_menten_qea_assumption}
  k_1 \phi_S \phi_E
  & \approx
      k_2 \phi_C.
\end{align}
This corresponds to the \emph{quasi-equilibrium approximations} (QEA). Note that the total concentration $E^0$ of enzyme molecules is conserved, i.e., $\phi_E+\phi_C = E^0$. Using this to eliminate $\phi_E$ in Equation \eqref{michaelis_menten_qea_assumption} one obtains $\phi_C = (\phi_S E^0)/ (\phi_S + K_M)$, where we defined the \emph{Michaelis-Menten constant} $K_M=k_2/k_1$. Let $v= \partial_t \phi_P = k_3 \phi_C$ be the production rate of the product $P$. According to Equation \eqref{michaelis_menten_res} this becomes 
\begin{align}\label{michaelis_menten_qea_result}
  v & =
    \frac{k_3 \phi_S E^0}{\phi_S + K_M},
\end{align}
which is the well-known Michaelis-Menten equation \cite{michaelis1913kinetik}.

\begin{figure} [t]
\centering
  \includegraphics[scale=1.5]{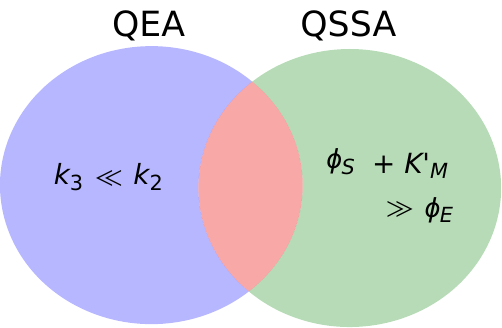} 
  \caption{\textbf{Visualisation of different time-scale separation regimes.} The figure visualises the two different time-scale separation regimes for the Michaelis-Menten reaction system with reactions in Equation \eqref{michaelis_menten}. The blue area represents the regime of parameter space where only the quasi-equilibrium assumption holds, the green area where only the quasi-steady-state assumption holds, and the red area where both assumptions hold. }
  \label{fig_time_scale_separation}
\end{figure}

Next, we illustrate the \emph{quasi-steady-state approximation} (QSSA). To this end, we assume that the complex is approximately in steady state:
\begin{align}\label{michaelis_menten_qssa_assumption}
  \partial_t \phi_C \approx 0. 
\end{align}
It can be shown that this is a reasonable approximation whenever $\phi_S + K_M' \gg \phi_E$ with $K_M' = (k_2 + k_3)/k_1$ \cite{segel1989quasi}. This corresponds to the complex $C$ and free enzyme $E$ being fast species, and the substrate $S$ being a slow species.

Using Equation \eqref{michaelis_menten_qssa_assumption} in Equation \eqref{michaelis_menten_res} and the conservation law $\phi_E+\phi_C = E^0$ we obtain 
\begin{align}\label{michaelis_menten_qssa_result}
  v & =
    \frac{k_3  \phi_S E^0}{\phi_S + K'_M}.
\end{align}
This is also called \emph{Briggs-Haldane kinetics} \cite{briggs1925note}. 
Note that Equation \eqref{michaelis_menten_qssa_result} has the same form as Equation \eqref{michaelis_menten_qea_result} but with a different constant $K_M'$, showing that the QEA and QSSA are generally not equivalent. Depending on the parameters, only one of the two conditions (or none) may be satisfied. This is illustrated in Figure \ref{fig_time_scale_separation}.

\textbf{The stochastic case.} For stochastic systems described by the CME, reductions based on time-scale separations are not as straightforward as in the deterministic case. Ideally, one would like to derive a reduced CME and/or a stochastic simulation algorithm for a reduced system. As it turns out, under the quasi-equilibrium assumption of fast and slow reactions it is indeed possible to derive a reduced CME \cite{goutsias2005quasiequilibrium}. In contrast, the quasi-steady-state assumption of fast and slow species does not necessarily allow to derive a reduced master equation since it leads to a non-Markovian process for the fast species \cite{janssen1989elimination2}. One can use a reduced master equation with non mass-action propensity functions that corresponds to the reduced deterministic system under the quasi-steady-state assumption. This heuristic master equation has been found to be accurate for some examples \cite{gonze2002deterministic,rao2003stochastic, sanft2011legitimacy}.
However, even if the quasi-steady-state assumption is valid for the deterministic system, the reduced master equation can be highly inaccurate \cite{thomas2011communication, kim2015relationship}. The validity of a reduction in the deterministic case does not generally imply a valid reduction in the stochastic case. The relationship between the two descriptions with respect to time-scale separation approximations has for example been studied in \cite{kim2015relationship}. In \cite{thomas2012slow} a reduced description of stochastic models under the quasi-steady-state assumption was derived by means of the linear noise approximation. 

It is however possible to derive reduced descriptions based on quasi-steady-state assumptions for a stochastic system if one requires stronger conditions on a system than the conditions needed to reduce the deterministic system. In \cite{cao2005slow}, for example, reactions are split into fast and slow reactions, and slow species are defined as those involved in slow reactions only and fast species as those participating in at least one fast reaction and any number of slow reactions. As mentioned before, the fast variables conditioned on the slow ones are not Markovian. The authors in \cite{cao2005slow} solve this problem by approximating the fast process by a virtual fast process that is affected by fast reactions only and therefore Markovian.

A large number of other approximation methods for CME type systems based on time-scale separations have been developed in recent years, see for example \cite{goutsias2005quasiequilibrium, salis2005equation, weinan2005nested, samant2005overcoming, weinan2007nested, gomez2008enhanced, pigolotti2008coarse, chevalier2009rigorous, cotter2011constrained, kang2013separation, bortolussi2015efficient}. Most of these methods split up the full CME into a CME for the fast variables conditioned on the slow reactions/species, and a CME for the slow reactions/variables with marginalised fast variables. On the time scale of the slow dynamics, the conditional CME of the fast dynamics is assumed to quickly reach steady state. Therefore, the CME of the slow dynamics only depends on the steady-state distribution of the fast dynamics. Most methods rely on SSA simulations of the slow dynamics and assume that the fast dynamics is in steady state \cite{salis2005equation, cao2005slow,
weinan2005nested, samant2005overcoming, weinan2007nested,cotter2011constrained}. 

For some systems, the reduced CMEs under time-scale separation have been solved exactly, for example for gene expression \cite{shahrezaei2008analytical, bokes2012multiscale, melykuti2014equilibrium}. In \cite{popovic2016geometric} a perturbative expansion based on time-scale separation for a two-stage gene expression model is derived that allows one to systematically include higher order corrections to the solution.

The studies mentioned so far mainly rely on stochastic simulations of the reduced equations or on analytic solutions for special cases. An alternative approach is to combine time-scale separation reductions with other approximations, such as approximating the fast variables by the chemical Langevin equation \cite{haseltine2002approximate,salis2005accurate, cotter2011constrained}, deterministic rate equations \cite{haseltine2002approximate}, or moment closure methods \cite{gomez2008enhanced}. Other methods combine time-scale separation with the finite state projection algorithm \cite{pelevs2006reduction} (c.f.~Section \ref{sec_other_approx_state_space}), tau-leaping \cite{puchalka2004bridging}, or the linear noise approximation \cite{pahlajani2011stochastic, thomas2012slow, thomas2012rigorous}. An adiabatic approximation  derived form a stochastic path integral description has been developed in \cite{sinitsyn2009adiabatic}. 
Other methods derive reduced models for systems that allow a partition in more than two typical time scales, see for example \cite{burrage2004multi, weinan2007nested}.  In \cite{thomas2014phenotypic} a conditional linear noise approximation for gene expression systems with finite, slow promoter states has been derived.

\subsubsection{Hybrid methods}\label{sec_other_approx_hybrid}

In many biochemical reaction networks of practical interest no time-scale separation assumptions apply. However, often some species occur in low and others in high copy numbers, which motivates the combination of different simulation and/or approximation methods for these two groups of species, similar to the time-scale separation case. One possibility would be to partition species into a discretely modelled group and a continuously modelled group, and accordingly reactions into discrete reactions involving discrete species and continuous reactions that do not involve discrete species. 
 In this case species that are involved in both continuous and discrete reactions describe diffusion-jump processes. 
Such methods are broadly referred to as \emph{hybrid methods}. Note that many of the methods discussed in Section \ref{sec_other_approx_time} fall under this definition. We give here an overview of hybrid methods that do not rely on separation of time scales, although  a clear distinction between the two cases is often not possible. 

Consider the case where we split the species of a system into a low abundance group which we model discretely using the SSA and a high abundance group which we model continuously using deterministic rate equations or chemical Langevin equations. Accordingly, we group reactions into discrete reactions involving discrete species and continuous reactions that do not involve discrete species. Between two discrete reaction events, the continuous variables follow rate equations or chemical Langevin equations which can be solved numerically or simulated in a standard way. The propensity functions of the discrete reactions, however, may depend on the continuous variables and hence depend on time in such a hybrid approach. This is akin to the case of extrinsic noise discussed in Section \ref{sec_extrinsic_noise}, where a discrete system's propensities where assumed to depend on some external stochastic process. Correspondingly, the discrete system of our hybrid system can not be simulated by a standard SSA. As in the extrinsic noise case, one possibility is to numerically integrate the time-dependent propensity functions over time until a target value is reached and the corresponding reaction occurs. The discrete system is then updated accordingly which may also change the propensity functions of the continuous reactions. 

A key challenge of such an approach is to decide which species should be modelled discretely and which continuously, in particular since some species may fluctuate strongly during a simulation. More sophisticated algorithms  therefore partition species and reactions adaptively during simulation. Many different methods addressing these and other issues have been developed in the literature, see for example \cite{vasudeva2004adaptive,alfonsi2004exact, takahashi2004multi,neogi2004dynamic, bentele2004general, harris2006partitioned, hellander2007hybrid,crudu2009hybrid, menz2012hybrid, ganguly2015jump, duncan2015hybrid, safta2015hybrid, hepp2015adaptive}. They differ mainly in how the partitioning is conducted and in the simulation methods for the resulting reduced system. 
Simulation-based approaches include combination of the SSA for the low abundance species with tau-leaping \cite{harris2006partitioned}, chemical Langevin equations \cite{neogi2004dynamic, crudu2009hybrid, ganguly2015jump, duncan2015hybrid,angius2015approximate}, the chemical Fokker-Planck equation \cite{safta2015hybrid}
and deterministic rate equation approximations\cite{vasudeva2004adaptive, alfonsi2004exact,takahashi2004multi,hellander2007hybrid, hepp2015adaptive}  for the abundance species. Other approaches split species into more than two sets \cite{bentele2004general}.

Since many of these methods are heuristic, it is not straightforward to assess their performance or to prove their convergence to the exact system in some limit. In some cases, this is possible, however. For example, error bounds for some hybrid methods  have been derived \cite{jahnke2011reduced, jahnke2012error, ganguly2015jump}. In \cite{crudu2012convergence} the convergence of different types of hybrid methods to the exact system have been studied, and in \cite{kang2013separation} criteria have been developed for the convergence of a CME to discrete-deterministic hybrid approximations. An error analysis of various methods has recently been conducted in \cite{cotter2016error}.

While simulation-based hybrid methods are often orders of magnitude more efficient than the standard SSA, they all still rely on stochastic simulations for the low abundance species and some of them also for the high abundance species. They therefore can still become computationally expensive depending on the studied system. Some methods aim at circumventing expensive simulations by applying additional approximations to the reduced systems. In \cite{hasenauer2014method}, for example, the dynamics of the large abundance species is formulated in terms of moment equations conditional on the low abundance (discrete) species. For non-linear systems these can be closed by means of moment closure approximations (c.f.~Section \ref{sec_stochkin_ma}). This method amounts to the solution of a differential algebraic equation which is difficult to solve. The authors in \cite{hasenauer2014method} propose several simulation based algorithms. This method has shown to give accurate approximations to the distributions and moments of some gene systems, but its implementation is non-trivial and requires additional approximations and/or simulations.  In \cite{soltani2015conditional} a simpler conditional moment closure has been proposed for two-state gene systems. Here, the conditional moment equations are closed by means of the second-order normal closure or derivative matching (c.f.~Section \ref{sec_stochkin_ma}). This method amounts to solving a coupled set of ordinary differential equations. In \cite{smith2015model} a hybrid method is developed where the abundance species is described by the (non-conditional) rate equations. This is a simplification of the previously mentioned methods that model the abundance species by rate equations conditional on the low abundance species. However, this simplification allows one to derive analytic solutions of the CME for the low abundance species for certain systems \cite{smith2015model}.

\section{Comparison of approximation methods} \label{sec_comparison}

In the previous section we gave a detailed introduction to three popular approximation methods of the CME: the chemical Langevin equation (CLE, Section \ref{sec_stochkin_cle}), the system size expansion (Section \ref{sec_sse}) and moment closure approximations (Section \ref{sec_stochkin_ma}). These methods have been successfully used in many applications in the literature, but there exist only very few studies comparing their accuracy. It thus remains unclear how the different methods compare to each other. 

In Sections \ref{sec_stochkin_cle} - \ref{sec_stochkin_ma} we discussed advantages and disadvantages of the different methods. Here, we perform a numerical case study to enable the reader to understand the differences between the methods. First, we study an enzyme reaction system of the Michaelis-Menten type in Section \ref{sec_comparison_enzyme} which can be viewed as a catalysed degradation of a spontaneously produced protein. We then extend the model in Section \ref{sec_comparison_bursts}  by including transcription of mRNA and translation of the protein from mRNA, which allows for bursts in protein production. Finally in Section \ref{sec_comparison_conclsuion} we summarise the results and give an overview of advantages and disadvantages of the different methods.

\noindent
\textbf{Implementation.} Implementation details for the different methods are: 

\begin{itemize}
\item \textbf{CLE}: we simulate the CLE using the Euler-Maruyama algorithm introduced in Section \ref{sec_euler_maruyama}. As pointed out in Section \ref{sec_stochkin_cle}, the CLE is traditionally defined for real variables, but suffers from the occurrences of square roots of negative expressions for which it is not well-defined. We therefore implement two versions of the CLE here: (i) a real-valued implementation which keeps the variables positive by rejecting Euler-Maruyama steps that lead to negative variables. We term this version ``CLE-R''; (ii) the recently proposed complex CLE which is defined for complex variables and has been shown to give real-valued moments in the ensemble average \cite{Schnoerr2014}. We term this version ``CLE-C''. 

The Euler-Maruyama algorithm requires a step size $d t$ for time discretisation. For the simulation of steady-state moments or distributions, we take a total number of $M$ samples from a single long trajectory at time steps  separated by $\Delta t$. The chosen values for these parameters will be given for each of the examples. 

\item \textbf{System size expansion and stochastic simulation algorithm}: we use the stand-alone software package iNA \cite{thomas2012intrinsic} for both the system size expansion and the stochastic simulation algorithm (Section \ref{sec_ssa}). We study the zeroth order system size expansion, i.e.,  the linear noise approximation (termed ``LNA'', Section \ref{subsec_lna}), as well as the first order corrections to the mean and variance (both termed ``SSE-1'') given in Equations \eqref{sse_emre_moments} and \eqref{sse_ios}, respectively. 

\item \textbf{Moment closure approximations}: we study here the second-order normal moment closure (termed ``2MA'') introduced in Section \ref{sec_stochkin_ma}, which is probably the most commonly used one in the literature. For its implementation we use the Mathematica software package MOCA \cite{Schnoerr2015}.

\end{itemize}

\begin{figure} []
\centering
  \includegraphics[scale=0.65]{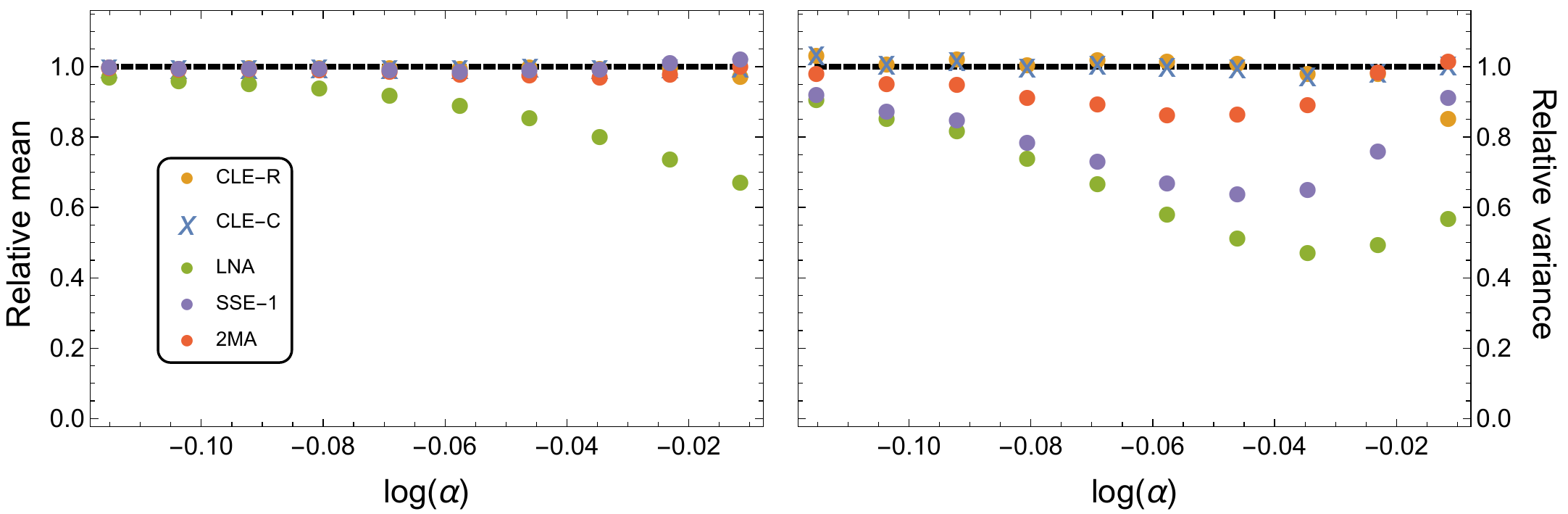} 
  \caption{\textbf{Steady-state mean and variance of protein for enzyme system in Equation \eqref{michaelis_menten2}.} The figures show the steady-state mean (left) and variance (right) of the protein $S$ of the enzyme system in Equation \eqref{michaelis_menten2}, as a function of the saturation parameter $\alpha = k_0 \Omega / (E^0 k_3) $ on logarithmic scale. The values obtained by the approximation methods are normalised by the result obtained from stochastic simulations using the SSA, which means that the dashed black line corresponds to the exact result (up to sampling error). 
  The parameters used are $E^0=2500, k_1 = 40, k_2 = 8, k_3 = 60, \Omega=1$ and we vary $k_0$ according to the shown values of $\alpha$. The simulation parameters for the two CLE implementations are $d t= 10^{-5}, \Delta t =1 $ and $M=10^5$.  }
  \label{fig_enzyme_steady_state}
\end{figure}

\begin{figure} [t]
\centering
  \includegraphics[scale=0.6]{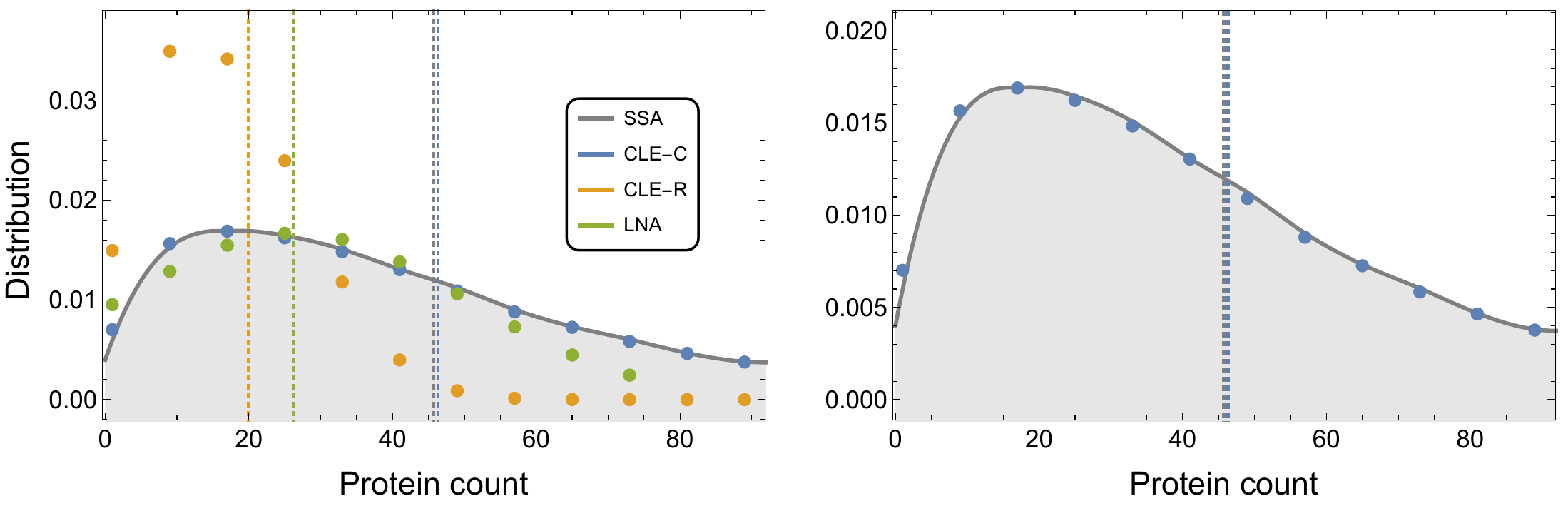} 
  \caption{\textbf{Steady-state distribution of protein for enzyme system in Equation \eqref{michaelis_menten2}.} Left: distribution obtained using the CLE-R, CLE-C, LNA and SSA. Right: only CLE-C and SSA. The vertical lines indicate the corresponding steady-state mean values. The mean values of the 2MA and SSE-1 are very close to the SSA result and are not shown here. The used parameters are $E^0=60, k_0=115, k_1 = 3.5, k_2 = 2, k_3 = 2, \Omega=1$. Simulation parameters for the CLE-R and CLE-C are $d t= 10^{-5}, \Delta t =1 $ and $M=10^5$.}
  \label{fig_enzyme_pdf}
\end{figure}

\subsection{Enzymatic protein degradation} \label{sec_comparison_enzyme}

Let us consider the well-known Michaelis-Menten system with reactions
\begin{align}\label{michaelis_menten2}
  \varnothing \xrightarrow{\quad k_0 \quad } S, \quad S+E  \xrightleftharpoons[k_2] {\quad k_1 \quad } C \xrightarrow{\quad k_3 \quad } E+P.
\end{align}
We studied this system without the first reaction in Section \ref{sec_other_approx_time} in the context of time-scale separations in a deterministic setting. A substrate molecule $S$ is created spontaneously and binds reversibly to a free enzyme molecule $E$ to form a complex $C$, which catalyses the substrate into a product molecule $P$. We consider $S$ to be a protein in the following.

Here, we are interested in the accuracy of the different approximation methods as compared to exact stochastic simulations of the corresponding stochastic system. Note that the total number of enzymes $E^0$ is conserved. The system has a steady state in the protein numbers if and only if $\alpha \equiv k_0 \Omega / (E^0 k_3) < 1$, which  means that the input rate must be smaller than the maximum turnover rate. The parameter $\alpha$ can hence be viewed as a saturation factor.

Figure \ref{fig_enzyme_steady_state} shows the relative mean and variance of the protein $S$ as a function of $\alpha$ for a system with a large number of total enzymes, $E^0 = 2500$. We find that both the real and complex CLE implementations (CLE-R and CLE-C), the first order system size expansion (SSE-1) and the second-order normal moment closure (2MA) give good approximations for the mean value. Only the LNA, which corresponds to the deterministic rate equations for mean values, shows significant deviations from the exact result. For the variance we observe larger deviations for all methods, with the LNA again being the least accurate. The two CLE implementations show the best performance, being more accurate than the 2MA, which in turn is more accurate than the SSE-1. Note that the CLE-R and the CLE-C give very similar results, meaning that the inclusion of a rejecting boundary has a negligible effect in this case. 

Overall, we find that the approximation methods give rise to larger errors for larger values of $\alpha$, which can be explained as follows. As mentioned before, the system only possesses a steady state for $\alpha < 1$ and becomes unstable for $\alpha >1$. We therefore expect larger fluctuations for values of $\alpha$ close to unity. Moreover, since most of the enzymes are in the complex state $C$ in this limit, we expect a skewed distribution for the substrate. 
In the other limit, $\alpha \to 0$,  most enzymes are in the free state $E$, which reduces the non-linear effect of the bimolecular reaction $S+E \to C$. We therefore expect the approximation methods to perform well in this limit, and to lead to larger errors for $\alpha \to 1$. This is exactly what we observe in Figure \ref{fig_enzyme_steady_state}.

Next, we study the steady-state probability distribution of the protein as predicted by the SSA, CLE-R, CLE-C and the LNA. Note that the CLE-C gives complex-valued samples. To obtain a distribution in real variables, we take the real parts of these samples.
We reduce the total number of enzymes to  $E^0 = 60$ to study small molecule number effects. The results are shown in Figure \ref{fig_enzyme_pdf}. We find that the LNA strongly underestimates the true mean value (obtained using the SSA) and does not reproduce the distribution very accurately. However, surprisingly, the real-valued CLE (CLE-R), does even worse than the LNA, both in terms of mean value and distribution. The complex valued CLE (CLE-C) on the other hand, predicts the true mean value with a negligible error, and even captures the highly skewed distribution very well (right panel of Figure \ref{fig_enzyme_pdf}). This demonstrates that a naive fixing of the boundary problem of the CLE (CLE-R) can lead to highly inaccurate results.

\begin{figure} []
\centering
  \includegraphics[scale=0.6]{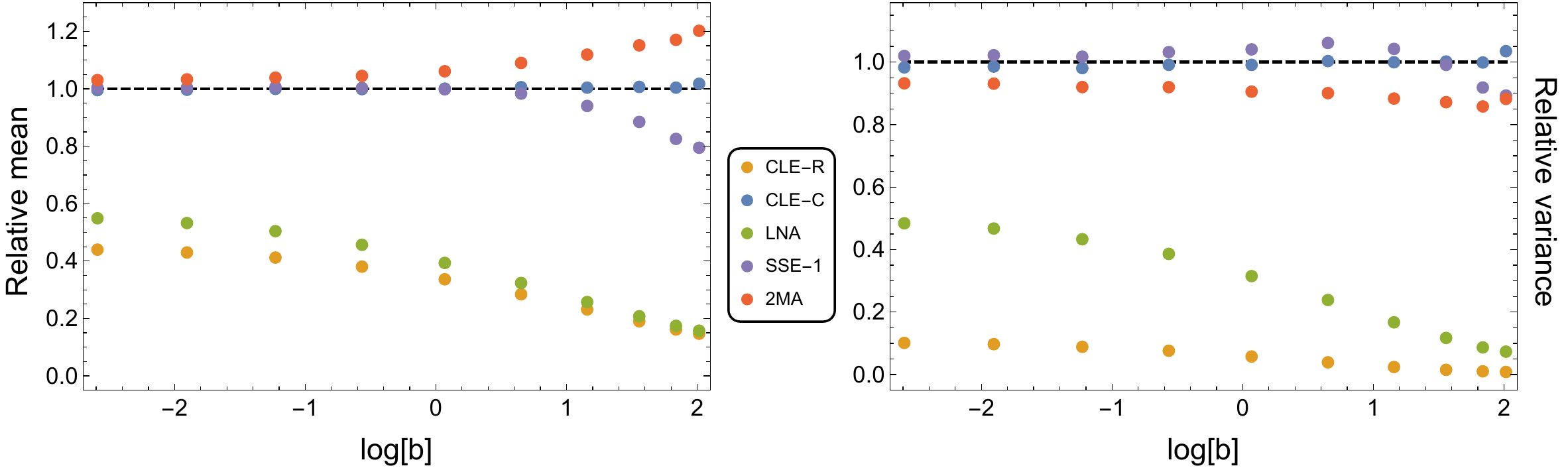} 
  \caption{\textbf{Steady-state mean and variance of protein for extended system in Equation \eqref{bursty_protein_production}.} The figures show the steady-state mean (left) and variance (right) of the protein $S$ of the enzyme system in \eqref{bursty_protein_production}, as a function of the burst size $b=k_s/k_{dM}$ on logarithmic scale. The values obtained by the approximation methods are normalised by the result obtained from stochastic simulations (SSA), which means that the dashed black line corresponds to the exact result (up to simulation error).
The used parameters are $E^0=60, k_s = 1.5,  k_1 = 3.5, k_2 = 2, k_3 = 2, \Omega=1$. $k_0$ and $k_{dM}$ are varied to vary the burst size $b=k_S / k_{dM}$ such that the average protein production rate $k_0 k_s/k_{dM}=115$ is held constant. Simulation parameters for the CLE-R and CLE-C are $d t= 10^{-4}, \Delta t =10 $ and $M=10^4$.}
  \label{fig_gene_moments}
\end{figure}

\begin{figure} []
\centering
  \includegraphics[scale=0.65]{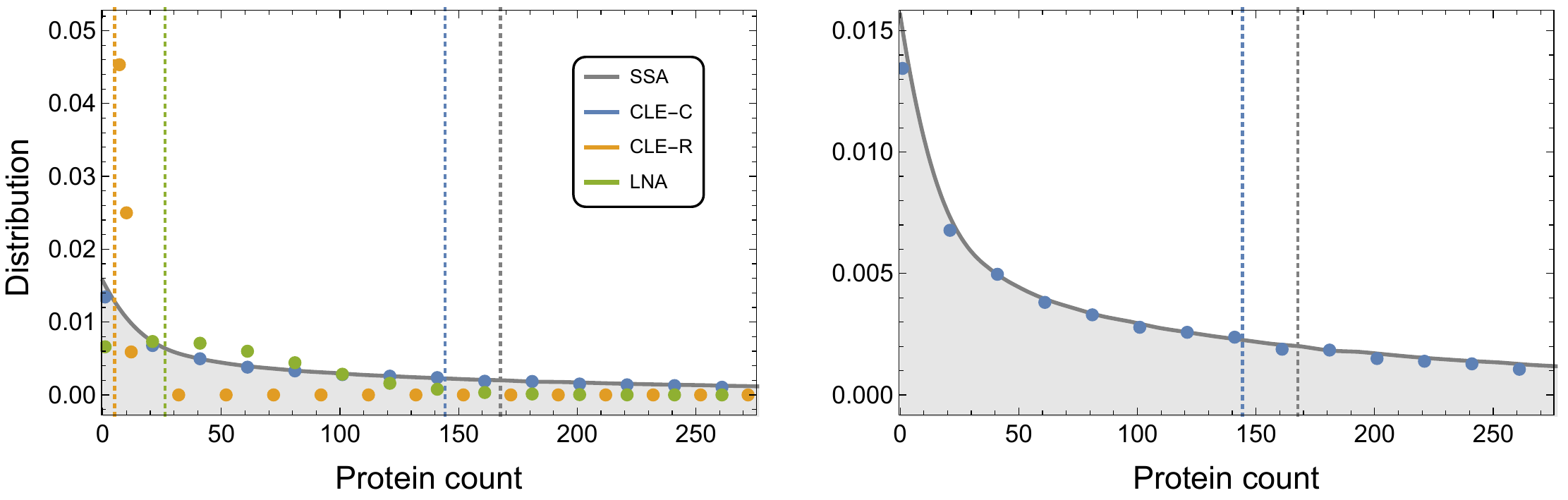} 
  \caption{\textbf{Steady-state distribution of protein for extended system in Equation \eqref{bursty_protein_production}.} Left: distribution obtained using the CLE-R, CLE-C, LNA and SSA. Right: only CLE-C and SSA. The vertical lines indicate the corresponding steady-state mean values. The used parameters are $E^0=60, k_s = 1.5, k_0=15.33, k_{dM}=0.2, k_1 = 3.5, k_2 = 2, k_3 = 2, \Omega=1$, corresponding to a burst size of $b=k_s/k_{dM}= 7.5$. Simulation parameters for the CLE-R and CLE-C are $d t= 10^{-4}, \Delta t =10 $ and $M=10^4$.}
  \label{fig_gene_pdf}
\end{figure}

\subsection{Bursty protein production} \label{sec_comparison_bursts}

We now extend the system in Equation \eqref{michaelis_menten2} by considering the protein $S$ to be produced in a gene expression motif, i.e., via transcription of mRNA $M$ followed by translation to protein:
\begin{equation}
\begin{split}
\label{bursty_protein_production}
  & \varnothing \xrightarrow{\quad k_0 \quad } M, \quad M \xrightarrow{\quad k_{dM} \quad } \varnothing,\quad M \xrightarrow{\quad k_s \quad } M + S, \\
  & S+E  \xrightleftharpoons[k_2] {\quad k_1 \quad } C \xrightarrow{\quad k_3 \quad } E+P.
\end{split}
\end{equation}
Depending on the parameters $k_{dM}$ and $k_s$ this system can give rise to bursts in the production of protein $S$, namely whenever each produced mRNA molecule produces on average several proteins during its lifetime. The average number of proteins per mRNA molecule is given by $b=k_s/k_{dM}$ and also called ``burst size'' \cite{thattai2001intrinsic}. Large $b$ correspond to large bursts which in turn lead to large fluctuations.  

Here, we are interested in the accuracy of the different approximation methods as a function of the burst size $b$, since we expect larger errors for larger fluctuations. To isolate the effect of the burst size $b$, we vary $k_0$ and $k_{dM}$ such that the average production rate of protein $k_0 k_s/k_{dM}=115$ in steady-state conditions is held constant and equal to the production rate used in the previous section in Figure \ref{fig_enzyme_pdf}.

Figure \ref{fig_gene_moments} shows the mean and variance of the protein $S$ as a function of $b$ as predicted by the different approximation methods. Similar to the previous section in Figure \ref{fig_enzyme_steady_state} we find that the LNA performs worse than the other methods for the mean and the variance, with the exception of the real-valued CLE (CLE-R) which performs even worse. Similarly to Figure \ref{fig_enzyme_pdf}, we find here that the complex-valued CLE (CLE-C) performs surprisingly well, demonstrating again the significance of the boundary problem of the CLE. In fact, the CLE-C performs significantly better than all the other methods, including the 2MA and SSE-1.  The latter two perform similarly and both significantly better than the LNA. One should keep in mind, however, that the 2MA, SSE-1 and LNA are computationally significantly more efficient than the CLE-C since they do not rely on sampling.

Next, we consider the steady-state distribution of the protein for a large burst size of $b=7.5$. Figure \ref{fig_gene_pdf} shows the results predicted by the stochastic simulation algorithm, the two CLE implementations, and the LNA.  We find that the CLE-R and LNA give highly inaccurate results, both for the mean value and the distribution, with even larger deviations than for the simpler system in the previous section in Figure \ref{fig_enzyme_pdf}. This is to be expected, since the system now leads to large bursts in protein production, while the average production rate of protein and the rate constants of the enzymatic reactions are the same as in Figure \ref{fig_enzyme_pdf}. Similar to the previous section, we find here that the complex-valued CLE gives a highly accurate approximation for the distribution, despite the large skewness of the later. Once again we find that the boundary problem of the CLE can lead to highly inaccurate results if not treated carefully.

\subsection{Discussion} \label{sec_comparison_conclsuion}

\textbf{Numerical results.}  In this section we gave a numerical comparison of the CLE, the LNA, the normal moment closure to second order (2MA) as well as the next leading order corrections to mean and variance of the system size expansion beyond the LNA (SSE-1). We implemented two versions of the CLE: a real-valued version with rejecting boundary and a complex-valued version. We considered an enzymatic protein degradation system (Section \ref{sec_comparison_enzyme}) and an extension thereof including bursty protein production (Section  \ref{sec_comparison_bursts}).

One important observation we made is the large discrepancy between the two CLE implementations: while we found a negligible difference in the case of large total enzyme numbers (Figure \ref{fig_enzyme_steady_state}), the complex version was significantly more accurate in the case of smaller total enzyme numbers (Figures \ref{fig_enzyme_pdf}, \ref{fig_gene_moments} and \ref{fig_gene_pdf}). Crucially, in the latter cases, the real-valued CLE performed worse than all the other methods, while the complex-valued CLE performed better than all the other methods. This illustrates the boundary problem of the CLE, and how significant inaccuracies can arise when fixing it in a naive way. 

Another important observation is that the LNA was throughout found to be less accurate than all the other methods (except the CLE-R). This result is not very surprising, since the LNA corresponds to the deterministic rate equations on the mean level, and gives the zeroth order fluctuations (in terms of the system size expansion) about the deterministic mean for the variance. The SSE-1 includes the leading order corrections to both the mean and variance. Similarly, the 2MA and CLE capture effects of fluctuations on the mean. It is hence not very surprising that these methods perform better than the LNA. 

In terms of the mean and variance (Figures \ref{fig_enzyme_steady_state} and \ref{fig_gene_moments}), we found that the SSE-1 and 2MA performed similar to each other, except for the variance in Figure \ref{fig_enzyme_steady_state} where 2MA was more accurate. Surprisingly however, the CLE-C was found to be significantly more accurate than the 2MA and SSE-1 in most cases. 

In terms of distributions (Figures \ref{fig_enzyme_pdf} and \ref{fig_gene_pdf}) we found the CLE-C to be highly accurate, even though the distributions were highly skewed. The LNA, which predicts a Gaussian distribution, was obviously not able to capture these skewed distributions. 

\textbf{Advantages and disadvantages.} 
The advantage of the CLE over the other methods is that it gives approximations of the process and distributions in contrast to moment closure methods and the system size expansion. Moment closure approximations give only approximations to the first few moments of a process. The system size expansion in principle predicts distributions. However, closed-form solutions for higher orders beyond the LNA have so far only been derived for one-dimensional systems \cite{thomas2015approximate}. It is not clear if the same is possible for multi-species systems. The higher orders of the system size expansion have therefore mainly been used to approximate the moments of a process. If one is interested in approximating the whole process or its distributions, the CLE is therefore a useful method. 

Suppose now that we are only interested in the moments of a process. In the numerical case study performed before, we found the CLE to be more accurate than the LNA, the SSE-1 and the 2MA. However, the CLE is computationally significantly more expensive than the other methods. While the CLE requires a large number of stochastic simulations to obtain the moments of a process, the other methods only require the numerical solution of a finite set of ODEs and are hence typically orders of magnitude faster. Moreover, if defined for real-valued variables, the CLE suffers from a boundary problem at zero molecule numbers, and real-valued modifications lead to inaccurate results. The boundary problem is solved by extending the CLE to complex-valued variables, which is however less efficient to simulate \cite{Schnoerr2014}. 
Due to these reasons, it seems preferable to use the system size expansion or moment closure approximations if one is only interested in the moments of a process.

Next, the question arises if the system size expansion or moment closure approximations are preferable. While the system size expansion is a systematic expansion in a small parameter, moment closure approximations are an \emph{ad hoc} approximation. This fact makes the system size expansion more appealing, since it is guaranteed to be accurate for large system volumes. The same cannot generally be expected to be true for moment closure approximations. On the other hand, the system size expansion has the disadvantage that it is not applicable to systems that are  deterministically multi-stable, a limitation not shared by moment closure methods. Moreover, higher order corrections of the system size expansion are significantly harder to derive and implement than higher order moment closure methods. Current software packages implementing the system size expansion only implement two orders beyond the LNA for the mean, and one order beyond the LNA for the covariance \cite{thomas2012intrinsic,kazeroonian2016cerena}. Moment closure approximations, on the other hand, are implemented to various orders \cite{Schnoerr2015, fan2016means, kazeroonian2016cerena}. Due to these reasons, it depends on the problem at hand as to decide which method is preferable.

\section{Inference}\label{sec_inference}

So far, we have focussed on the {\it forward problem} of approximating marginal distributions of a fully specified process. Such distributions depend naturally on the parametrisation of the process: it is not uncommon for e.g. steady-state distributions to exhibit qualitatively different behaviours depending on the specific value of reaction propensities. In many concrete applications, the model parameters may only be known approximately: direct measurements of kinetic reaction parameters are difficult to obtain, and, even in cases when good estimates are available, {\it in vivo} parameters of a concrete system embedded in a cell may be influenced by a plethora of additional factors, leading to significant uncertainty. It is therefore of considerable interest to also address the {\it inverse problem}: using (noisy) observations of a system to constrain the uncertainty over model parameters and/ or predictions. This is a well-studied problem in statistics and machine learning: we give here a brief review of recent developments within the {\it Bayesian} approach to solving this inverse problem, with particular attention to methodologies which have employed the approximation methods described earlier.

\subsection{General problem formulation}

The setup we will consider is the following: we consider a stochastic process $p(\mathbf{x}_{0:T}\vert\theta)$ as a measure over the space of trajectories $\mathbf{x}_{0:T}$ of the system in the time interval $[0,T]$, with $\mathbf{x}_t$ representing the value of the state variable at time $t$. In the case of a CME system, such trajectories will be piecewise constant functions from $[0,T]$ onto a discrete space, while for a continuous approximation (e.g. CLE or LNA) $\mathbf{x}_t$ will be real-valued. The stochastic process is assumed to depend on a set of kinetic parameters $\theta$, whose {\it a priori} uncertainty is captured by a prior distribution $p(\theta)$. Additionally, we assume the existence of a measurement process which associates each trajectory $\mathbf{x}_{0:T}$ with an observed random variable $\mathbf{y}$; in the simplest case, the observed variable $\mathbf{y}$ may just be thought of as the state of the system at a particular set of time points, corrupted by random observation noise. We account for such experimental errors through an {\it observation model} encoded in a probability distribution $p(\mathbf{y}\vert\mathbf{x}_{0:T})$,  i.e., the likelihood of a measurement given the true state of the system, which may depend on additional parameters  (here omitted for notational conciseness). We restrict our interest here to the case where the observations are in the form of a time series of state variable observations with independent and identically distributed noise, i.e., measurements of all or of a subgroup of the species in the system. Hence, the observation vector will take the form $\mathbf{y}=(y_0,\ldots,y_T)$ for some discrete time points $0,...,T$. More general cases where the observations take other forms, such as continuous-time constraints or penalties over particular areas of the state space, are treated for example in \cite{bortolussi2013learning,cseke2013approximate}. Also, we consider all observations to come from a single trajectory, or from replicate trajectories with the same (unknown) kinetic parameter values; in other words, we exclude the case where parameters can also be variable, e.g. to accommodate extrinsic noise between cells \cite{zechner2014scalable}.

The general inference problem is then the problem of computing the joint posterior measure $p(\mathbf{x}_{0:T},\theta\vert\mathbf{y})$ over the set of trajectories and parameters of the system. This is formally obtained by applying Bayes' rule \begin{equation}
p(\mathbf{x}_{0:T},\theta\vert\mathbf{y})\propto p(\mathbf{y}\vert\mathbf{x}_{0:T})p(\mathbf{x}_{0:T}\vert\theta)p(\theta).\label{Bayes}\end{equation}
This joint posterior provides information both about the parameters and about the state of the system at all time points during the specific trajectory for which data was collected. In most systems of interest, computation of the normalisation constant of the posterior distribution is analytically intractable, due to the requirement of performing very large sums/ high dimensional integrals. Much of the research in Bayesian statistics and machine learning is therefore focussed on computing efficient approximations to posterior distributions: such approximations can be analytic, usually obtained by variational minimisation of a divergence functional, or sampling based \cite{bishop2006pattern}. This latter class has received particular attention in recent years, and is predicated on constructing a Markov chain which has the required posterior as an invariant distribution. This implies that, asymptotically, the Markov chain will sample from the correct posterior distribution, enabling Monte Carlo computation of any desired statistic ({\it Markov chain Monte Carlo, MCMC}).

\subsection{The Forward-Backward algorithm}

Inference in dynamical systems is based on the fundamental factorisation of the single time posterior marginal
\begin{equation}
    p(\mathbf{x}_t | \mathbf{y}_{0}, \ldots \mathbf{y}_T)\propto p(\mathbf{x}_t\vert\mathbf{y}_{i\le t})p(\mathbf{y}_{i>t}\vert\mathbf{x}_t).\label{factorisation}
\end{equation}
This factorisation, which is a simple consequence of the Markovian assumption and the product rule of probability \cite{bishop2006pattern}, states that the posterior probability at time $t$ is a product of the posterior $p(\mathbf{x}_t\vert\mathbf{y}_{i\le t})$ based on the data seen so far up to time $t$ (the so called {\it filtering distribution}) and the likelihood $p(\mathbf{y}_{i>t}\vert\mathbf{x}_t)$ of future data conditioned on the current state.
The factors in equation \eqref{factorisation} can be computed iteratively using the celebrated {\it Forward-Backward algorithm} \cite{rabiner1986introduction}. 

The forward part, which is also referred to as \emph{filtering} \cite{kalman1960new}, works as follows. Assume that we know the \emph{posterior} $p(x_{i-1} | y_{i-1}, \ldots y_0)$ at time step ${i-1}$, and that we can solve the system forward in time (i.e.,  solve the CME) to obtain the transition probability $p(\mathbf{x}_i | \mathbf{x}_{i-1})$ and hence  the \emph{predictive distribution } $p(\mathbf{x}_i | \mathbf{y}_{i-1}, \ldots \mathbf{y}_0) = \int d \mathbf{x}_{i-1} p(\mathbf{x}_i | \mathbf{x}_{i-1}) p(\mathbf{x}_{i-1} | \mathbf{y}_{i-1}, \ldots \mathbf{y}_0)$.  The posterior of time step $i$ is then obtained by taking the measurement $\mathbf{y}_i$ at time point $i$ into account by means of the the \emph{Bayesian measurement update}
\begin{equation}\label{bayesian_update}
\begin{split}
  p(\mathbf{x}_i | \mathbf{y}_i, \ldots \mathbf{y}_0)
  & = 
    \frac{p(\mathbf{y}_i | \mathbf{x}_i,  \mathbf{y}_{i-1}, \ldots \mathbf{y}_0) p(\mathbf{x}_i | \mathbf{y}_{i-1}, \ldots \mathbf{y}_0)}{p(\mathbf{y}_i |  \mathbf{y}_{i-1}, \ldots \mathbf{y}_0)} \\
  & = 
    \frac{p(\mathbf{y}_i | \mathbf{x}_i) p(\mathbf{x}_i | \mathbf{y}_{i-1}, \ldots \mathbf{y}_0)}
    {p(\mathbf{y}_i |  \mathbf{y}_{i-1}, \ldots \mathbf{y}_0)},
\end{split}
\end{equation}
where we have used the Markov property to obtain the second line. The filtering procedure thus comprises iteratively solving the process between measurements and performing the Bayesian measurement update in Equation \eqref{bayesian_update} and yields the filtering distribution $p(\mathbf{x}_t\vert\mathbf{y}_{i\le t})$. 

The second term in Equation \eqref{factorisation} can be obtained by the Backward algorithm,  a recursion procedure similar to filtering (but running backward in time) \cite{rabiner1986introduction}. Furthermore, a modification of the algorithm, the so called {\it forward filtering/ backward sampling} algorithm \cite{fruhwirth1994data}, can be used to draw sample trajectories from the posterior process, which can then be used in MCMC approaches for joint state and parameter inference.

\subsection{Parameter inference} 

Suppose we are not interested in state inference but  only  in the parameters of the system. In this case we do not need to compute the posterior marginal in Equation \eqref{factorisation}. Rather, it is sufficient to either run the forward algorithm (filtering) or the backward algorithm, since each of them independently deliver the marginal likelihood $p(\mathbf{y})$. The backward algorithm computes the likelihood $p(\mathbf{y}_{i>t}\vert\mathbf{x}_t)$ of future data conditioned on the current state, and $p(\mathbf{y})$ is simply the end result of the recursion algorithm at time $0$. 

To see that the forward algorithm also allows to compute the likelihood  $p(\mathbf{y})$, note that due to the Markov property the latter can be written as 
\begin{equation}\label{likelihood}
\begin{split}
  p(\mathbf{y}) = p(\mathbf{y}_0) \prod_{i=1}^T p(\mathbf{y}_i | \mathbf{y}_{i-1}, \ldots, \mathbf{y}_0).
\end{split}
\end{equation}
We find the factors on the r.h.s. of Equation \eqref{likelihood} are just the normalisations factors of the Bayesian updates of the filtering procedure in Equation \eqref{likelihood}. 

Optimising $p(\mathbf{y})=p(\mathbf{y} | \theta)$ with respect to the parameters yields asymptotically consistent parameter estimates, also called \emph{maximum likelihood estimate}. In a Bayesian framework, one would combine the likelihood with a parameter prior $p(\theta)$ to give the posterior over the parameters  according to Bayes' law 
\begin{equation}
\begin{split}
  p(\theta| \mathbf{y}) \propto p(\mathbf{y}| \theta) p(\theta),
\end{split}
\end{equation}
which also allows to quantify uncertainty of the inferred parameter values.

\subsection{Computational methods for Bayesian inference in stochastic chemical reaction networks}

\subsubsection{Methods for general networks}

The primary difficulty in applying the forward-backward approach to inference in chemical reaction networks is the requirement for forward integrability of the system dynamics: calculating the transition probabilities $p(\mathbf{x}_i | \mathbf{x}_{i-1})$ requires solving the CME, which is generally not possible analytically. Some approaches resort to numerical integration of the equations, including the variational approach of \cite{opper2008variational} and the uniformisation sampler of \cite{rao2013fast}. These approaches can be effective for closed systems with low molecular numbers; however their application to open systems invariably requires an artificial truncation of the state space, introducing a bias which is hard to quantify. Truncations are also used in the recent work of \cite{georgoulas2015unbiased}; however, here a random truncation scheme guarantees unbiasedness of the results, as well as leading to substantial computational savings. Other approaches that can handle open systems either introduce additional latent auxiliary variables (such as the number of reactions in the time interval as in \cite{boys2008bayesian}), or resort to a sequential Monte-Carlo scheme which relies on multiple simulations from different initial conditions (particle filtering, \cite{Golightly2011}). Both such schemes incur potentially large computational overheads.

The computationally intensive nature of inference methodologies adopting a microscopic system description constitutes a formidable obstacle to inference in large-scale reaction networks. This has justified a considerable interest in the use of mesoscopic approximations for inference. One of the earliest attempts \cite{Golightly2005} relied on the chemical Langevin equation approximation to the CME (Section \ref{sec_stochkin_cle}); this provides a more efficient inference scheme compared to the auxiliary variable approach of \cite{boys2008bayesian}, however the computational costs remain high due to the need to compute transition probabilities for non-linear diffusion processes. In this light, the LNA provides a more promising avenue, since the sufficient statistics of the (Gaussian) single-time marginals can be efficiently computed by integrating a system of ordinary differential equations. Several authors have therefore  proposed  inference schemes which integrate the LNA approximation \cite{Ruttor2009,stathopoulos2013markov,Fearnhead2014}. Moment closure approximations provide an alternative approximation scheme with similar computational complexity to the LNA, however they do not generally compute a marginal distribution, rather only a few moments of a generally unknown distribution. Their use for time series inference is therefore limited to second-order normal moment closure schemes, where a Gaussian approximation is taken. This approach has been proposed in \cite{Milner2013}, where it has been shown to yield accurate results with modest computational overheads. In \cite{cseke2015expectation} the second-order normal moment closure has been combined with the chemical Langevin equation and integrated into an expectation-propagation algorithm for intractable likelihoods and continuous-time constraints. 
Another interesting opportunity for moment-based inference is offered by flow-cytometry data: here, simultaneous measurements of millions of cells enable an empirical characterisation of the marginal moments directly (albeit potentially corrupted by extrinsic noise), which can then be fitted to a moment-approximation of the CME \cite{Zechner2012}. 

\subsubsection{Inference for gene expression data}

The inference methods mentioned above do not assume any knowledge about a given system. While this makes them in principle applicable to any type of reaction network, more efficient and/or accurate methods can often be employed by including a priori knowledge in a model. Gene expression systems constitute a particularly important example where this is often the case. Cells typically possess only one or very few copies of a gene. Proteins and mRNA molecules on the other hand often occur at copy numbers that are orders of magnitude larger. Such systems are therefore often suitable for hybrid methods (c.f., Section \ref{sec_other_approx_hybrid}) that model some of the species as discrete variables (e.g. the genes) and others as continuous variables (e.g. the mRNA and proteins). 

While it is often not straightforward to integrate  hybrid methods into inference schemes, significant progress has recently been made in this respect. In \cite{stimberg2011inference}, for instance, the different promoter states of a gene are modelled as a change-point process which drives  a linear SDE representing the protein dynamics, and an efficient MCMC inference method is developed.  The same model has been integrated into a variational inference method which additionally allows transcriptional feedback  in \cite{ocone2011reconstructing, ocone2013hybrid}. In \cite{sherlock2014bayesian}, a particle MCMC scheme is developed based on a hybrid method that approximates the continuous variables by the linear noise approximation (c.f. Section \ref{subsec_lna}). More recently a different hybrid approach combining several types of approximations has been integrated into an efficient Bayesian inference scheme in \cite{hey2015stochastic}. 

\subsection{Summary}

This bird eye survey of inference for stochastic chemical reaction networks highlights the diversity of statistical research in the area. Such diversity can appear baffling to the outsider, and a major problem for the greater diffusion of these ideas is the lack of standard software tools. Inference tools for stochastic chemical reaction networks often form a small subsection of software tools for parameter estimation of deterministic methods \cite{liepe2010abc,chen2010calibayes}, and there haven't been systematic comparisons of various inference schemes that investigate the relative merits of the different algorithms on a number of relevant examples. Furthermore, inference approaches inevitably construct an approximation of a posterior distribution; while much statistics research investigates the convergence properties of these approximations, it remains entirely unclear how inference errors combine with approximation errors when inference schemes are deployed on approximate dynamics. Due to these reasons, we have chosen not to include an explicit numerical comparison between inference methods in this tutorial, as we feel this would deserve a separate review on its own.

\section{Conclusions}\label{sec_conclusions}

Recent years have seen an explosion of experimental studies revealing the crucial role that stochastic fluctuations in chemical reaction networks play for living cells.  Driven by these discoveries a plethora of methods for the mathematical and statistical analysis of such systems has evolved.  The goal of this review is to give a self-contained introduction to the field of modelling for stochastic chemical kinetics. Moreover, it introduces key approximation and inference methods for this field and gives an overview of recent developments. 

The Chemical Master Equation (CME) constitutes the accepted non-spatial description of stochastic chemical networks. Recent years have seen a burst of analysis and approximation methods based on the CME. We gave here an introduction to the CME modelling framework and discussed stochastic simulation and analytic solution methods. Next, we introduced various approximation methods with particular focus on the chemical Langevin equation, the system size expansion and moment closure methods.
We also gave an introduction to time-scale separation based approximations, as well as hybrid methods and reviewed the existing literature. Finally, we gave an introduction to the problem of statistical inference from experimental data in a Bayesian framework, and reviewed existing methods. The presentation is aimed to be a self-contained introduction for scientists from different disciplines.

In a numerical case study we compared the chemical Langevin equation, the zeroth order system size expansion (linear noise approximation, LNA), the first-order corrections of the system size expansion to mean and covariance (SSE-1) and the second-order normal moment closure (2MA) with exact results obtained using stochastic simulations.  In terms of moments, we found that a naive real-valued implementation of the CLE (CLE-R) enforcing positive concentrations was less accurate than all the other methods. A complex version of the CLE (CLE-C), in contrast, was found to be the most accurate of all methods. The SSE-1 and 2MA performed similar two each other and significantly better than the LNA. In terms of steady-state distributions, we compared the CLE-R, the CLE-C and the LNA with exact results obtained using the SSA. We found that the CLE-R and the LNA were not able to accurately capture the distributions, the LNA being more accurate than the CLE-R, however. The CLE-C, in contrast, gave accurate approximations even for highly skewed distributions.

The CME is a valid description of systems that are well-mixed and sufficiently dilute, i.e., the diffusion of particles constitutes the fastest time scale of the system and the total volume of all molecules in the model is much smaller than the system volume. While these assumptions are valid in some cases, it turns out that they are not met by many biological systems. Whenever this is the case, models need to be employed that take spatial positions and diffusion of particles into account. The main family of such models goes under the name of \emph{stochastic reaction-diffusion processes} (SRDPs), Markovian models where independent particles diffuse in space and react whenever they come in contact (or sufficiently close). The evolution equation for the marginal probabilities of an SRDP, the spatial analogue of the CME, is complicated by the fact that the number of particles varies in time, and needs to be defined on an infinite-dimensional Fock space \cite{Doi1976,Doi1976b}. Solving, or even approximating, such an equation is essentially impossible, and hence SRDPs are mainly analysed in an algorithmic way: each particle is simulated performing Brownian motion in continuous space and chemical reactions between particles happen stochastically under certain rules \cite{Erban2009}. This is computationally extremely expensive and significant effort has been spent in improved simulation methods \cite{Zon2005, Donev2010}. An alternative modelling framework is given by the \emph{reaction-diffusion master equation} (RDME) which coarse-grains an SRDP by assuming a compartmentalisation of space and locally homogeneous conditions within each compartment \cite{Isaacson2008}. While simulations in this framework are generally more efficient than in the continuous case, they are typically still expensive \cite{Fu2014}, generally significantly more expensive than in the non-spatial CME case. 
More importantly, the RDME is not a systematic discretisation of an SRDP, in the sense that its continuum limit generally does not lead to the original SRDP in two or more spatial dimensions \cite{Isaacson2008}, because bimolecular reactions become vanishingly infrequent in the continuum limit. 

Due to these reasons, analysis methodologies for spatial models are much less developed than in the non-spatial CME case. Some studies investigating spatial stochastic phenomena and comparing the SRDP and RDME approaches include \cite{Erban2009,isaacson2009reaction, lipkova2011analysis,hellander2012reaction,mahmutovic2012lost,isaacson2013convergent, taylor2014deriving}. In contrast to the CME case, very few studies have attempted analytical approximations for SRDPs. However, some progress has been made in recent years in this respect. In \cite{scott2010approximating} and \cite{asllani2013linear}, for example, the linear noise approximation was extended to spatial systems, and in \cite{smith2016analytical} higher orders of the system size expansion for effective one-species systems. Even fewer studies have addressed inference for SRDPs. Only a handful of studies have approached the issue of inference working directly with the SRDP or RDME framework, using likelihood-free sampling methods \cite{Holmes2012} or variational approximations \cite{Dewar2010}.  In \cite{Ruttor2010}, the linear noise approximation has been integrated into an inference scheme for SRDPs. This method is however limited to certain classes of one-dimensional systems. More recently, in \cite{Schnoerr2016} it was shown that SRDPs can be approximated by spatio-temporal point processes, a popular class of models from statistics \cite{Cressie2011}. This was used to derive an efficent inference algorithm for general SRDPs.

In summary, we have presented an introduction to modelling, approximation and inference methods for stochastic chemical kinetics based on the Chemical Master Equation. We hope that this review will help  scientists from other disciplines to dive into this exciting field, and that it will stimulate research in the presented areas.

\section*{Acknowledgments}
This work was supported by the Biotechnology and Biological Sciences Research Council  [BB/F017073/1]; the Leverhulme Trust [RPG-2013-171]; and the European Research Council [MLCS 306999]. The authors thank Elco Bakker, Lucia Bandiera, Lito Papaxenopoulou and Michael Rule for useful comments on a draft of this article.

{\small
\bibliography{jpa_review}

\begin{thebibliography}{100}

\bibitem{eldar2010functional}
Avigdor Eldar and Michael~B Elowitz.
\newblock Functional roles for noise in genetic circuits.
\newblock {\em Nature}, 467(7312):167--173, 2010.

\bibitem{Elowitz2002}
Michael~B Elowitz, Arnold~J Levine, Eric~D Siggia, and Peter~S Swain.
\newblock Stochastic gene expression in a single cell.
\newblock {\em Science}, 297(5584):1183--1186, 2002.

\bibitem{Gillespie1992}
Daniel~T Gillespie.
\newblock A rigorous derivation of the chemical master equation.
\newblock {\em Physica A: Statistical Mechanics and its Applications},
  188(1):404--425, 1992.

\bibitem{Gillespie1976}
Daniel~T Gillespie.
\newblock A general method for numerically simulating the stochastic time
  evolution of coupled chemical reactions.
\newblock {\em Journal of Computational Physics}, 22(4):403--434, 1976.

\bibitem{Gillespie1977}
Daniel~T Gillespie.
\newblock Exact stochastic simulation of coupled chemical reactions.
\newblock {\em The Journal of Physical Chemistry}, 81(25):2340--2361, 1977.

\bibitem{turner2004stochastic}
Thomas~E Turner, Santiago Schnell, and Kevin Burrage.
\newblock Stochastic approaches for modelling in vivo reactions.
\newblock {\em Computational biology and chemistry}, 28(3):165--178, 2004.

\bibitem{el2005stochastic}
Hana El~Samad, Mustafa Khammash, Linda Petzold, and Dan Gillespie.
\newblock Stochastic modelling of gene regulatory networks.
\newblock {\em International Journal of Robust and Nonlinear Control},
  15(15):691--711, 2005.

\bibitem{li2008algorithms}
Hong Li, Yang Cao, Linda~R Petzold, and Daniel~T Gillespie.
\newblock Algorithms and software for stochastic simulation of biochemical
  reacting systems.
\newblock {\em Biotechnology Progress}, 24(1):56--61, 2008.

\bibitem{pahle2009biochemical}
J{\"u}rgen Pahle.
\newblock Biochemical simulations: stochastic, approximate stochastic and
  hybrid approaches.
\newblock {\em Briefings in Bioinformatics}, 10(1):53--64, 2009.

\bibitem{Mauch2011}
Sean Mauch and Mark Stalzer.
\newblock Efficient formulations for exact stochastic simulation of chemical
  systems.
\newblock {\em IEEE/ACM Transactions on Computational Biology and
  Bioinformatics (TCBB)}, 8(1):27--35, 2011.

\bibitem{Gillespie2013}
Daniel~T Gillespie, Andreas Hellander, and Linda~R Petzold.
\newblock Perspective: {S}tochastic algorithms for chemical kinetics.
\newblock {\em The Journal of Chemical Physics}, 138(17):170901, 2013.

\bibitem{szekely2014stochastic}
Tamas Szekely and Kevin Burrage.
\newblock Stochastic simulation in systems biology.
\newblock {\em Computational and structural biotechnology journal},
  12(20):14--25, 2014.

\bibitem{nicolis1977self}
Gregoire Nicolis, Ilya Prigogine, et~al.
\newblock {\em Self-organization in nonequilibrium systems}, volume 191977.
\newblock Wiley, New York, 1977.

\bibitem{VanKampen2007}
Nicolaas~G van Kampen.
\newblock {\em Stochastic processes in physics and chemistry}, volume~1.
\newblock Elsevier, 1992.

\bibitem{Gardiner2009}
Crispin~W Gardiner.
\newblock {\em Handbook of {S}tochastic {M}ethods}, volume~3.
\newblock Springer Berlin, 1985.

\bibitem{goutsias2013markovian}
John Goutsias and Garrett Jenkinson.
\newblock Markovian dynamics on complex reaction networks.
\newblock {\em Physics Reports}, 529(2):199--264, 2013.

\bibitem{weber2016master}
Markus~F Weber and Erwin Frey.
\newblock Master equations and the theory of stochastic path integrals.
\newblock {\em arXiv preprint arXiv:1609.02849}, 2016.

\bibitem{blythe2007stochastic}
Richard~A Blythe and Alan~J McKane.
\newblock Stochastic models of evolution in genetics, ecology and linguistics.
\newblock {\em Journal of Statistical Mechanics: Theory and Experiment},
  2007(07):P07018, 2007.

\bibitem{datta2010jump}
Samik Datta, Gustav~W Delius, and Richard Law.
\newblock A jump-growth model for predator--prey dynamics: derivation and
  application to marine ecosystems.
\newblock {\em Bulletin of Mathematical Biology}, 72(6):1361--1382, 2010.

\bibitem{black2012stochastic}
Andrew~J Black and Alan~J McKane.
\newblock Stochastic formulation of ecological models and their applications.
\newblock {\em Trends in Ecology \& Evolution}, 27(6):337--345, 2012.

\bibitem{bartlett1949some}
MS~Bartlett.
\newblock Some evolutionary stochastic processes.
\newblock {\em Journal of the Royal Statistical Society. Series B
  (Methodological)}, 11(2):211--229, 1949.

\bibitem{bailey1950simple}
Norman~TJ Bailey.
\newblock A simple stochastic epidemic.
\newblock {\em Biometrika}, pages 193--202, 1950.

\bibitem{keeling2008methods}
Matthew~James Keeling and Joshua~V Ross.
\newblock On methods for studying stochastic disease dynamics.
\newblock {\em Journal of The Royal Society Interface}, 5(19):171--181, 2008.

\bibitem{black2010stochasticity}
Andrew~J Black and Alan~J McKane.
\newblock Stochasticity in staged models of epidemics: quantifying the dynamics
  of whooping cough.
\newblock {\em Journal of The Royal Society Interface}, 7(49):1219--1227, 2010.

\bibitem{jenkinson2012numerical}
Garrett Jenkinson and John Goutsias.
\newblock Numerical integration of the master equation in some models of
  stochastic epidemiology.
\newblock {\em PloS One}, 7(5):e36160, 2012.

\bibitem{pastor2015epidemic}
Romualdo Pastor-Satorras, Claudio Castellano, Piet Van~Mieghem, and Alessandro
  Vespignani.
\newblock Epidemic processes in complex networks.
\newblock {\em Reviews of Modern Physics}, 87(3):925, 2015.

\bibitem{weidlich1991physics}
Wolfgang Weidlich.
\newblock Physics and social science? {T}he approach of synergetics.
\newblock {\em Physics Reports}, 204(1):1--163, 1991.

\bibitem{weidlich2002sociodynamics}
Wolfgang Weidlich.
\newblock Sociodynamics -- a systematic approach to mathematical modelling in
  the social sciences.
\newblock {\em NONLINEAR PHENOMENA IN COMPLEX SYSTEMS-MINSK-}, 5(4):479--487,
  2002.

\bibitem{bonabeau2002agent}
Eric Bonabeau.
\newblock Agent-based modeling: {M}ethods and techniques for simulating human
  systems.
\newblock {\em Proceedings of the National Academy of Sciences}, 99(suppl
  3):7280--7287, 2002.

\bibitem{ohira1993master}
Toru Ohira and Jack~D Cowan.
\newblock Master-equation approach to stochastic neurodynamics.
\newblock {\em Physical Review E}, 48(3):2259, 1993.

\bibitem{el2009master}
Sami El~Boustani and Alain Destexhe.
\newblock A master equation formalism for macroscopic modeling of asynchronous
  irregular activity states.
\newblock {\em Neural Computation}, 21(1):46--100, 2009.

\bibitem{benayoun2010avalanches}
Marc Benayoun, Jack~D Cowan, Wim van Drongelen, and Edward Wallace.
\newblock Avalanches in a stochastic model of spiking neurons.
\newblock {\em PLoS Computational Biology}, 6(7):e1000846, 2010.

\bibitem{buice2010systematic}
Michael~A Buice, Jack~D Cowan, and Carson~C Chow.
\newblock Systematic fluctuation expansion for neural network activity
  equations.
\newblock {\em Neural Computation}, 22(2):377--426, 2010.

\bibitem{wallace2011emergent}
Edward Wallace, Marc Benayoun, Wim Van~Drongelen, and Jack~D Cowan.
\newblock Emergent oscillations in networks of stochastic spiking neurons.
\newblock {\em Plos One}, 6(5):e14804, 2011.

\bibitem{leen2012stochastic}
Todd~K Leen and Robert Friel.
\newblock Stochastic perturbation methods for spike-timing-dependent
  plasticity.
\newblock {\em Neural computation}, 24(5):1109--1146, 2012.

\bibitem{goychuk2015stochastic}
Igor Goychuk and Andriy Goychuk.
\newblock Stochastic {W}ilson--{C}owan models of neuronal network dynamics with
  memory and delay.
\newblock {\em New Journal of Physics}, 17(4):045029, 2015.

\bibitem{alberts1995molecular}
Bruce Alberts, Dennis Bray, Julian Lewis, Martin Raff, Keith Roberts, James~D
  Watson, and AV~Grimstone.
\newblock Molecular {B}iology of the {C}ell (3rd edn).
\newblock {\em Trends in Biochemical Sciences}, 20(5):210--210, 1995.

\bibitem{swain2002intrinsic}
Peter~S Swain, Michael~B Elowitz, and Eric~D Siggia.
\newblock Intrinsic and extrinsic contributions to stochasticity in gene
  expression.
\newblock {\em Proceedings of the National Academy of Sciences},
  99(20):12795--12800, 2002.

\bibitem{austin2006gene}
DW~Austin, MS~Allen, JM~McCollum, RD~Dar, JR~Wilgus, GS~Sayler, NF~Samatova,
  CD~Cox, and ML~Simpson.
\newblock Gene network shaping of inherent noise spectra.
\newblock {\em Nature}, 439(7076):608--611, 2006.

\bibitem{Ozbudak2002}
Ertugrul~M Ozbudak, Mukund Thattai, Iren Kurtser, Alan~D Grossman, and
  Alexander van Oudenaarden.
\newblock Regulation of noise in the expression of a single gene.
\newblock {\em Nature Genetics}, 31(1):69--73, 2002.

\bibitem{colman2005regulated}
Alejandro Colman-Lerner, Andrew Gordon, Eduard Serra, Tina Chin, Orna Resnekov,
  Drew Endy, C~Gustavo Pesce, and Roger Brent.
\newblock Regulated cell-to-cell variation in a cell-fate decision system.
\newblock {\em Nature}, 437(7059):699--706, 2005.

\bibitem{rosenfeld2005gene}
Nitzan Rosenfeld, Jonathan~W Young, Uri Alon, Peter~S Swain, and Michael~B
  Elowitz.
\newblock Gene regulation at the single-cell level.
\newblock {\em Science}, 307(5717):1962--1965, 2005.

\bibitem{bar2006noise}
Arren Bar-Even, Johan Paulsson, Narendra Maheshri, Miri Carmi, Erin O'Shea,
  Yitzhak Pilpel, and Naama Barkai.
\newblock Noise in protein expression scales with natural protein abundance.
\newblock {\em Nature genetics}, 38(6):636--643, 2006.

\bibitem{Taniguchi2010}
Yuichi Taniguchi, Paul~J Choi, Gene-Wei Li, Huiyi Chen, Mohan Babu, Jeremy
  Hearn, Andrew Emili, and X~Sunney Xie.
\newblock Quantifying {E}. coli proteome and transcriptome with single-molecule
  sensitivity in single cells.
\newblock {\em Science}, 329(5991):533--538, 2010.

\bibitem{brehm2004single}
Byron~F Brehm-Stecher and Eric~A Johnson.
\newblock Single-cell microbiology: tools, technologies, and applications.
\newblock {\em Microbiology and Molecular Biology Reviews}, 68(3):538--559,
  2004.

\bibitem{maheshri2007living}
Narendra Maheshri and Erin~K O'Shea.
\newblock Living with noisy genes: how cells function reliably with inherent
  variability in gene expression.
\newblock {\em Annual Review of Biophysics and Biomolecular Structure},
  36:413--434, 2007.

\bibitem{Balazsi2011}
G{\'a}bor Bal{\'a}zsi, Alexander van Oudenaarden, and James~J Collins.
\newblock Cellular decision making and biological noise: from microbes to
  mammals.
\newblock {\em Cell}, 144(6):910--925, 2011.

\bibitem{guldberg1864studies}
Cato~M Guldberg and Peter Waage.
\newblock Studies concerning affinity.
\newblock {\em CM Forhandlinger: Videnskabs-Selskabet i Christiana}, 35, 1864.

\bibitem{waage1864experiments}
Peter Waage.
\newblock Experiments for determining the affinity law.
\newblock {\em Forhandlinger: Videnskabs-Selskabet i Christiana}, 92, 1864.

\bibitem{guldberg1864concerning}
Cato~Maximilian Guldberg.
\newblock Concerning the laws of chemical affinity.
\newblock {\em CM Forhandlinger: Videnskabs-Selskabet i Christiana}, 111, 1864.

\bibitem{guldberg1879concerning}
Cato~Maxilian Guldberg and Peter Waage.
\newblock Concerning chemical affinity.
\newblock {\em Erdmann's Journal f\"{u}r Praktische Chemie}, 127:69--114, 1879.

\bibitem{klipp2008systems}
Edda Klipp, Ralf Herwig, Axel Kowald, Christoph Wierling, and Hans Lehrach.
\newblock {\em Systems biology in practice: concepts, implementation and
  application}.
\newblock John Wiley \& Sons, 2008.

\bibitem{Delbruck1940}
Max Delbr{\"u}ck.
\newblock Statistical fluctuations in autocatalytic reactions.
\newblock {\em The Journal of Chemical Physics}, 8(1):120--124, 1940.

\bibitem{bartholomay1958stochastic}
Anthony~F Bartholomay.
\newblock Stochastic models for chemical reactions: {I}. {T}heory of the
  unimolecular reaction process.
\newblock {\em The Bulletin of Mathematical Biophysics}, 20(3):175--190, 1958.

\bibitem{mcquarrie1963kinetics}
Donald~A McQuarrie.
\newblock Kinetics of small systems. {I}.
\newblock {\em The Journal of Chemical Physics}, 38(2):433--436, 1963.

\bibitem{ishida1964stochastic}
Kenji Ishida.
\newblock Stochastic model for bimolecular reaction.
\newblock {\em The Journal of Chemical Physics}, 41(8):2472--2478, 1964.

\bibitem{mcquarrie1964kinetics}
Donald~A McQuarrie, CJ~Jachimowski, and ME~Russell.
\newblock Kinetics of small systems. {II}.
\newblock {\em The Journal of Chemical Physics}, 40(10):2914--2921, 1964.

\bibitem{gillespie2009diffusional}
Daniel~T Gillespie.
\newblock A diffusional bimolecular propensity function.
\newblock {\em The Journal of Chemical Physics}, 131(16):164109, 2009.

\bibitem{GrimaNewman2012}
R~Grima, Deena~R Schmidt, and Timothy~J Newman.
\newblock Steady-state fluctuations of a genetic feedback loop: {A}n exact
  solution.
\newblock {\em The Journal of Chemical Physics}, 137(3):035104, 2012.

\bibitem{Guerriero2011}
Maria~Luisa Guerriero, Alexandra Pokhilko, Aurora~Pi{\~n}as Fern{\'a}ndez,
  Karen~J Halliday, Andrew~J Millar, and Jane Hillston.
\newblock Stochastic properties of the plant circadian clock.
\newblock {\em Journal of The Royal Society Interface}, page rsif20110378,
  2011.

\bibitem{Anderson2007}
David~F Anderson.
\newblock A modified next reaction method for simulating chemical systems with
  time dependent propensities and delays.
\newblock {\em The Journal of Chemical Physics}, 127(21):214107, 2007.

\bibitem{Shahrezaei2008b}
Vahid Shahrezaei, Julien~F Ollivier, and Peter~S Swain.
\newblock Colored extrinsic fluctuations and stochastic gene expression.
\newblock {\em Molecular Systems Biology}, 4(1), 2008.

\bibitem{caravagna2013interplay}
Giulio Caravagna, Giancarlo Mauri, and Alberto d'Onofrio.
\newblock The interplay of intrinsic and extrinsic bounded noises in
  biomolecular networks.
\newblock {\em PLoS One}, 8(2):e51174, 2013.

\bibitem{voliotis2016stochastic}
Margaritis Voliotis, Philipp Thomas, Ramon Grima, and Clive~G Bowsher.
\newblock Stochastic simulation of biomolecular networks in dynamic
  environments.
\newblock {\em PLoS Comput Biol}, 12(6):e1004923, 2016.

\bibitem{al2011computing}
Awad~H Al-Mohy and Nicholas~J Higham.
\newblock Computing the action of the matrix exponential, with an application
  to exponential integrators.
\newblock {\em SIAM Journal on Scientific Computing}, 33(2):488--511, 2011.

\bibitem{moler2003nineteen}
Cleve Moler and Charles Van~Loan.
\newblock Nineteen dubious ways to compute the exponential of a matrix,
  twenty-five years later.
\newblock {\em SIAM Review}, 45(1):3--49, 2003.

\bibitem{Darvey1966}
IG~Darvey and PJ~Staff.
\newblock Stochastic {A}pproach to {F}irst-{O}rder {C}hemical {R}eaction
  {K}inetics.
\newblock {\em The Journal of Chemical Physics}, 44(3):990--997, 1966.

\bibitem{gardiner1977poisson}
CW~Gardiner and S~Chaturvedi.
\newblock The poisson representation. i. a new technique for chemical master
  equations.
\newblock {\em Journal of Statistical Physics}, 17(6):429--468, 1977.

\bibitem{heuett2006grand}
William~J Heuett and Hong Qian.
\newblock Grand canonical markov model: a stochastic theory for open
  nonequilibrium biochemical networks.
\newblock {\em The Journal of Chemical Physics}, 124(4):044110, 2006.

\bibitem{Jahnke2007}
Tobias Jahnke and Wilhelm Huisinga.
\newblock Solving the chemical master equation for monomolecular reaction
  systems analytically.
\newblock {\em Journal of Mathematical Biology}, 54(1):1--26, 2007.

\bibitem{Whittle1986}
Peter Whittle.
\newblock {\em Systems in stochastic equilibrium}.
\newblock John Wiley \& Sons, Inc., 1986.

\bibitem{van1976equilibrium}
NG~Van~Kampen.
\newblock The equilibrium distribution of a chemical mixture.
\newblock {\em Physics Letters A}, 59(5):333--334, 1976.

\bibitem{Anderson2010}
David~F Anderson, Gheorghe Craciun, and Thomas~G Kurtz.
\newblock Product-form stationary distributions for deficiency zero chemical
  reaction networks.
\newblock {\em Bulletin of Mathematical Biology}, 72(8):1947--1970, 2010.

\bibitem{Horn1972}
Fritz Horn and Roy Jackson.
\newblock General mass action kinetics.
\newblock {\em Archive for Rational Mechanics and Analysis}, 47(2):81--116,
  1972.

\bibitem{Feinberg1995}
Martin Feinberg.
\newblock The existence and uniqueness of steady states for a class of chemical
  reaction networks.
\newblock {\em Archive for Rational Mechanics and Analysis}, 132(4):311--370,
  1995.

\bibitem{voituriez2005corrections}
R~Voituriez, M~Moreau, and G~Oshanin.
\newblock Corrections to the law of mass action and properties of the
  asymptotic $ t=\infty $ state for reversible diffusion-limited reactions.
\newblock {\em The Journal of Chemical Physics}, 122:084103, 2005.

\bibitem{gerstung2009noisy}
Moritz Gerstung, Jens Timmer, and Christian Fleck.
\newblock Noisy signaling through promoter logic gates.
\newblock {\em Physical Review E}, 79(1):011923, 2009.

\bibitem{thomas2010stochastic}
Philipp Thomas, Arthur~V Straube, and Ramon Grima.
\newblock Stochastic theory of large-scale enzyme-reaction networks: Finite
  copy number corrections to rate equation models.
\newblock {\em The Journal of Chemical Physics}, 133(19):195101, 2010.

\bibitem{petrosyan2014nonequilibrium}
KG~Petrosyan and Chin-Kun Hu.
\newblock Nonequilibrium lyapunov function and a fluctuation relation for
  stochastic systems: Poisson-representation approach.
\newblock {\em Physical Review E}, 89(4):042132, 2014.

\bibitem{iyer2014mixed}
Srividya Iyer-Biswas and Ciriyam Jayaprakash.
\newblock Mixed poisson distributions in exact solutions of stochastic
  autoregulation models.
\newblock {\em Physical Review E}, 90(5):052712, 2014.

\bibitem{sugar2014self}
Istv{\'a}n Sug{\'a}r and Istv{\'a}n Simon.
\newblock Self-regulating genes. exact steady state solution by using poisson
  representation.
\newblock {\em Open Physics}, 12(9):615--627, 2014.

\bibitem{Schnoerr2016}
David Schnoerr, Ramon Grima, and Guido Sanguinetti.
\newblock Cox process representation and inference for stochastic
  reaction-diffusion processes.
\newblock {\em Nature Communications}, 7:11729, 2016.

\bibitem{laurenzi2000analytical}
Ian~J Laurenzi.
\newblock An analytical solution of the stochastic master equation for
  reversible bimolecular reaction kinetics.
\newblock {\em The Journal of Chemical Physics}, 113(8):3315--3322, 2000.

\bibitem{kumar2014exact}
Niraj Kumar, Thierry Platini, and Rahul~V Kulkarni.
\newblock Exact distributions for stochastic gene expression models with
  bursting and feedback.
\newblock {\em Physical Review Letters}, 113(26):268105, 2014.

\bibitem{aranyi1976full}
P~Ar{\'a}nyi and J~T{\'o}th.
\newblock A full stochastic description of the {M}ichaelis-{M}enten reaction
  for small systems.
\newblock {\em Acta biochimica et biophysica; Academiae Scientiarum
  Hungaricae}, 12(4):375--388, 1976.

\bibitem{staff1970stochastic}
PJ~Staff.
\newblock A stochastic development of the reversible {M}ichaelis-{M}enten
  mechanism.
\newblock {\em Journal of Theoretical Biology}, 27(2):221--232, 1970.

\bibitem{Schnoerr2014}
David Schnoerr, Guido Sanguinetti, and Ramon Grima.
\newblock The complex chemical {L}angevin equation.
\newblock {\em The Journal of Chemical Physics}, 141(2):024103, 2014.

\bibitem{mather2010correlation}
William~H Mather, Natalie~A Cookson, Jeff Hasty, Lev~S Tsimring, and Ruth~J
  Williams.
\newblock Correlation resonance generated by coupled enzymatic processing.
\newblock {\em Biophysical Journal}, 99(10):3172--3181, 2010.

\bibitem{Kramers1940}
Hendrik~A Kramers.
\newblock Brownian motion in a field of force and the diffusion model of
  chemical reactions.
\newblock {\em Physica}, 7(4):284--304, 1940.

\bibitem{Moyal1949}
Jos{\' e}~E Moyal.
\newblock Stochastic processes and statistical physics.
\newblock {\em Journal of the Royal Statistical Society. Series B
  (Methodological)}, 11(2):150--210, 1949.

\bibitem{gillespie2000chemical}
Daniel~T Gillespie.
\newblock The chemical {L}angevin equation.
\newblock {\em The Journal of Chemical Physics}, 113(1):297--306, 2000.

\bibitem{Melykuti2010}
Bence Melykuti, Kevin Burrage, and Konstantinos~C Zygalakis.
\newblock Fast stochastic simulation of biochemical reaction systems by
  alternative formulations of the chemical {L}angevin equation.
\newblock {\em The Journal of Chemical Physics}, 132(16):164109, 2010.

\bibitem{Kloeden1992}
Peter~E Kloeden and Eckhard Platen.
\newblock {\em Numerical {S}olution of {S}tochastic {D}ifferential
  {E}quations}, volume~1.
\newblock Springer Berlin, 1992.

\bibitem{burrage2000numerical}
Kevin Burrage, Pamela Burrage, and Taketomo Mitsui.
\newblock Numerical solutions of stochastic differential
  equations--implementation and stability issues.
\newblock {\em Journal of Computational and Applied Mathematics},
  125(1):171--182, 2000.

\bibitem{higham2001algorithmic}
Desmond~J Higham.
\newblock An algorithmic introduction to numerical simulation of stochastic
  differential equations.
\newblock {\em SIAM Review}, 43(3):525--546, 2001.

\bibitem{burrage2004numerical}
Kevin Burrage, PM~Burrage, and Tianhai Tian.
\newblock Numerical methods for strong solutions of stochastic differential
  equations: an overview.
\newblock {\em Proceedings of The Royal Society of London A: Mathematical,
  Physical and Engineering Sciences}, 460(2041):373--402, 2004.

\bibitem{Kurtz}
Thomas~G Kurtz.
\newblock Limit theorems and diffusion approximations for density dependent
  {M}arkov chains.
\newblock In {\em Stochastic Systems: Modeling, Identification and
  Optimization, {I}}, pages 67--78. Springer, 1976.

\bibitem{Dana2011}
Saswati Dana and Soumyendu Raha.
\newblock Physically consistent simulation of mesoscale chemical kinetics:
  {T}he non-negative {FIS}-$\alpha$ method.
\newblock {\em Journal of Computational Physics}, 230(24):8813--8834, 2011.

\bibitem{Wilkie2008}
Joshua Wilkie and Yin~M Wong.
\newblock Positivity preserving chemical {L}angevin equations.
\newblock {\em Chemical Physics}, 353(1):132--138, 2008.

\bibitem{Szpruch2009}
Lukasz Szpruch and Desmond~J Higham.
\newblock Comparing hitting time behavior of {M}arkov jump processes and their
  diffusion approximations.
\newblock {\em Multiscale Modeling \& Simulation}, 8(2):605--621, 2010.

\bibitem{liao2015tensor}
Shuohao Liao, Tom{\'a}{\v{s}} Vejchodsk{\`y}, and Radek Erban.
\newblock Tensor methods for parameter estimation and bifurcation analysis of
  stochastic reaction networks.
\newblock {\em Journal of The Royal Society Interface}, 12(108):20150233, 2015.

\bibitem{biancalani2014noise}
Tommaso Biancalani, Louise Dyson, and Alan~J McKane.
\newblock Noise-induced bistable states and their mean switching time in
  foraging colonies.
\newblock {\em Physical review letters}, 112(3):038101, 2014.

\bibitem{Duncan2015b}
Andrew Duncan, Shuohao Liao, Tom{\'a}{\v{s}} Vejchodsk{\`y}, Radek Erban, and
  Ramon Grima.
\newblock Noise-induced multistability in chemical systems: {D}iscrete versus
  continuum modeling.
\newblock {\em Physical Review E}, 91(4):042111, 2015.

\bibitem{beccuti2014analysis}
Marco Beccuti, Enrico Bibbona, Andras Horvath, Roberta Sirovich, Alessio
  Angius, and Gianfranco Balbo.
\newblock Analysis of petri net models through stochastic differential
  equations.
\newblock In {\em International Conference on Applications and Theory of Petri
  Nets and Concurrency}, pages 273--293. Springer, 2014.

\bibitem{VanKampen1976}
Nicolaas~G van Kampen.
\newblock The expansion of the master equation.
\newblock {\em Advance in Chemical Physics}, 34:245--309, 1976.

\bibitem{Thomas2012b}
Philipp Thomas, Hannes Matuschek, and Ramon Grima.
\newblock Computation of biochemical pathway fluctuations beyond the linear
  noise approximation using i{N}{A}.
\newblock In {\em IEEE International Conference on Bioinformatics and
  Biomedicine (BIBM)}, pages 1--5, 2012.

\bibitem{Grima2010}
Ramon Grima.
\newblock An effective rate equation approach to reaction kinetics in small
  volumes: {T}heory and application to biochemical reactions in nonequilibrium
  steady-state conditions.
\newblock {\em The Journal of Chemical Physics}, 133(3):035101, 2010.

\bibitem{thomas2015approximate}
Philipp Thomas and Ramon Grima.
\newblock Approximate probability distributions of the master equation.
\newblock {\em Physical Review E}, 92(1):012120, 2015.

\bibitem{mckane2007amplified}
Alan~J McKane, James~D Nagy, Timothy~J Newman, and Marianne~O Stefanini.
\newblock Amplified biochemical oscillations in cellular systems.
\newblock {\em Journal of Statistical Physics}, 128(1-2):165--191, 2007.

\bibitem{pahle2012biochemical}
J{\"u}rgen Pahle, Joseph~D Challenger, Pedro Mendes, and Alan~J McKane.
\newblock Biochemical fluctuations, optimisation and the linear noise
  approximation.
\newblock {\em BMC systems biology}, 6(1):1, 2012.

\bibitem{challenger2013synchronization}
Joseph~D Challenger and Alan~J McKane.
\newblock Synchronization of stochastic oscillators in biochemical systems.
\newblock {\em Physical Review E}, 88(1):012107, 2013.

\bibitem{bianca2015evaluation}
C~Bianca and A~Lemarchand.
\newblock Evaluation of reaction fluxes in stationary and oscillating
  far-from-equilibrium biological systems.
\newblock {\em Physica A: Statistical Mechanics and its Applications},
  438:1--16, 2015.

\bibitem{hufton2016intrinsic}
Peter~G Hufton, Yen~Ting Lin, Tobias Galla, and Alan~J McKane.
\newblock Intrinsic noise in systems with switching environments.
\newblock {\em Physical Review E}, 93(5):052119, 2016.

\bibitem{thomas2013reliable}
Philipp Thomas, Hannes Matuschek, and Ramon Grima.
\newblock How reliable is the linear noise approximation of gene regulatory
  networks?
\newblock {\em BMC genomics}, 14(Suppl 4):S5, 2013.

\bibitem{dauxois2009enhanced}
Thierry Dauxois, Francesca Di~Patti, Duccio Fanelli, and Alan~J McKane.
\newblock Enhanced stochastic oscillations in autocatalytic reactions.
\newblock {\em Physical Review E}, 79(3):036112, 2009.

\bibitem{cianci2012analytical}
Claudia Cianci, Francesca Di~Patti, Duccio Fanelli, and Luigi Barletti.
\newblock Analytical study of non {G}aussian fluctuations in a stochastic
  scheme of autocatalytic reactions.
\newblock {\em The European Physical Journal Special Topics}, 212(1):5--22,
  2012.

\bibitem{scott2012non}
M~Scott.
\newblock Non-linear corrections to the time-covariance function derived from a
  multi-state chemical master equation.
\newblock {\em IET Systems Biology}, 6(4):116--124, 2012.

\bibitem{thomas2013signatures}
Philipp Thomas, Arthur~V Straube, Jens Timmer, Christian Fleck, and Ramon
  Grima.
\newblock Signatures of nonlinearity in single cell noise-induced oscillations.
\newblock {\em Journal of Theoretical Biology}, 335:222--234, 2013.

\bibitem{Pawula1967}
RF~Pawula.
\newblock Approximation of the linear {B}oltzmann equation by the
  {F}okker-{P}lanck equation.
\newblock {\em Physical Review}, 162(1):186, 1967.

\bibitem{leen2011perturbation}
Todd~K Leen and Robert Friel.
\newblock Perturbation theory for stochastic learning dynamics.
\newblock In {\em Neural Networks (IJCNN), The 2011 International Joint
  Conference on}, pages 2031--2038. IEEE, 2011.

\bibitem{Goodman1953}
Leo~A Goodman.
\newblock Population growth of the sexes.
\newblock {\em Biometrics}, 9(2):212--225, 1953.

\bibitem{Whittle1957}
Peter Whittle.
\newblock On the use of the normal approximation in the treatment of stochastic
  processes.
\newblock {\em Journal of the Royal Statistical Society. Series B
  (Methodological)}, pages 268--281, 1957.

\bibitem{Gomez2007}
Carlos~A Gomez-Uribe and George~C Verghese.
\newblock Mass fluctuation kinetics: {C}apturing stochastic effects in systems
  of chemical reactions through coupled mean-variance computations.
\newblock {\em The Journal of Chemical Physics}, 126(2):024109, 2007.

\bibitem{Nasell2003}
Ingemar N{\aa}sell.
\newblock An extension of the moment closure method.
\newblock {\em Theoretical Population Biology}, 64(2):233--239, 2003.

\bibitem{Keeling2000}
Matt~J Keeling.
\newblock Multiplicative moments and measures of persistence in ecology.
\newblock {\em Journal of Theoretical Biology}, 205(2):269--281, 2000.

\bibitem{Edwin1988}
Edwin~L Crow and Kunio Shimizu.
\newblock {\em Lognormal distributions: {T}heory and applications}, volume~88.
\newblock M. Dekker New York, 1988.

\bibitem{Hespanha2008}
Joao Hespanha.
\newblock Moment closure for biochemical networks.
\newblock In {\em 3rd International Symposium on Communications, Control and
  Signal Processing}, pages 142--147. IEEE, 2008.

\bibitem{singh2006lognormal}
Abhyudai Singh and Joao~Pedro Hespanha.
\newblock Lognormal moment closures for biochemical reactions.
\newblock In {\em IEEE Conference on Decision and Control}, pages 2063--2068.
  IEEE, 2006.

\bibitem{Grima2012}
Ramon Grima.
\newblock A study of the accuracy of moment-closure approximations for
  stochastic chemical kinetics.
\newblock {\em The Journal of Chemical Physics}, 136(15):154105, 2012.

\bibitem{Ale2013}
Angelique Ale, Paul Kirk, and Michael~PH Stumpf.
\newblock A general moment expansion method for stochastic kinetic models.
\newblock {\em The Journal of Chemical Physics}, 138(17):174101, 2013.

\bibitem{Schnoerr2014b}
David Schnoerr, Guido Sanguinetti, and Ramon Grima.
\newblock Validity conditions for moment closure approximations in stochastic
  chemical kinetics.
\newblock {\em The Journal of Chemical Physics}, 141(8):084103, 2014.

\bibitem{Schnoerr2015}
David Schnoerr, Guido Sanguinetti, and Ramon Grima.
\newblock Comparison of different moment-closure approximations for stochastic
  chemical kinetics.
\newblock {\em The Journal of Chemical Physics}, 143(18):185101, 2015.

\bibitem{Hespanha2011}
Abhyudai Singh and Joao~P Hespanha.
\newblock Approximate moment dynamics for chemically reacting systems.
\newblock {\em IEEE Transactions on Automatic Control}, 56(2):414--418, 2011.

\bibitem{Lakatos2015}
Eszter Lakatos, Angelique Ale, Paul~DW Kirk, and Michael~PH Stumpf.
\newblock Multivariate moment closure techniques for stochastic kinetic models.
\newblock {\em The Journal of Chemical Physics}, 143(9):094107, 2015.

\bibitem{shannon2001mathematical}
Claude~Elwood Shannon.
\newblock A mathematical theory of communication.
\newblock {\em ACM SIGMOBILE Mobile Computing and Communications Review},
  5(1):3--55, 2001.

\bibitem{andreychenko2015model}
Alexander Andreychenko, Linar Mikeev, and Verena Wolf.
\newblock Model reconstruction for moment-based stochastic chemical kinetics.
\newblock {\em ACM Transactions on Modeling and Computer Simulation (TOMACS)},
  25(2):12, 2015.

\bibitem{andreychenko2015distribution}
Alexander Andreychenko, Luca Bortolussi, Ramon Grima, Philipp Thomas, and
  Verena Wolf.
\newblock Distribution approximations for the chemical master equation:
  comparison of the method of moments and the system size expansion.
\newblock {\em arXiv preprint arXiv:1509.09104}, 2015.

\bibitem{smadbeck2013closure}
Patrick Smadbeck and Yiannis~N Kaznessis.
\newblock A closure scheme for chemical master equations.
\newblock {\em Proceedings of the National Academy of Sciences},
  110(35):14261--14265, 2013.

\bibitem{ramsey2005dizzy}
Stephen Ramsey, David Orrell, and Hamid Bolouri.
\newblock Dizzy: stochastic simulation of large-scale genetic regulatory
  networks.
\newblock {\em Journal of Bioinformatics and Computational Biology},
  3(02):415--436, 2005.

\bibitem{hoops2006copasi}
Stefan Hoops, Sven Sahle, Ralph Gauges, Christine Lee, J{\"u}rgen Pahle,
  Natalia Simus, Mudita Singhal, Liang Xu, Pedro Mendes, and Ursula Kummer.
\newblock {COPASI}--a complex pathway simulator.
\newblock {\em Bioinformatics}, 22(24):3067--3074, 2006.

\bibitem{sanft2011stochkit2}
Kevin~R Sanft, Sheng Wu, Min Roh, Jin Fu, Rone~Kwei Lim, and Linda~R Petzold.
\newblock Stoch{K}it2: software for discrete stochastic simulation of
  biochemical systems with events.
\newblock {\em Bioinformatics}, 27(17):2457--2458, 2011.

\bibitem{maarleveld2013stochpy}
Timo~R Maarleveld, Brett~G Olivier, and Frank~J Bruggeman.
\newblock Stoch{P}y: a comprehensive, user-friendly tool for simulating
  stochastic biological processes.
\newblock {\em PloS One}, 8(11):e79345, 2013.

\bibitem{thomas2012intrinsic}
Philipp Thomas, Hannes Matuschek, and Ramon Grima.
\newblock Intrinsic noise analyzer: a software package for the exploration of
  stochastic biochemical kinetics using the system size expansion.
\newblock {\em PloS One}, 7(6):e38518, 2012.

\bibitem{Hespanha2007}
Joao~P Hespanha.
\newblock Stoch{D}yn{T}ools - a {MATLAB} toolbox to compute moment dynamics for
  stochastic networks of bio-chemical reactions, 2006.

\bibitem{Gillespie2009}
Colin~S Gillespie.
\newblock Moment-closure approximations for mass-action models.
\newblock {\em Systems Biology, IET}, 3(1):52--58, 2009.

\bibitem{fan2016means}
Sisi Fan, Quentin Geissmann, Eszter Lakatos, Saulius Lukauskas, Angelique Ale,
  Ann~C Babtie, Paul~DW Kirk, and Michael~PH Stumpf.
\newblock {MEANS}: python package for {M}oment {E}xpansion {A}pproximation,
  i{N}ference and {S}imulation.
\newblock {\em Bioinformatics}, page btw229, 2016.

\bibitem{kazeroonian2016cerena}
Atefeh Kazeroonian, Fabian Fr{\"o}hlich, Andreas Raue, Fabian~J Theis, and Jan
  Hasenauer.
\newblock {CERENA}: {C}h{E}mical {RE}action {N}etwork {A}nalyzer--{A} {T}oolbox
  for the {S}imulation and {A}nalysis of {S}tochastic {C}hemical {K}inetics.
\newblock {\em PloS One}, 11(1):e0146732, 2016.

\bibitem{munsky2006finite}
Brian Munsky and Mustafa Khammash.
\newblock The finite state projection algorithm for the solution of the
  chemical master equation.
\newblock {\em The Journal of Chemical Physics}, 124(4):044104, 2006.

\bibitem{dinh2016understanding}
Khanh~N Dinh and Roger~B Sidje.
\newblock Understanding the finite state projection and related methods for
  solving the chemical master equation.
\newblock {\em Physical biology}, 13(3):035003, 2016.

\bibitem{gillespie2001approximate}
Daniel~T Gillespie.
\newblock Approximate accelerated stochastic simulation of chemically reacting
  systems.
\newblock {\em The Journal of Chemical Physics}, 115(4):1716--1733, 2001.

\bibitem{rathinam2003stiffness}
Muruhan Rathinam, Linda~R Petzold, Yang Cao, and Daniel~T Gillespie.
\newblock Stiffness in stochastic chemically reacting systems: The implicit
  tau-leaping method.
\newblock {\em The Journal of Chemical Physics}, 119(24):12784--12794, 2003.

\bibitem{cao2007adaptive}
Yang Cao, Daniel~T Gillespie, and Linda~R Petzold.
\newblock Adaptive explicit-implicit tau-leaping method with automatic tau
  selection.
\newblock {\em The Journal of Chemical Physics}, 126(22):224101, 2007.

\bibitem{tian2004binomial}
Tianhai Tian and Kevin Burrage.
\newblock Binomial leap methods for simulating stochastic chemical kinetics.
\newblock {\em The Journal of Chemical Physics}, 121(21):10356--10364, 2004.

\bibitem{chatterjee2005binomial}
Abhijit Chatterjee, Dionisios~G Vlachos, and Markos~A Katsoulakis.
\newblock Binomial distribution based $\tau$-leap accelerated stochastic
  simulation.
\newblock {\em The Journal of Chemical Physics}, 122(2):024112, 2005.

\bibitem{peng2007efficient}
Xinjun Peng, Wen Zhou, and Yifei Wang.
\newblock Efficient binomial leap method for simulating chemical kinetics.
\newblock {\em The Journal of Chemical Physics}, 126(22):224109, 2007.

\bibitem{leier2008generalized}
Andr{\'e} Leier, Tatiana~T Marquez-Lago, and Kevin Burrage.
\newblock Generalized binomial $\tau$-leap method for biochemical kinetics
  incorporating both delay and intrinsic noise.
\newblock {\em The Journal of Chemical Physics}, 128(20):205107, 2008.

\bibitem{pettigrew2007multinomial}
Michel~F Pettigrew and Haluk Resat.
\newblock Multinomial tau-leaping method for stochastic kinetic simulations.
\newblock {\em The Journal of Chemical Physics}, 126(8):084101, 2007.

\bibitem{cai2007k}
Xiaodong Cai and Zhouyi Xu.
\newblock K-leap method for accelerating stochastic simulation of coupled
  chemical reactions.
\newblock {\em The Journal of Chemical Physics}, 126(7):74102--74102, 2007.

\bibitem{anderson2008incorporating}
David~F Anderson.
\newblock Incorporating postleap checks in tau-leaping.
\newblock {\em The Journal of Chemical Physics}, 128(5):054103, 2008.

\bibitem{gillespie2008simulation}
Daniel~T Gillespie.
\newblock Simulation methods in systems biology.
\newblock In {\em Formal Methods for Computational Systems Biology}, volume
  5016, pages 125--167. Springer, 2008.

\bibitem{bowen1963singular}
JR~Bowen, Andreas Acrivos, and AK~Oppenheim.
\newblock Singular perturbation refinement to quasi-steady state approximation
  in chemical kinetics.
\newblock {\em Chemical Engineering Science}, 18(3):177--188, 1963.

\bibitem{janssen1989elimination}
JAM Janssen.
\newblock The elimination of fast variables in complex chemical reactions.
  {II}. {M}esoscopic level (reducible case).
\newblock {\em Journal of Statistical Physics}, 57(1-2):171--185, 1989.

\bibitem{heinrich2012regulation}
Reinhart Heinrich and Stefan Schuster.
\newblock {\em The regulation of cellular systems}.
\newblock Springer Science \& Business Media, 2012.

\bibitem{michaelis1913kinetik}
Leonor Michaelis and Maud~L Menten.
\newblock Die {K}inetik der {I}nvertinwirkung.
\newblock {\em Biochemische Zeitschrift}, 49(333-369):352, 1913.

\bibitem{segel1989quasi}
Lee~A Segel and Marshall Slemrod.
\newblock The quasi-steady-state assumption: a case study in perturbation.
\newblock {\em SIAM Review}, 31(3):446--477, 1989.

\bibitem{briggs1925note}
George~Edward Briggs and John Burdon~Sanderson Haldane.
\newblock A note on the kinetics of enzyme action.
\newblock {\em Biochemical Journal}, 19(2):338, 1925.

\bibitem{goutsias2005quasiequilibrium}
John Goutsias.
\newblock Quasiequilibrium approximation of fast reaction kinetics in
  stochastic biochemical systems.
\newblock {\em The Journal of Chemical Physics}, 122(18):184102, 2005.

\bibitem{janssen1989elimination2}
JAM Janssen.
\newblock The elimination of fast variables in complex chemical reactions.
  {III}. {M}esoscopic level (irreducible case).
\newblock {\em Journal of Statistical Physics}, 57(1-2):187--198, 1989.

\bibitem{gonze2002deterministic}
Didier Gonze, Jos{\'e} Halloy, and Albert Goldbeter.
\newblock Deterministic versus stochastic models for circadian rhythms.
\newblock {\em Journal of Biological Physics}, 28(4):637--653, 2002.

\bibitem{rao2003stochastic}
Christopher~V Rao and Adam~P Arkin.
\newblock Stochastic chemical kinetics and the quasi-steady-state assumption:
  application to the gillespie algorithm.
\newblock {\em The Journal of Chemical Physics}, 118(11):4999--5010, 2003.

\bibitem{sanft2011legitimacy}
Kevin~R Sanft, Daniel~T Gillespie, and Linda~R Petzold.
\newblock Legitimacy of the stochastic {M}ichaelis-{M}enten approximation.
\newblock {\em IET Systems Biology}, 5(1):58--69, 2011.

\bibitem{thomas2011communication}
Philipp Thomas, Arthur~V Straube, and Ramon Grima.
\newblock Communication: limitations of the stochastic quasi-steady-state
  approximation in open biochemical reaction networks.
\newblock {\em The Journal of Chemical Physics}, 135(18):181103, 2011.

\bibitem{kim2015relationship}
Jae~K Kim, Kre{\v{s}}imir Josi{\'c}, and Matthew~R Bennett.
\newblock The relationship between stochastic and deterministic quasi-steady
  state approximations.
\newblock {\em BMC Systems Biology}, 9(1):87, 2015.

\bibitem{thomas2012slow}
Philipp Thomas, Arthur~V Straube, and Ramon Grima.
\newblock The slow-scale linear noise approximation: an accurate, reduced
  stochastic description of biochemical networks under timescale separation
  conditions.
\newblock {\em BMC Systems Biology}, 6(1):39, 2012.

\bibitem{cao2005slow}
Yang Cao, Daniel~T Gillespie, and Linda~R Petzold.
\newblock The slow-scale stochastic simulation algorithm.
\newblock {\em The Journal of Chemical Physics}, 122(1):014116, 2005.

\bibitem{salis2005equation}
Howard Salis and Yiannis~N Kaznessis.
\newblock An equation-free probabilistic steady-state approximation: dynamic
  application to the stochastic simulation of biochemical reaction networks.
\newblock {\em The Journal of Chemical Physics}, 123(21):214106, 2005.

\bibitem{weinan2005nested}
E~Weinan, Di~Liu, and Eric Vanden-Eijnden.
\newblock Nested stochastic simulation algorithm for chemical kinetic systems
  with disparate rates.
\newblock {\em The Journal of Chemical Physics}, 123(19):194107, 2005.

\bibitem{samant2005overcoming}
A~Samant and DG~Vlachos.
\newblock Overcoming stiffness in stochastic simulation stemming from partial
  equilibrium: a multiscale {M}onte {C}arlo algorithm.
\newblock {\em The Journal of Chemical Physics}, 123(14):144114, 2005.

\bibitem{weinan2007nested}
E~Weinan, Di~Liu, and Eric Vanden-Eijnden.
\newblock Nested stochastic simulation algorithms for chemical kinetic systems
  with multiple time scales.
\newblock {\em Journal of Computational Physics}, 221(1):158--180, 2007.

\bibitem{gomez2008enhanced}
Carlos~A G{\'o}mez-Uribe, George~C Verghese, and Abraham~R Tzafriri.
\newblock Enhanced identification and exploitation of time scales for model
  reduction in stochastic chemical kinetics.
\newblock {\em The Journal of Chemical Physics}, 129(24):244112, 2008.

\bibitem{pigolotti2008coarse}
Simone Pigolotti and Angelo Vulpiani.
\newblock Coarse graining of master equations with fast and slow states.
\newblock {\em The Journal of Chemical Physics}, 128(15):154114, 2008.

\bibitem{chevalier2009rigorous}
Michael~W Chevalier and Hana El-Samad.
\newblock A rigorous framework for multiscale simulation of stochastic cellular
  networks.
\newblock {\em The Journal of Chemical Physics}, 131(5):054102, 2009.

\bibitem{cotter2011constrained}
Simon~L Cotter, Konstantinos~C Zygalakis, Ioannis~G Kevrekidis, and Radek
  Erban.
\newblock A constrained approach to multiscale stochastic simulation of
  chemically reacting systems.
\newblock {\em The Journal of Chemical Physics}, 135(9):094102, 2011.

\bibitem{kang2013separation}
Hye-Won Kang, Thomas~G Kurtz, et~al.
\newblock Separation of time-scales and model reduction for stochastic reaction
  networks.
\newblock {\em The Annals of Applied Probability}, 23(2):529--583, 2013.

\bibitem{bortolussi2015efficient}
Luca Bortolussi, Dimitrios Milios, and Guido Sanguinetti.
\newblock Efficient stochastic simulation of systems with multiple time scales
  via statistical abstraction.
\newblock In {\em International Conference on Computational Methods in Systems
  Biology}, pages 40--51. Springer, 2015.

\bibitem{shahrezaei2008analytical}
Vahid Shahrezaei and Peter~S Swain.
\newblock Analytical distributions for stochastic gene expression.
\newblock {\em Proceedings of the National Academy of Sciences},
  105(45):17256--17261, 2008.

\bibitem{bokes2012multiscale}
Pavol Bokes, John~R King, Andrew~TA Wood, and Matthew Loose.
\newblock Multiscale stochastic modelling of gene expression.
\newblock {\em Journal of Mathematical Biology}, 65(3):493--520, 2012.

\bibitem{melykuti2014equilibrium}
Bence M{\'e}lyk{\'u}ti, Joao~P Hespanha, and Mustafa Khammash.
\newblock Equilibrium distributions of simple biochemical reaction systems for
  time-scale separation in stochastic reaction networks.
\newblock {\em Journal of The Royal Society Interface}, 11(97):20140054, 2014.

\bibitem{popovic2016geometric}
Nikola Popovi{\'c}, Carsten Marr, and Peter~S Swain.
\newblock A geometric analysis of fast-slow models for stochastic gene
  expression.
\newblock {\em Journal of Mathematical Biology}, 72(1-2):87--122, 2016.

\bibitem{haseltine2002approximate}
Eric~L Haseltine and James~B Rawlings.
\newblock Approximate simulation of coupled fast and slow reactions for
  stochastic chemical kinetics.
\newblock {\em The Journal of Chemical Physics}, 117(15):6959--6969, 2002.

\bibitem{salis2005accurate}
Howard Salis and Yiannis Kaznessis.
\newblock Accurate hybrid stochastic simulation of a system of coupled chemical
  or biochemical reactions.
\newblock {\em The Journal of Chemical Physics}, 122(5):054103, 2005.

\bibitem{pelevs2006reduction}
Slaven Pele{\v{s}}, Brian Munsky, and Mustafa Khammash.
\newblock Reduction and solution of the chemical master equation using time
  scale separation and finite state projection.
\newblock {\em The Journal of Chemical Physics}, 125(20):204104, 2006.

\bibitem{puchalka2004bridging}
Jacek Pucha{\l}ka and Andrzej~M Kierzek.
\newblock Bridging the gap between stochastic and deterministic regimes in the
  kinetic simulations of the biochemical reaction networks.
\newblock {\em Biophysical Journal}, 86(3):1357--1372, 2004.

\bibitem{pahlajani2011stochastic}
Chetan~D Pahlajani, Paul~J Atzberger, and Mustafa Khammash.
\newblock Stochastic reduction method for biological chemical kinetics using
  time-scale separation.
\newblock {\em Journal of Theoretical Biology}, 272(1):96--112, 2011.

\bibitem{thomas2012rigorous}
Philipp Thomas, Ramon Grima, and Arthur~V Straube.
\newblock Rigorous elimination of fast stochastic variables from the linear
  noise approximation using projection operators.
\newblock {\em Physical Review E}, 86(4):041110, 2012.

\bibitem{sinitsyn2009adiabatic}
NA~Sinitsyn, Nicolas Hengartner, and Ilya Nemenman.
\newblock Adiabatic coarse-graining and simulations of stochastic biochemical
  networks.
\newblock {\em Proceedings of the National Academy of Sciences},
  106(26):10546--10551, 2009.

\bibitem{burrage2004multi}
Kevin Burrage, Tianhai Tian, and Pamela Burrage.
\newblock A multi-scaled approach for simulating chemical reaction systems.
\newblock {\em Progress in Biophysics and Molecular Biology}, 85(2):217--234,
  2004.

\bibitem{thomas2014phenotypic}
Philipp Thomas, Nikola Popovi{\'c}, and Ramon Grima.
\newblock Phenotypic switching in gene regulatory networks.
\newblock {\em Proceedings of the National Academy of Sciences},
  111(19):6994--6999, 2014.

\bibitem{vasudeva2004adaptive}
Karan Vasudeva and Upinder~S Bhalla.
\newblock Adaptive stochastic-deterministic chemical kinetic simulations.
\newblock {\em Bioinformatics}, 20(1):78--84, 2004.

\bibitem{alfonsi2004exact}
Aur{\'e}lien Alfonsi, Eric Cances, Gabriel Turinici, Barbara Di~Ventura, and
  Wilhelm Huisinga.
\newblock Exact simulation of hybrid stochastic and deterministic models for
  biochemical systems.
\newblock {\em Rr-5435, INRIA-Rocquencourt}, 2004.

\bibitem{takahashi2004multi}
Kouichi Takahashi, Kazunari Kaizu, Bin Hu, and Masaru Tomita.
\newblock A multi-algorithm, multi-timescale method for cell simulation.
\newblock {\em Bioinformatics}, 20(4):538--546, 2004.

\bibitem{neogi2004dynamic}
Natasha~A Neogi.
\newblock Dynamic partitioning of large discrete event biological systems for
  hybrid simulation and analysis.
\newblock In {\em Hybrid Systems: Computation and Control}, pages 463--476.
  Springer, 2004.

\bibitem{bentele2004general}
Martin Bentele and Roland Eils.
\newblock General stochastic hybrid method for the simulation of chemical
  reaction processes in cells.
\newblock In {\em Computational Methods in Systems Biology}, pages 248--251.
  Springer, 2004.

\bibitem{harris2006partitioned}
Leonard~A Harris and Paulette Clancy.
\newblock A ``partitioned leaping'' approach for multiscale modeling of
  chemical reaction dynamics.
\newblock {\em The Journal of Chemical Physics}, 125(14):144107, 2006.

\bibitem{hellander2007hybrid}
Andreas Hellander and Per L{\"o}tstedt.
\newblock Hybrid method for the chemical master equation.
\newblock {\em Journal of Computational Physics}, 227(1):100--122, 2007.

\bibitem{crudu2009hybrid}
Alina Crudu, Arnaud Debussche, and Ovidiu Radulescu.
\newblock Hybrid stochastic simplifications for multiscale gene networks.
\newblock {\em BMC Systems Biology}, 3(1):1, 2009.

\bibitem{menz2012hybrid}
Stephan Menz, Juan~C Latorre, Christof Schutte, and Wilhelm Huisinga.
\newblock Hybrid {S}tochastic--{D}eterministic {S}olution of the {C}hemical
  {M}aster {E}quation.
\newblock {\em Multiscale Modeling \& Simulation}, 10(4):1232--1262, 2012.

\bibitem{ganguly2015jump}
Arnab Ganguly, Derya Altintan, and Heinz Koeppl.
\newblock Jump-diffusion approximation of stochastic reaction dynamics: error
  bounds and algorithms.
\newblock {\em Multiscale Modeling \& Simulation}, 13(4):1390--1419, 2015.

\bibitem{duncan2015hybrid}
Andrew Duncan, Radek Erban, and Konstantinos Zygalakis.
\newblock Hybrid framework for the simulation of stochastic chemical kinetics.
\newblock {\em arXiv preprint arXiv:1512.03988}, 2015.

\bibitem{safta2015hybrid}
Cosmin Safta, Khachik Sargsyan, Bert Debusschere, and Habib~N Najm.
\newblock Hybrid discrete/continuum algorithms for stochastic reaction
  networks.
\newblock {\em Journal of Computational Physics}, 281:177--198, 2015.

\bibitem{hepp2015adaptive}
Benjamin Hepp, Ankit Gupta, and Mustafa Khammash.
\newblock Adaptive hybrid simulations for multiscale stochastic reaction
  networks.
\newblock {\em The Journal of Chemical Physics}, 142(3):034118, 2015.

\bibitem{angius2015approximate}
Alessio Angius, Gianfranco Balbo, Marco Beccuti, Enrico Bibbona, Andras
  Horvath, and Roberta Sirovich.
\newblock Approximate analysis of biological systems by hybrid switching jump
  diffusion.
\newblock {\em Theoretical Computer Science}, 587:49--72, 2015.

\bibitem{jahnke2011reduced}
Tobias Jahnke.
\newblock On reduced models for the chemical master equation.
\newblock {\em Multiscale Modeling \& Simulation}, 9(4):1646--1676, 2011.

\bibitem{jahnke2012error}
Tobias Jahnke and Michael Kreim.
\newblock Error bound for piecewise deterministic processes modeling stochastic
  reaction systems.
\newblock {\em Multiscale Modeling \& Simulation}, 10(4):1119--1147, 2012.

\bibitem{crudu2012convergence}
Alina Crudu, Arnaud Debussche, Aur{\'e}lie Muller, Ovidiu Radulescu, et~al.
\newblock Convergence of stochastic gene networks to hybrid piecewise
  deterministic processes.
\newblock {\em The Annals of Applied Probability}, 22(5):1822--1859, 2012.

\bibitem{cotter2016error}
Simon~L Cotter and Radek Erban.
\newblock Error analysis of diffusion approximation methods for multiscale
  systems in reaction kinetics.
\newblock {\em SIAM Journal on Scientific Computing}, 38(1):B144--B163, 2016.

\bibitem{hasenauer2014method}
J~Hasenauer, V~Wolf, A~Kazeroonian, and FJ~Theis.
\newblock Method of conditional moments ({MCM}) for the chemical master
  equation.
\newblock {\em Journal of Mathematical Biology}, 69(3):687--735, 2014.

\bibitem{soltani2015conditional}
Mohammad Soltani, Cesar~Augusto Vargas-Garcia, and Abhyudai Singh.
\newblock Conditional {M}oment {C}losure {S}chemes for {S}tudying {S}tochastic
  {D}ynamics of {G}enetic {C}ircuits.
\newblock {\em IEEE Transactions on Biomedical Circuits and Systems},
  9(4):518--526, 2015.

\bibitem{smith2015model}
Stephen Smith, Claudia Cianci, and Ramon Grima.
\newblock Model reduction for stochastic chemical systems with abundant
  species.
\newblock {\em The Journal of Chemical Physics}, 143(21):214105, 2015.

\bibitem{thattai2001intrinsic}
Mukund Thattai and Alexander Van~Oudenaarden.
\newblock Intrinsic noise in gene regulatory networks.
\newblock {\em Proceedings of the National Academy of Sciences},
  98(15):8614--8619, 2001.

\bibitem{bortolussi2013learning}
Luca Bortolussi and Guido Sanguinetti.
\newblock Learning and designing stochastic processes from logical constraints.
\newblock In {\em QEST}, volume 8054, pages 89--105. Springer, 2013.

\bibitem{cseke2013approximate}
Botond Cseke, Manfred Opper, and Guido Sanguinetti.
\newblock Approximate inference in latent {G}aussian-{M}arkov models from
  continuous time observations.
\newblock In {\em Advances in Neural Information Processing Systems}, pages
  971--979, 2013.

\bibitem{zechner2014scalable}
Christoph Zechner, Michael Unger, Serge Pelet, Matthias Peter, and Heinz
  Koeppl.
\newblock Scalable inference of heterogeneous reaction kinetics from pooled
  single-cell recordings.
\newblock {\em Nature Methods}, 11(2):197--202, 2014.

\bibitem{bishop2006pattern}
Christopher~M Bishop.
\newblock Pattern recognition and machine learning.
\newblock {\em Machine Learning}, 2006.

\bibitem{rabiner1986introduction}
Lawrence Rabiner and B~Juang.
\newblock An introduction to hidden {M}arkov models.
\newblock {\em ieee assp magazine}, 3(1):4--16, 1986.

\bibitem{kalman1960new}
Rudolph~Emil Kalman.
\newblock A new approach to linear filtering and prediction problems.
\newblock {\em Journal of basic Engineering}, 82(1):35--45, 1960.

\bibitem{fruhwirth1994data}
Sylvia Fr{\"u}hwirth-Schnatter.
\newblock Data augmentation and dynamic linear models.
\newblock {\em Journal of Time Series Analysis}, 15(2):183--202, 1994.

\bibitem{opper2008variational}
Manfred Opper and Guido Sanguinetti.
\newblock Variational inference for {M}arkov jump processes.
\newblock In {\em Advances in Neural Information Processing Systems}, pages
  1105--1112, 2008.

\bibitem{rao2013fast}
Vinayak Rao and Yee~Whye Teh.
\newblock Fast {MCMC} sampling for {M}arkov jump processes and extensions.
\newblock {\em The Journal of Machine Learning Research}, 14(1):3295--3320,
  2013.

\bibitem{georgoulas2015unbiased}
Anastasis Georgoulas, Jane Hillston, and Guido Sanguinetti.
\newblock Unbiased {B}ayesian inference for population {M}arkov jump processes
  via random truncations.
\newblock {\em Statistics and Computing}, 2016.

\bibitem{boys2008bayesian}
Richard~J Boys, Darren~J Wilkinson, and Thomas~BL Kirkwood.
\newblock Bayesian inference for a discretely observed stochastic kinetic
  model.
\newblock {\em Statistics and Computing}, 18(2):125--135, 2008.

\bibitem{Golightly2011}
Andrew Golightly and Darren~J Wilkinson.
\newblock Bayesian parameter inference for stochastic biochemical network
  models using particle {M}arkov chain {M}onte {C}arlo.
\newblock {\em Interface Focus}, page rsfs20110047, 2011.

\bibitem{Golightly2005}
Andrew Golightly and Darren~J Wilkinson.
\newblock Bayesian inference for stochastic kinetic models using a diffusion
  approximation.
\newblock {\em Biometrics}, 61(3):781--788, 2005.

\bibitem{Ruttor2009}
Andreas Ruttor and Manfred Opper.
\newblock Efficient statistical inference for stochastic reaction processes.
\newblock {\em Physical Review Letters}, 103(23):230601, 2009.

\bibitem{stathopoulos2013markov}
Vassilios Stathopoulos and Mark~A Girolami.
\newblock Markov chain {M}onte {C}arlo inference for {M}arkov jump processes
  via the linear noise approximation.
\newblock {\em Philosophical Transactions of the Royal Society of London A:
  Mathematical, Physical and Engineering Sciences}, 371(1984):20110541, 2013.

\bibitem{Fearnhead2014}
Paul Fearnhead, Vasilieos Giagos, and Chris Sherlock.
\newblock Inference for reaction networks using the linear noise approximation.
\newblock {\em Biometrics}, 70(2):457--466, 2014.

\bibitem{Milner2013}
Peter Milner, Colin~S Gillespie, and Darren~J Wilkinson.
\newblock Moment closure based parameter inference of stochastic kinetic
  models.
\newblock {\em Statistics and Computing}, 23(2):287--295, 2013.

\bibitem{cseke2015expectation}
Botond Cseke, David Schnoerr, Manfred Opper, and Guido Sanguinetti.
\newblock Expectation propagation for continuous time stochastic processes.
\newblock {\em Journal of Physics A: Mathematical and Theoretical},
  49(49):494002, 2016.

\bibitem{Zechner2012}
Christoph Zechner, Jakob Ruess, Peter Krenn, Serge Pelet, Matthias Peter, John
  Lygeros, and Heinz Koeppl.
\newblock Moment-based inference predicts bimodality in transient gene
  expression.
\newblock {\em Proceedings of the National Academy of Sciences},
  109(21):8340--8345, 2012.

\bibitem{stimberg2011inference}
Florian Stimberg, Manfred Opper, Guido Sanguinetti, and Andreas Ruttor.
\newblock Inference in continuous-time change-point models.
\newblock In {\em Advances in Neural Information Processing Systems}, pages
  2717--2725, 2011.

\bibitem{ocone2011reconstructing}
Andrea Ocone and Guido Sanguinetti.
\newblock Reconstructing transcription factor activities in hierarchical
  transcription network motifs.
\newblock {\em Bioinformatics}, 27(20):2873--2879, 2011.

\bibitem{ocone2013hybrid}
Andrea Ocone, Andrew~J Millar, and Guido Sanguinetti.
\newblock Hybrid regulatory models: a statistically tractable approach to model
  regulatory network dynamics.
\newblock {\em Bioinformatics}, 29(7):910--916, 2013.

\bibitem{sherlock2014bayesian}
Chris Sherlock, Andrew Golightly, and Colin~S Gillespie.
\newblock Bayesian inference for hybrid discrete-continuous stochastic kinetic
  models.
\newblock {\em Inverse Problems}, 30(11):114005, 2014.

\bibitem{hey2015stochastic}
Kirsty~L Hey, Hiroshi Momiji, Karen Featherstone, Julian~RE Davis, Michael~RH
  White, David~A Rand, and B{\"a}rbel Finkenst{\"a}dt.
\newblock A stochastic transcriptional switch model for single cell imaging
  data.
\newblock {\em Biostatistics}, 16(4):655--669, 2015.

\bibitem{liepe2010abc}
Juliane Liepe, Chris Barnes, Erika Cule, Kamil Erguler, Paul Kirk, Tina Toni,
  and Michael~PH Stumpf.
\newblock {ABC}-{S}ys{B}io--approximate {}bayesian computation in python with
  {GPU} support.
\newblock {\em Bioinformatics}, 26(14):1797--1799, 2010.

\bibitem{chen2010calibayes}
Yuhui Chen, Conor Lawless, Colin~S Gillespie, Jake Wu, Richard~J Boys, and
  Darren~J Wilkinson.
\newblock Cali{B}ayes and {BASIS}: integrated tools for the calibration,
  simulation and storage of biological simulation models.
\newblock {\em Briefings in Bioinformatics}, 11(3):278--289, 2010.

\bibitem{Doi1976}
Masao Doi.
\newblock Second quantization representation for classical many-particle
  system.
\newblock {\em Journal of Physics A: Mathematical and General}, 9(9):1465,
  1976.

\bibitem{Doi1976b}
Masao Doi.
\newblock Stochastic theory of diffusion-controlled reaction.
\newblock {\em Journal of Physics A: Mathematical and General}, 9(9):1479,
  1976.

\bibitem{Erban2009}
Radek Erban and S~Jonathan Chapman.
\newblock Stochastic modelling of reaction-diffusion processes: algorithms for
  bimolecular reactions.
\newblock {\em Physical Biology}, 6(4):046001, 2009.

\bibitem{Zon2005}
Jeroen~S van Zon and Pieter~R Ten~Wolde.
\newblock Simulating biochemical networks at the particle level and in time and
  space: {G}reen's function reaction dynamics.
\newblock {\em Physical Review Letters}, 94(12):128103, 2005.

\bibitem{Donev2010}
Aleksandar Donev, Vasily~V Bulatov, Tomas Oppelstrup, George~H Gilmer, Babak
  Sadigh, and Malvin~H Kalos.
\newblock A first-passage kinetic {M}onte {C}arlo algorithm for complex
  diffusion-reaction systems.
\newblock {\em Journal of Computational Physics}, 229(9):3214--3236, 2010.

\bibitem{Isaacson2008}
Samuel~A Isaacson.
\newblock Relationship between the reaction-diffusion master equation and
  particle tracking models.
\newblock {\em Journal of Physics A: Mathematical and Theoretical},
  41(6):065003, 2008.

\bibitem{Fu2014}
Jin Fu, Sheng Wu, Hong Li, and Linda~R Petzold.
\newblock The time dependent propensity function for acceleration of spatial
  stochastic simulation of reaction-diffusion systems.
\newblock {\em Journal of Computational Physics}, 274:524--549, 2014.

\bibitem{isaacson2009reaction}
Samuel~A Isaacson.
\newblock The reaction-diffusion master equation as an asymptotic approximation
  of diffusion to a small target.
\newblock {\em SIAM Journal on Applied Mathematics}, 70(1):77--111, 2009.

\bibitem{lipkova2011analysis}
Jana Lipkov{\'a}, Konstantinos~C Zygalakis, S~Jonathan Chapman, and Radek
  Erban.
\newblock Analysis of brownian dynamics simulations of reversible bimolecular
  reactions.
\newblock {\em SIAM Journal on Applied Mathematics}, 71(3):714--730, 2011.

\bibitem{hellander2012reaction}
Stefan Hellander, Andreas Hellander, and Linda Petzold.
\newblock Reaction-diffusion master equation in the microscopic limit.
\newblock {\em Physical Review E}, 85(4):042901, 2012.

\bibitem{mahmutovic2012lost}
Anel Mahmutovic, David Fange, Otto~G Berg, and Johan Elf.
\newblock Lost in presumption: stochastic reactions in spatial models.
\newblock {\em Nature methods}, 9(12):1163--1166, 2012.

\bibitem{isaacson2013convergent}
Samuel~A Isaacson.
\newblock A convergent reaction-diffusion master equation.
\newblock {\em The Journal of Chemical Physics}, 139(5):054101, 2013.

\bibitem{taylor2014deriving}
PR~Taylor, RE~Baker, and CA~Yates.
\newblock Deriving appropriate boundary conditions, and accelerating
  position-jump simulations, of diffusion using non-local jumping.
\newblock {\em Physical biology}, 12(1):016006, 2014.

\bibitem{scott2010approximating}
Matthew Scott, Francis~J Poulin, and Herbert Tang.
\newblock Approximating intrinsic noise in continuous multispecies models.
\newblock In {\em Proceedings of the Royal Society of London A: Mathematical,
  Physical and Engineering Sciences}, page rspa20100275. The Royal Society,
  2010.

\bibitem{asllani2013linear}
Malbor Asllani, Tommaso Biancalani, Duccio Fanelli, and Alan~J McKane.
\newblock The linear noise approximation for reaction-diffusion systems on
  networks.
\newblock {\em The European Physical Journal B}, 86(11):1--10, 2013.

\bibitem{smith2016analytical}
Stephen Smith, Claudia Cianci, and Ramon Grima.
\newblock Analytical approximations for spatial stochastic gene expression in
  single cells and tissues.
\newblock {\em Journal of The Royal Society Interface}, 13(118):20151051, 2016.

\bibitem{Holmes2012}
Geoffrey~R Holmes, Sean~R Anderson, Giles Dixon, Anne~L Robertson,
  Constantino~C Reyes-Aldasoro, Stephen~A Billings, Stephen~A Renshaw, and
  Visakan Kadirkamanathan.
\newblock Repelled from the wound, or randomly dispersed? {R}everse migration
  behaviour of neutrophils characterized by dynamic modelling.
\newblock {\em Journal of The Royal Society Interface}, page rsif20120542,
  2012.

\bibitem{Dewar2010}
Michael~A Dewar, Visakan Kadirkamanathan, Manfred Opper, and Guido Sanguinetti.
\newblock Parameter estimation and inference for stochastic reaction-diffusion
  systems: application to morphogenesis in {D}. melanogaster.
\newblock {\em BMC Systems Biology}, 4(1):1, 2010.

\bibitem{Ruttor2010}
Andreas Ruttor and Manfred Opper.
\newblock Approximate parameter inference in a stochastic reaction-diffusion
  model.
\newblock In {\em International Conference on Artificial Intelligence and
  Statistics}, volume~9, pages 669--676, 2010.

\bibitem{Cressie2011}
Noel Cressie and Christopher~K Wikle.
\newblock {\em Statistics for spatio-temporal data}.
\newblock John Wiley \& Sons, 2015.

\end{thebibliography}
\bibliographystyle{unsrt}
}

\end{document}